\documentclass[useAMS,usenatbib,usegraphicx]{mn2e}
\usepackage{pslatex,color,amsmath,amsfonts,amssymb,amsxtra,eufrak,url,caption}
\def\idm#1{{\mbox{\scriptsize #1}}}
\def\vec#1{{\pmb #1}}
\def\kepler{{\em Kepler}}
\def\Y{\langle Y \rangle}

\def\mean#1{\langle{}#1{}\rangle}
\def\tv#1{{\pmb #1}}

\newcommand{\au}{\mbox{au}} 
\newcommand{\msun}{\mbox{m}_{\odot}}

\newcommand{\mE}{\mbox{m}_{\oplus}}

\newcommand{\Mmean}{\mathcal{M}}
\newcommand{\tepoch}{t_0}

\newcommand{\Pc}{P_{\idm{2}}}

\newcommand{\yr}{\mbox{yr}}
\newcommand{\ab}{a_{\idm{1}}}

\newcommand{\eb}{e_{\idm{1}}}

\def\kepler{{\sc Kepler}}
\def\idm#1{{\mbox{\scriptsize #1}}}
\newcommand\Chi{{\chi^2_\nu}}

\title[The origin and dynamics of the Kepler-29 system]%
{The origin and 9:7 MMR dynamics of the Kepler-29 system}
\author[Migaszewski, Go\'zdziewski \& Panichi]%
{Cezary Migaszewski$^{1,2\star}$, Krzysztof Go\'zdziewski$^{2\star}$ \& Federico Panichi$^{1\star}$\\
$^{1}$Institute of Physics and CASA*, Faculty of Mathematics and Physics, University of Szczecin, Wielkopolska 15, PL-70-451 Szczecin, Poland\\
$^{2}$Centre for Astronomy, Faculty of Physics, Astronomy and Informatics, Nicolaus Copernicus University, Grudziadzka 5, PL-87-100 Toru\'n, Poland}
\begin{document}
%
\date{Accepted 2016 November 3. Received 2016 November 2; in original form 2016 September 28.}
\pagerange{\pageref{firstpage}--\pageref{lastpage}} \pubyear{2016}
\maketitle
\label{firstpage}

\captionsetup[figure]{labelfont=bf,font=small}

\begin{abstract}
We analyse the Transit Timing Variation (TTV) measurements of a~system of two
super-Earths detected as Kepler-29, in order to constrain the planets' masses
and orbital parameters. A dynamical analysis of the best-fitting configurations
constrains the masses to be $\sim 6$ and $\sim 5$ Earth masses for the inner and
the outer planets, respectively. The analysis also reveals that the system is
likely locked in the 9:7~mean motion resonance. However, a variety of orbital
architectures regarding eccentricities and the relative orientation of orbits is
permitted by the observations as well as by stability constraints. We attempt to
find configurations preferred by the planet formation scenarios as an
additional, physical constraint. We show that configurations with low
eccentricities and anti-aligned apsidal lines of the orbits are a natural and
most likely outcome of the convergent migration. However, we show that
librations of the critical angles are not necessary for the Kepler-29
system to be dynamically resonant, and such configurations may be formed on the
way of migration as well. We argue, on the other hand, that aligned
configurations with $e \gtrsim 0.03$ may be not consistent with the
migration scenario. 
\end{abstract}


%
\section{Introduction}
%

\let\thefootnote\relax\footnote{$^{\star}$Email: migaszewski@umk.pl (CM), chris@umk.pl (KG), federico.panichi@stud.usz.edu.pl (FP)}

The \kepler{} mission has lead to the discovery of a~few hundred multiple
planetary systems with super-Earth planets. Some of those systems are very
compact and exhibit orbital period ratios close to small rational numbers. This
may indicate their proximity to low-order mean motion resonances (MMRs)
\citep[e.g.][]{Fabrycky2013}. This is not yet a fully resolved issue, since most
of the \kepler{} systems are not sufficiently characterised, regarding both the planet's
masses and orbital architectures. Dynamical modelling of Transit Timing
Variation measurements \citep{Agol2005} or the photodynamical method
\citep{Carter2011}, which account for the mutual $N$-body interactions, are
common and usually the only approaches making it possible to model multiple
systems in the \kepler{} sample \citep{Rowe2015,Mullally2015,Holczer2016}.
A~further difficulty is a~relatively narrow time-window of observations and low
signal-to-noise ratio that typically lead to weakly constrained eccentricities
and longitudes of pericenter. Therefore, definite conclusions about orbital
architectures of such systems are hard to derive unless {\em a priori}
constraints are imposed, like requirements of dynamical stability and evolution
consistent with the planetary migration. Nevertheless, the TTV method is the
major technique making it possible to determine the dynamical masses, if
spectroscopic measurements could not be made for faint or/and chromospherically
active stars.

In this paper we aim to characterise the dynamical architecture of
Kepler-29 (KOI-738) planetary system detected by \cite{Fabrycky2012}. We use the
TTV measurements spanning 17 quarters of \kepler{} long-cadence photometric
lightcurves \citep{Rowe2015}. Kepler-29 is composed of two super-Earth planets
with dynamical masses of $4.69$ and $4.16$~Earth masses, respectively
\citep{JontofHutter2016}. Their stability analysis has been restricted to a
relatively short-term direct $N$-body integration for a few~Myrs. We focus
rather on qualitative dynamical analysis of this resonant or near-resonant
system and consider the planetary migration as a possible formation scenario.

The paper is structured as follows. Section~2 is devoted to the dynamical model
of a co-planar Kepler-29 system constrained by the TTV measurements in
\citep{Rowe2015}. We aim to obtain a comprehensive view of the parameter space
with two independent optimisation methods: the Markov Chain Monte Carlo sampling
as well as with genetic and evolutionary algorithms. We show the results of the
stability analysis of plausible configurations with the long-term direct
numerical integrations and with the fast indicator technique. We found that the
planets may be in 9:7~MMR, although its presence and behaviour of critical
angles, as well as the stability depend on {\em a priori} set eccentricity
distribution. Different geometric configurations with librating or rotating
critical angles are permitted by both the observational and dynamical
constraints. Therefore, in Section~3 we attempt to construct a~global, analytic
approximation of the system close to the 9:7~MMR, and to verify whether or not
the best-fitting models may be formed on the way of planetary migration
(Section~4). Conclusions are given in the last section.
%
\section{The TTV data model and optimisation}
%
The mathematical TTV model in this paper is essentially the same as in our
previous work devoted to the Kepler-60 system \citep{Gozdziewski2016}. The TTV
data set ${\cal D}$ for Kepler-29 consists of 187 measurements spanning quarters
Q1-Q17 ($\simeq 1450$~days), with reported mean uncertainties
$\mean{{\sigma}^{\idm{TTV}}_1} \simeq 0.009$~d and
$\mean{{\sigma}^{\idm{TTV}}_2}\simeq 0.011$~d, for the inner and outer planet,
respectively \citep{Rowe2015}. The uncertainties are significant, given the full
range of the TTV measurements spans $\sim 0.12$~d and $\sim 0.15$~d,
respectively. Such a TTV variability is relatively clear as compared to other
2-planet systems in the \kepler{} sample published in the \citep{Rowe2015}
catalogue. Recently, \cite{Holczer2016} performed a~new re-analysis of
the \kepler{} data and they report 162 measurements for Kepler-29 that cover both the
long-cadence and, when available, short-cadence \kepler{} light-curves. While we performed
the TTV analysis of this data set after it has been published, we did not make an
extensive use of the results. Our orbital models derived on the basis of the
\citep{Rowe2015} catalogue fit the TTV time-series in \cite{Holczer2016} as
well. We have got qualitatively the same distributions of the best-fitting
parameters.

Given a relatively low signal-to-noise TTV data, the inclinations
$I_{1,2}$ and the nodal longitudes $\Omega_{1,2}$ cannot be constrained.
Moreover, since we focus on a hypothesis that the Kepler-29 system could have
originated through the convergent planetary migration, we assume that it is
coplanar or close to coplanar, and we fix $I_{1,2}=90^{\circ}$ and
$\Omega_{1,2}=0^{\circ}$. As we want to cover possibly wide range of eccentricities
$e_{1,2}$ and for $e_{1,2} \sim 0$ the longitudes of pericenter
$\varpi_{1,2}$ are weakly
constrained, we introduce non-singular, osculating, astrocentric elements
$\left\{P_i, x_i, y_i, T_i \right\}$ instead of $\left\{a_i, e_i, \omega_i,
\Mmean_i \right\}$, $i=1,2$:
\[
P_i = 2\,\pi \sqrt{\frac{a_i^3}{k^2\,(m_0 + m_i)}}, \quad
T_i = \tepoch + 
\frac{P_i}{2\pi} \left( \Mmean_i^{(\idm{t})} - \Mmean_i \right),
\]
and $x_i = e_i \, \cos\varpi_i$, $y_i = e_i \, \sin\varpi_i$, where $k$ is the
Gauss gravitational constant, $\Mmean_i^{(\idm{t})}$ is the mean anomaly at the
epoch of the first transit $T_i$, and ${\cal M}_i$, $P_i$, $a_i$ are for the
mean anomaly, the orbital period and semi-major axis, at the osculating epoch
$\tepoch$ for each planet, respectively. We computed the transits moments with
the {\sc TTVFast} package \citep{Deck2014} and with our own codes for an
independent check.

The least-squares fit to the TTV data with their raw uncertainties results in
solutions having large $\Chi \sim 2$. The scatter of residuals is roughly
symmetric, however its magnitude is significant w.r.t. the TTV signal itself.
Here, we assume that the TTV uncertainties are Gaussian and independent, which
may be justified since {\em a posteriori} Lomb-Scargle periodograms of the
residuals of best-fitting models did not show apparent, isolated frequencies.
Therefore, the large $\Chi$ might be explained by underestimated uncertainties.
To correct for this factor, we optimised the maximum likelihood function ${\cal
L}$:
\begin{equation}
 \log {\cal L}({\cal D}|\tv{\xi}) =   
-\frac{1}{2}\sum_{i,t}\frac{{[\mbox{O}({\cal D})-\mbox{C}(\tv{\xi})]}_{i,t}^2}{\sigma_{i,t}^2}
     -\frac{1}{2}\sum_{i,t} \log {\sigma_{i,t}^2} 
     -\frac{1}{2} N \log{2\pi},
\label{eq:Lfun}
\end{equation}
where $(\mbox{O-C})_{i,t}$ is the (O-C) deviation of the observed $t$-th transit
moment of an $i$-th planet from its $N$-body ephemeris determined through a
model parameters vector $\tv{\xi}$, and $N$ is the number of TTV measurements
encoded as data set ${\cal D}$. This more general form of ${\cal L}$ makes it
possible to determine the free parameter $\sigma_f$ that scales the TTV
uncertainties $\sigma_{i,t}$ in quadrature, such that $\sigma_{i,t}^2
\rightarrow \sigma_{i,t}^2+\sigma_f^2$ results in $\Chi \sim 1$. 

Because values of ${\cal L}$ are non-intuitive for comparing solutions,
therefore we define a quasi-r.m.s. measure of the fits quality,  
$
 \log L = \log 0.2420 - \log {\cal L}/N,
$
expressed in days. For statistically optimal solutions $\chi^2/N \sim 1$,
therefore $L \sim \mean{\sigma}$ is a scatter of measurements around the
best-fitting model \citep[e.g.,][]{Baluev2009}. We observed that in fact $L$
remains close to the usual r.m.s. goodness-of-fit measure.

A quasi-global optimisation of the dynamical model relies on investigating the
space of $11$ free parameters $\tv{\xi}$, which are the osculating elements
$(P_i, T_i, x_i, y_i)$, dynamical masses $m_i$, $i=1,2$, as well as ``the error
floor'' $\sigma_f$, common for all TTV measurements.

The Markov Chain Monte Carlo (MCMC) technique is widely used by the photometric
community to determine the posterior probability distribution ${\cal
P}(\tv{\xi}|{\cal D})$ of model parameters $\tv{\xi}$, given the data set ${\cal
D}$: 
$
   {\cal P}(\tv{\xi}|{\cal D}) 
  \propto {\cal P}(\tv{\xi}) {\cal P}({\cal D}|\tv{\xi}),
$
where ${\cal P}(\tv{\xi})$ is the prior, and the sampling data distribution
${\cal P}({\cal D}|\tv{\xi}) \equiv \log{\cal L}({\cal D}|\tv{\xi})$. For most
of the parameters, priors have been set as uniform (or uniform improper) through
imposing parameters ranges available for the exploration, i.e., $P_{i}>0$~d,
$T_{i}>0$~d, $m_i \in [0.0001,30]~\mE$, $\sigma_f>0$~d.

Choosing priors for $(x,y)$--elements is a more subtle matter. We already know
\citep[e.g.,][]{Hadden2014,JontofHutter2016} that these parameters are
unconstrained and biased towards large eccentricities, contrary to the physical,
{\em a priori} determined quasi-circular architecture of the system. Therefore,
besides uniform priors for $\xi \equiv x_1, x_2, y_1$ and~$y_2$, i.e., $\xi \in
(-0.48,0.48)$, we also examined Gaussian priors imposed on these parameters,
which are determined through
$
 P(\xi) = \exp (-(\xi-\overline{\xi})^2/\sigma_{\xi}^2),
$
with the zero mean value $\overline\xi$, and a few variances $\sigma_{\xi}=0.05,
0.1, 0.25, 0.33$, respectively. This approach is similar to that one used by
\cite{JontofHutter2016}, who argue that the eccentricity distribution for
multiple planetary systems is not uniform \citep{Moorhead2011,
Kane2012,Plavchan2014,VanEylen2015,Hadden2014}.

In order to perform the MCMC analysis, we prepared {\tt Python} interfaces to model
functions written in {\tt Fortran 90} and we used excellent {\tt emcee} package
of the affine-invariant ensemble sampler \cite{Goodman2010}, kindly provided by
\cite{Foreman2014}. 
As a second approach to CPU-effective exploration of weakly constrained
parameters space, we maximised the $\log {\cal L}$ function with genetic and
evolutionary algorithms \citep[GEA from hereafter,][]{Charbonneau1995,Izzo2010}.
We set similar parameter bounds as in the MCMC experiments. The GEA parameter
surveys are very useful to select starting solutions for the MCMC analysis,
which makes the sampling of presumably multi-modal
distributions more CPU efficient.

In order to characterise the dynamical stability of the solutions, we use the
fast indicator technique, so called Mean Exponential Growth factor of Nearby
Orbits \citep[MEGNO or $\Y$,][]{Cincotta2003}, an incarnation of the Maximal
Lyapunov Characteristic Exponent (mLCE). Since the period ratio derived from
the transit data indicate a system close to the 9:7~MMR, we also investigate
critical angles of this resonance
\begin{eqnarray}
\label{eq:angle}
\phi_1 &=& 7 \lambda_1 - 9 \lambda_2 + 2 \varpi_1, \nonumber\\
\phi_2 &=& 7 \lambda_1 - 9 \lambda_2 + 2 \varpi_2, \\
\phi_3 &=& 7 \lambda_1 - 9 \lambda_2 + \varpi_1 + \varpi_2.\nonumber
\end{eqnarray}
We also used the refined Fourier frequency analysis
\citep{Laskar1993,Sidlichovsky1996} which makes it possible to determine
fundamental frequencies of the system. We focus on the 9:7 MMR, hence the proper
mean motions are determined through the modified Fourier transform (FMFT) of the
time series
$
\left\{ a_i(t) \exp[ \mbox{i} \lambda_i(t)] \right\},
$ 
where $a_i(t$) and $\lambda_i(t)$ are the osculating semi-major axis and the
mean longitude, respectively. These canonical astrocentric elements are defined
as geometrical elements inferred from the Poincar\'e
coordinates~\cite[e.g.][]{Morbidelli2002}, sometimes called the democratic
heliocentric coordinates. 
We use the canonical elements only for this analysis internally in the
code, while in the fitting process and throughout the paper, the initial
conditions are parametrised through the usual, two-body astrocentric
osculating Keplerian elements.
%
%
\subsection{The best-fitting configurations}
%
%
\begin{figure*}
\centerline{
\vbox{
\hspace{1mm}\hbox{
\includegraphics[width=0.5\textwidth]{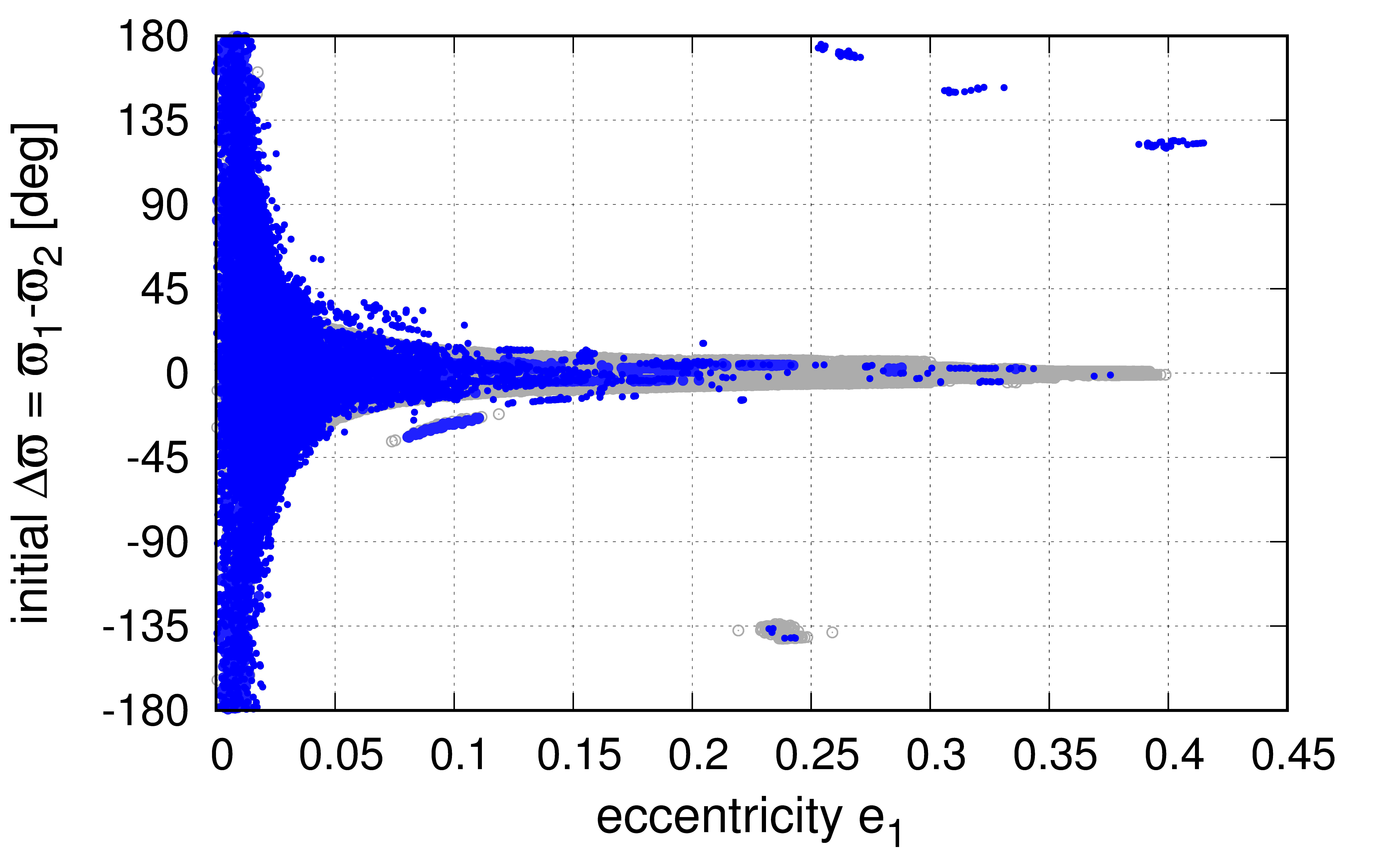}
\includegraphics[width=0.5\textwidth]{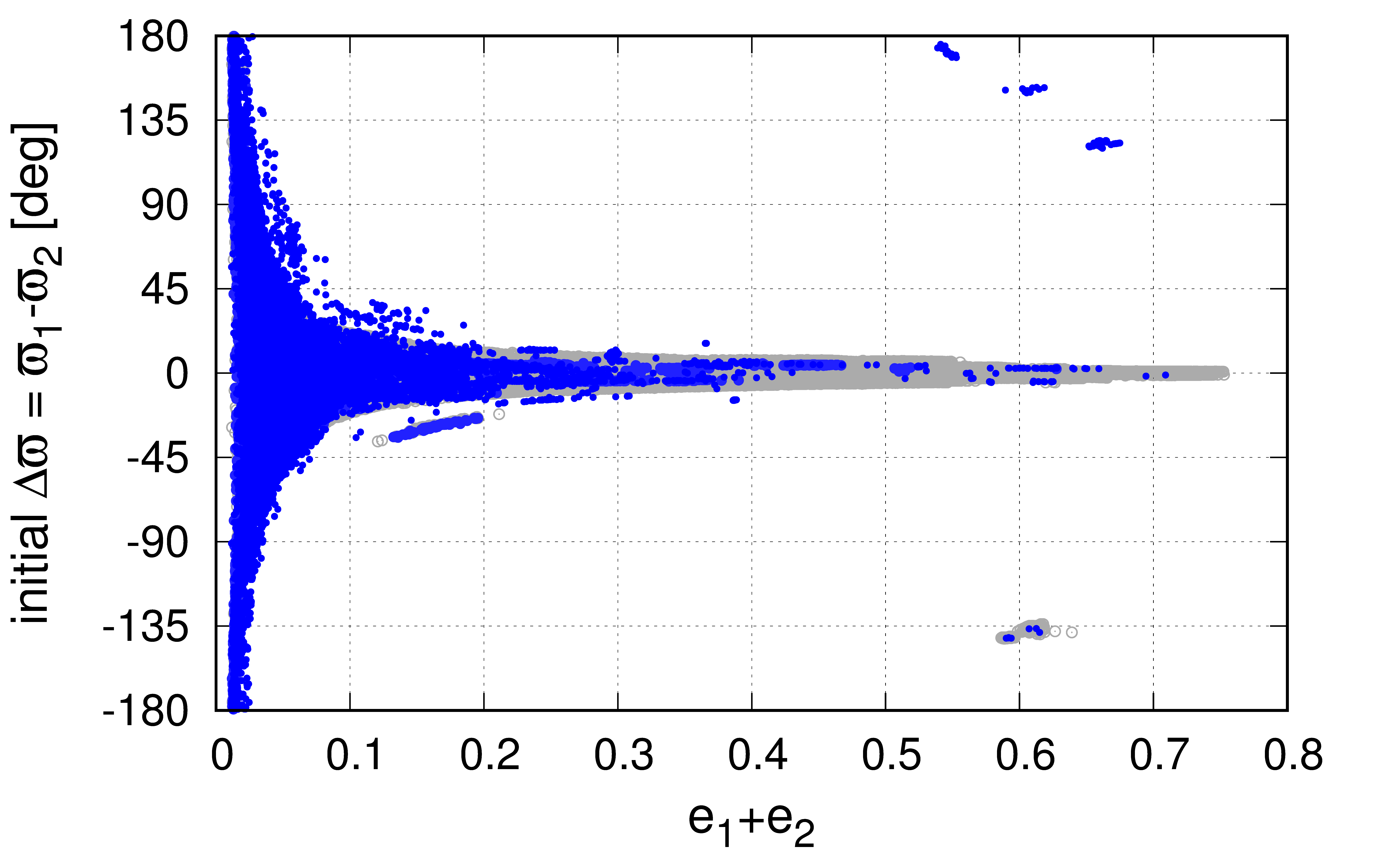}
}
}
}
\caption{
Solutions with $L<0.0145$~d (grey dots) derived through optimisation of the
maximum likelihood function with the GEA. Blue dots are for configurations that
result in MEGNO $\Y \sim 2$ integrated for 64~kyrs, indicating
regular solutions. Stable, high-eccentricity solutions are visible in small,
isolated ``clumps'', outside a region of $\Delta\varpi=0$.
}
\label{fig:fig1}
\end{figure*}

We first performed an extensive search with the GEA, collecting sets of $\sim
10^6$ solutions in each multi-CPU run. We did not impose any prior information
on the model parameters in this survey, however we should not expect that the
problem of unconstrained eccentricities could be avoided. Indeed, we found a
continuum of models with $L < L_\idm{min}= 0.0149$~d, well determined orbital
periods $P_i$ and transit epochs $T_i$, $i=1,2$. The error floor $\sigma_f \sim
0.01$~days is roughly uniform for all these solutions. The $(x_i,
y_i)$-parameters transformed to the $(e_i,\varpi_i)$-elements ($i=1,2$) form
a~pin-like structure in the $(e_1, \Delta\varpi)$- and $(e_1+e_2,
\Delta\varpi)$-planes, as shown in Fig.~\ref{fig:fig1}. The two panels look
similar, because $e_1 \approx e_2$ (see below). When the eccentricities reach
moderate values up to $\sim 0.05$, the TTV models are found close to
$\Delta\varpi=0$--axis. A similar effect has been observed for other Kepler
systems \citep{JontofHutter2016}. It is not clear whether the apsidal alignment
could be physical in the presence of planetary migration. As shown in
\citep{Xiang-Gruess2015} aligned configurations for second order MMRs can be
formed through migration, however that happens for the eccentricities
significantly different one from another, not for $e_1 \approx e_2$ like for
Kepler-29. Aligned configurations studied in the cited paper remind systems in
2:1~MMR that move, during the migration, along a branch of stable periodic
configuration and eventually change the libration centre of $\Delta\varpi$ from
$\pi$ to $0$ \citep[e.g.,][]{Ferraz-Mello2003}. When $e_1 \approx e_2$ much more
likely, and naturally emerging are resonant configurations with anti-aligned
apsides, which are also present in Fig.~\ref{fig:fig1} for {\em small}
eccentricities. 

Figure~\ref{fig:fig1} illustrates also the results of the stability analysis for
the models gathered. Due to relatively large masses of $\sim 6$~Earth masses,
and close orbits, significant mutual perturbations could be possible. Therefore,
for all solutions with $L<L_\idm{min}$ we computed their MEGNO signatures for
64~kyrs ($\sim 1.8\times 10^6$ outermost orbits). Such an interval well
covers the characteristic short-term dynamical time-scale associated with the
9:7~MMR. In the case of the Kepler-29, dynamically stable models may be found in
similarly wide range of the parameter plane. Curiously, stable solutions
that fit the TTV data exist for $e_{1,2}$ as large as 0.3-0.4. Besides
the pin-like structure, we also found isolated islands of high-eccentricity
solutions beyond the $\Delta\varpi=0$ axis. These models could be rather
associated with $\Delta\varpi$ librating around $\pi$.
\begin{figure*}
\centerline{\includegraphics[width=1.\textwidth]{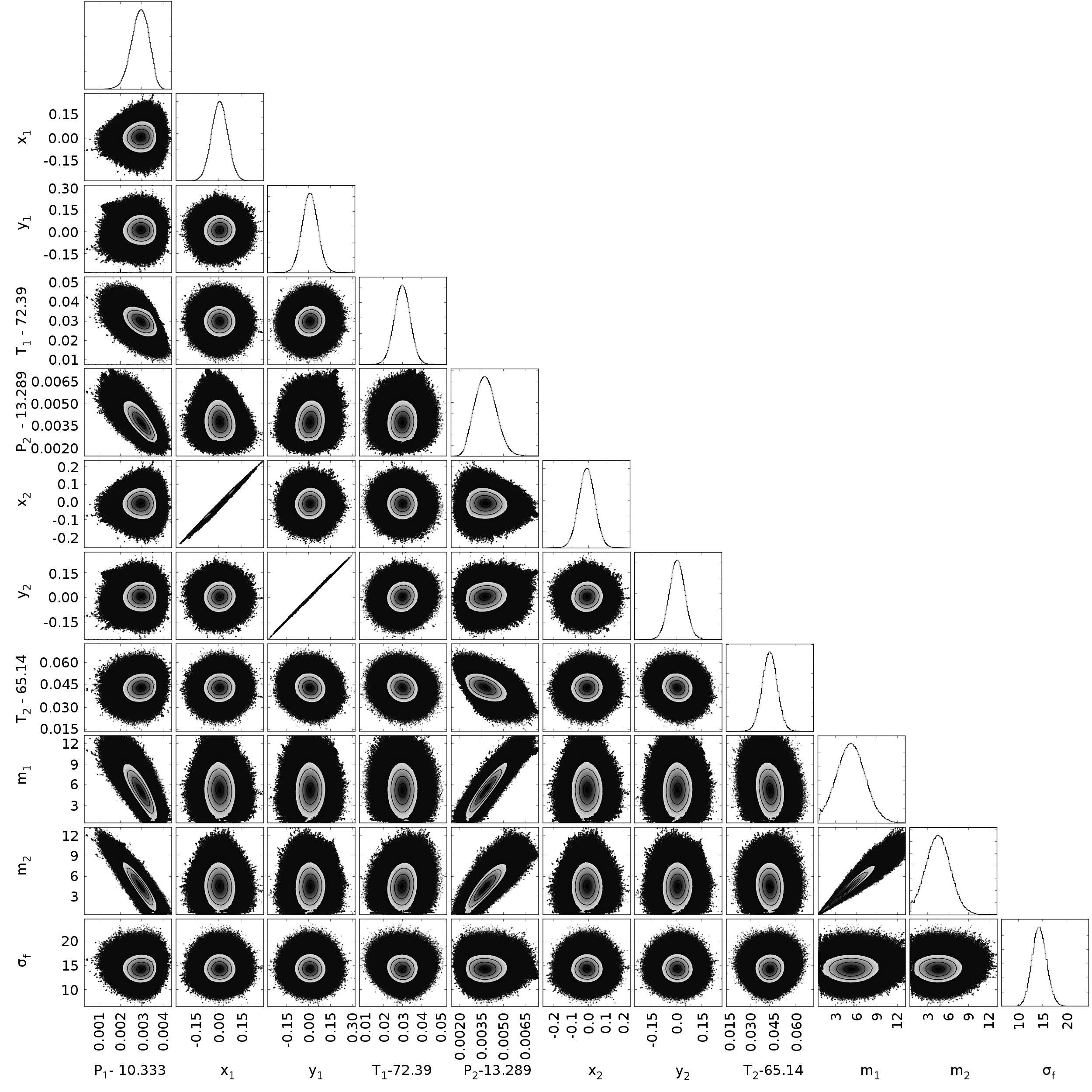}}
\caption{
One-- and two--dimensional projections of the posterior probability distribution
for all free parameters of the TTV model. The MCMC chain length is equivalent to
1,024,000 iterations initiated with 256 different instances in a small ball
centred at the best-fitting model found with the GEA. The $(x_{1,2},y_{1,2})$ prior is
Gaussian with the zero mean and the variance equal to 0.10. Parameters $T_i$
and $P_i$ are expressed in days, masses $m_i$ are expressed in Earth
masses ($i=1,2$), and the uncertainty correction term $\sigma_f$ is given in minutes.
Indices $1$ and $2$ are for the inner and outer planet, respectively. Contours
are for the 16th, 50th and 84th percentile of samples in the posterior
distribution. We removed about of 10\% initial, burn-out samples.
}
\label{fig:fig2}
\end{figure*}

\begin{table}
\caption{
Orbital parameters of a representative TTV model of the Kepler-29 
system. The osculating epoch is $t_0$=KBJD+64.0~days. The configuration is coplanar
with inclinations $I=90^{\circ}$ and nodal longitudes $\Omega=0^{\circ}$.
Mass of the star is $1.0\,\msun$ \citep{Rowe2015}. 
{Elements $(x_1, x_2)$ and $(y_1, y_2)$ are strongly correlated pairwise (see Fig.~\ref{fig:fig2}).
}
}
\label{tab:tab1}
\centerline{
\begin{tabular}{l r r}
\hline\hline
planet           & Kepler-29\,b & Kepler-29\,c \\
\hline\hline
\smallskip
$m_p\,[\mE]$     &  5.7$_{-1.9}^{+1.6}$   &   4.9$_{-1.6}^{+1.4}$ \\
\smallskip
$P\,$[d]         & 10.33585$_{-0.00032}^{+0.00039}$ 
                 & 13.29292$_{-0.00069}^{+0.00058}$  \\
$x\equiv e \cos\varpi$   & 0.0046 $\pm$ 0.062     & -0.009 $\pm$ 0.05  \\
$y\equiv e \sin\varpi$   & 0.0154 $\pm$ 0.062     &  0.008 $\pm$ 0.05  \\
$T\,$[d]         & 72.4200 $\pm$ 0.0041    
                 & 65.1830 $\pm$ 0.0048 \\
\hline
$a\,[\au]$       &  0.0928613    &   0.1098205  \\
$e$              &  0.0160    &   0.01169  \\
$\omega\,$[deg]       &  73.44    &   139.94   \\
${\cal M}\,$[deg]    &    -96.18  &  96.99    \\
\hline
$\sigma_f$\,[d]  & \multicolumn{2}{c}{0.009 $\pm$ 0.001} \\
$L$\,[d]         & \multicolumn{2}{c}{0.0145} \\
\hline
\end{tabular}
}
\end{table}

Therefore, even if the stability constraints are considered, we actually cannot
choose a~``proper'' or ``best-fitting'' configuration of the system. More tight 
constraints are required. At the first step, such constraints may be imposed by
a statistical eccentricity distribution expected for compact {\sc Kepler}
systems. 

We preformed MCMC experiments to account for the physical limits of $(x,y)$
elements. The results are illustrated in Fig.~\ref{fig:fig2}. It shows one-- and
two--dimensional projections of the posterior probability distribution for
Gaussian prior set to $(x,y)$ with zero mean and the variance equal to 0.1.
Computations were performed in multi-CPU environment, making it possible to
evaluate as much as 1,024,000 iterations to avoid the auto-correlation effect.
Each run composed of 256 {\tt emcee} ``walkers'' initiated in a~small ball
around low-$L$ solutions found in the GEA search. The posterior is uni-modal and
centred close to $(x,y) \simeq 0$ with masses $\sim 6~\mE$ and $\sim 5~\mE$ for
the inner and outer planet, respectively. The posterior distribution does not
change its character, i.e., relatively well determined peaks when the
$(x,y)$-prior has the variance set to 0.05, 0.1, 0.25 and 0.33, yet the masses
are strongly correlated. If the $(x,y)$-priors are uniform, masses and
$(x,y)$-elements are not constrained.

For all $(x,y)$ Gaussian priors, we found strong linear correlations between
pairs of $(x_1,x_2)$ and $(y_1,y_2)$. These linear correlations could mean
a~tight alignment of apsidal lines which is a common dynamical feature of the
low-order MMRs. This is likely a general effect discussed by
\cite{JontofHutter2016} for the first order MMRs. It may be explained by the
evolution of eccentricity vectors $[e_i \cos\varpi_i, e_i\sin\varpi_i]$, which
are not constrained individually, but their components are tightly correlated.

The orbital period ratio in the best-fitting solutions is very
close to 9/7, indicating a possible MMR. We searched for signatures of the 9:7
MMR by computing amplitudes of the critical angles of solutions sampled in the
MCMC experiments. When a candidate solution showed $L<0.0149$~d, we numerically
integrated the $N$-body equations of motion, and the full amplitudes of all
critical angles have been determined for $60$~yrs ($\sim 2000 P_2)$. The
amplitudes are expressed through
\[
\theta_\idm{min} = \min(\max \sin\phi-\min\sin\phi, \max\cos\phi-\min\cos\phi),
\]
where $\phi$ is for any of the three critical arguments of the 9:7~MMR, $\phi_i$, $i=1,2,3$ (Eq.~\ref{eq:angle}).

The results are illustrated in Fig.~\ref{fig:fig3} for three planes of orbital
parameters selected from samples with $L<0.0149$~d (grey dots). The grey area is
filled, as guaranteed by the MCMC sampling, hence we may be confident that the
search covers all the relevant parameter space. Solutions with
$\theta_\idm{min}<1.53$ (i.e., with  full amplitude
$\sim 1.53$ of $\cos$ or $\sin$)  are marked with blue filled circles. This experiment shows
that solutions with anti-aligned apsides are preferred for small eccentricities
$\simeq 0.01$, while models with librating critical angles and aligned apsides
are found mostly for moderate and large eccentricities.

Curiously, the distribution of models in the ($e_1,e_2$)--plane forms a strip
having a sharp ``cut-off'' at small eccentricities region ($e_1+e_2 \sim 0.01$).
This effect could be explained by measurable, mutual interactions of the
planets. Since the total angular momentum must be conserved, it implies
eccentricities variations in anti-phase. 

%
\subsection{The resonant character of the system}
%
%
We computed 2-dim dynamical maps in the neighbourhood of two representative
best-fitting solutions to visualise the dynamical structure of the 9:7~MMR.
First of these solutions is the result of the MCMC sampling for the
variance of $x_{1,2}, y_{1,2}$ set to $0.1$. Its parameters are given in
Tab.~\ref{tab:tab1} and the synthetic TTV signals are presented in
Fig.~\ref{fig:fig4}. The top-row of Fig.~\ref{fig:fig5} shows dynamical maps in
the $(\ab, \eb)$--plane for this model. All other orbital elements are kept at
their best-fitting values. For each initial condition at the grid, the $N$-body
equations of motion where integrated up to $36~$kyr. This corresponds to $\sim
10^6 \times \Pc$, sufficient to detect short-term chaotic motions for the MMRs
instability time-scale \citep[e.g.][]{Gozdziewski2014}. As the dynamical maps
show, the observational uncertainties of $a_1$ and $a_2$ $\simeq 10^{-5}\,\au$
are much smaller than the width of the 9:7~MMR, see Fig.~\ref{fig:fig5}. The
best-fitting solution is found, somehow ironically, just in a narrow unstable
structure which may be identified with the separatrix. The 9:7 MMR structure may
be even better visible in the right-hand panel of Fig.~\ref{fig:fig5} which is
for the frequency map of the system and shows deviations of the ratio of mean
motions $f_2/f_1 \equiv n_2/n_1$ (fundamental frequencies) from the exact 9/7
value. The 9:7 MMR spans the middle part of the map, depicted as wide
grey/yellow strip of regular motions limited by vertical separatrices. In the
middle of this strip, a boomerang-like structure appears with $n_2/n_1$
deviations from the 9/7 ratio as small as $10^{-6}$ and sharp borders coinciding
with curved, narrow separatrices identified in the MEGNO map.

We found that the critical angles $\phi_{1,2,3}$ are not fully adequate
signatures of the 9:7~MMR since they may librate with large full amplitudes
reaching~$2\pi$, even in the boomerang-like region characterised with
almost exact 9/7 ratio of the orbital periods.

Yet a highly ordered evolution of $(\Delta\varpi, \phi_1)$ during first
36~kyr ($\sim 10^6$ outermost periods) is illustrated in the top row
of Fig.~\ref{fig:fig6} for three initial conditions selected at maps in
Fig.~\ref{fig:fig5}, with the same elements as the best-fitting model,
besides changed $e_1$.  The left-hand panel is for $e_1=0.006$ (below the
lower separatrix of the boomerang-like structure), $e_1=0.019$ (close to the
nominal solution, between the separatrices), and $e_1=0.026$ (above the
upper separatrix).  In all these cases a clear, resonant behaviour of the
system is apparent which we understand here as a strong correlation of
the critical angles rather than low-amplitude librations of these critical
arguments.  The primary indication of the presence of the resonance are the
dynamical maps, Fig.~\ref{fig:fig5} and the particular, vertical
structure, which is common for MMRs in the $(a,e)$-plane.

There is also a clear difference between evolution of the critical angles
inside and outside the 9:7 MMR structure.  To show this, we integrated
two configurations with $a_1 = 0.09280\,\au$ and $a_2 = 0.09298\,\au$ (the
left and the right-hand panels of the bottom row of Fig.~\ref{fig:fig6},
respectively) that are located outside the resonance.  In contrast to highly
correlated behaviour of the angles inside the vertical structure,
illustrated in the top row of Fig.~\ref{fig:fig6}, in both these cases the
evolution of the angles is not ordered in the sense explained above.  Angle
$\phi_1$ can be equal to $0$ or $\pi$ when $\Delta\varpi$ equals $0$.  The
middle panel in the bottom row illustrates the resonant behaviour, however
the initial $a_1 = 0.09293\,\au$, so the system is very close to the
separatrix between resonant and non-resonant regions.  Although the
system evolves almost in a whole $(\Delta\varpi, \phi_1)$-plane, similarly
to the top row of Fig.~\ref{fig:fig6}, the angles are synchronised and when
$\Delta\varpi = 0$ or~$\pi$, $\phi_1$ cannot be $0$.

We also found clear semi-major axes oscillations, expected for
systems in MMR, whose amplitude is a few times larger in the MMR region, when compared to
solutions beyond the MMR structure (not shown here).  Moreover, the curved,
thin
chaotic borders encompassing the boomerang-like structure inside the MMR
could be identified with separatrices of secondary resonances of the frequency of
oscillation of the semi-major axes (the resonant frequency) with the frequency of
librations of the secular angle $\Delta\varpi$
\citep{Morbidelli1993,Michtchenko2001}.

The bottom row in Fig.~\ref{fig:fig5} is for the dynamical maps computed for
MCMC derived models with the eccentricity priors set to $0.25$. This prior leads
to systematically larger osculating eccentricities, however the respective
posterior distributions looks similarly as in Fig.~\ref{fig:fig2}. Curiously,
the best-fitting solution remains ``glued'' to the unstable separatrix of the
boomerang-like structure.

A problem of constraining the dynamical model of Kepler-29 is finally
illustrated in Fig.~\ref{fig:fig7} which shows a MEGNO map for osculating
elements from a~small island of stable solutions around $(e_1\simeq 0.23,
\Delta\varpi \simeq -135^{\circ}$), see Fig.~\ref{fig:fig1}. Each point in this
$\mean{Y}$ map has been integrated for 64~kyrs, which guarantees the Lagrange
stability for 10-100 times longer interval, hence for $\sim 10$~Myrs. The tested
initial condition is found in an island in a~kind of archipelago with
eccentricities as large as $0.7$. Figure~\ref{fig:fig1} displays a~few isolated
models of this type with large eccentricities.

\section{Periodic orbits}

In the previous sections we could not determine any unique model of the Kepler-29 system
with both the observational and dynamical constraints.
Therefore we aim to impose additional constraints through the most likely
convergent planetary migration of the system in the past. As it has been shown
\citep{Beauge2003b,Beauge2006,Hadjidemetriou2006,Migaszewski2015}, systems of
two planets that undergo convergent migration evolve along families of periodic
orbits. Although the cited papers are devoted to first order MMRs, one could
expect that also for a second order 
MMR, like 9:7, periodic orbits play an
important role. Therefore we seek for families of periodic orbits of 9:7~MMR and
show where the configurations that fit the TTV data locate with respect to them.

\begin{center}
\begin{minipage}[l]{\linewidth}
\includegraphics[width=1.\textwidth]{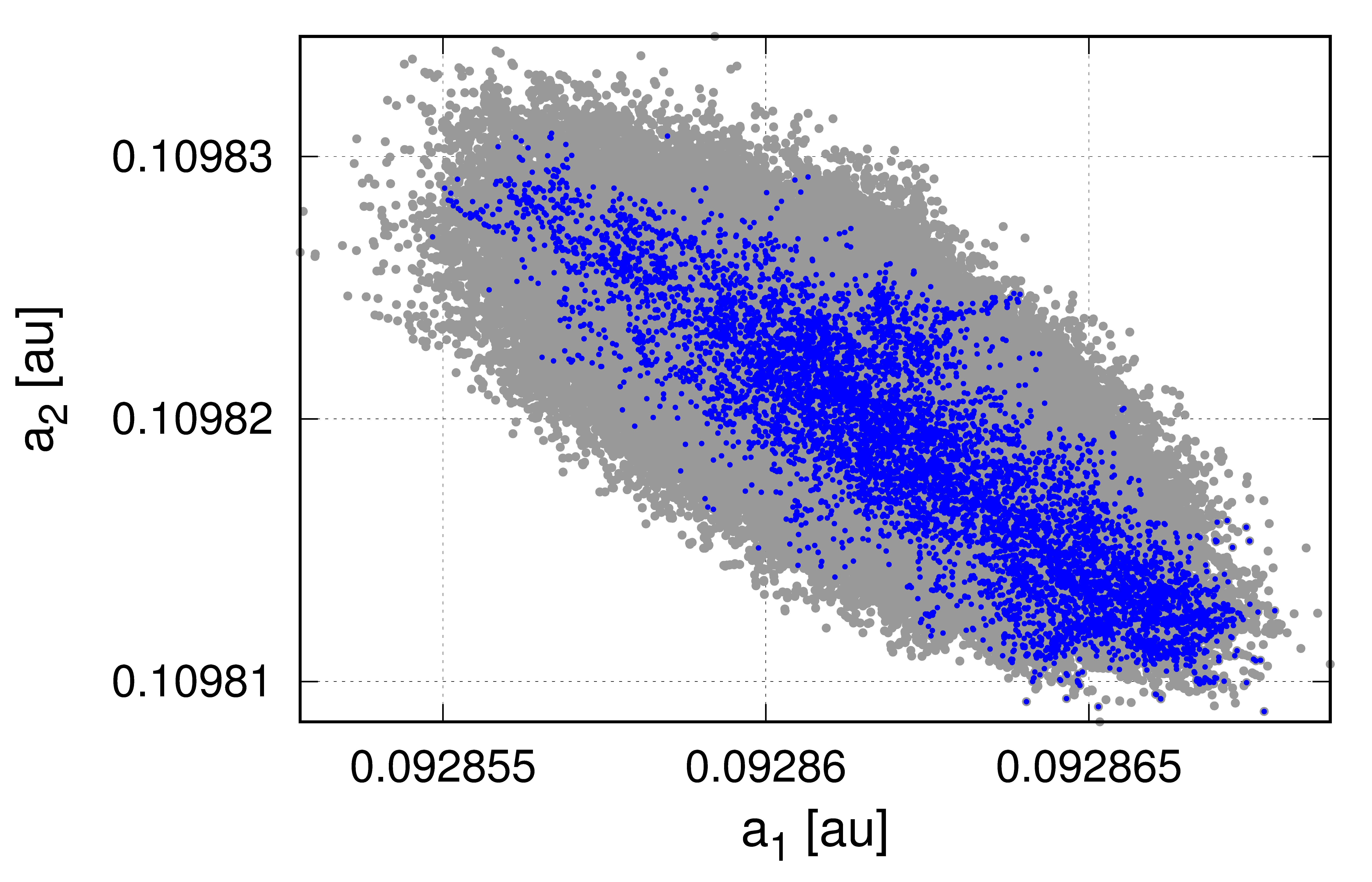} 
\includegraphics[width=1.\textwidth]{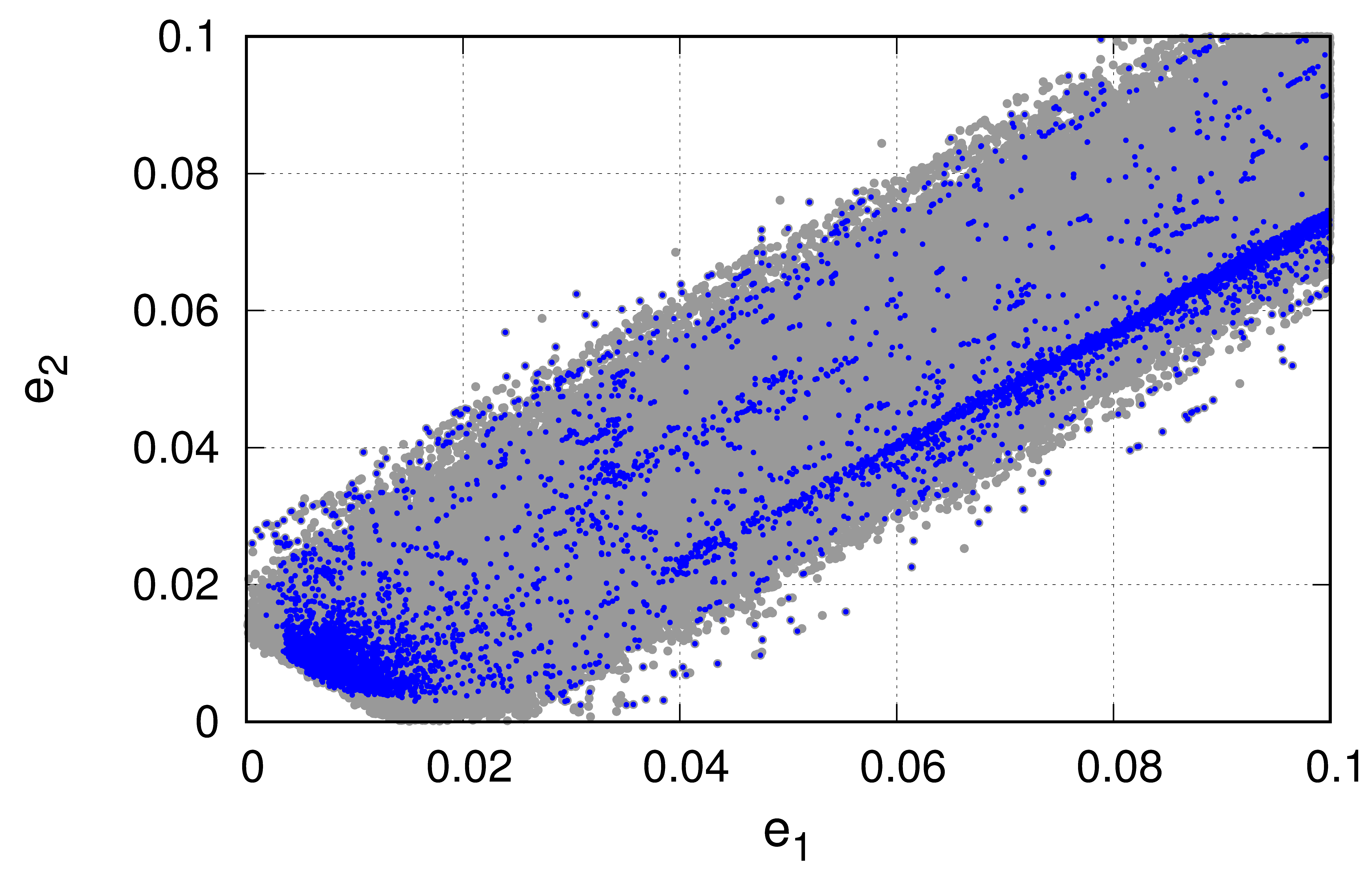} 
\includegraphics[width=1.\textwidth]{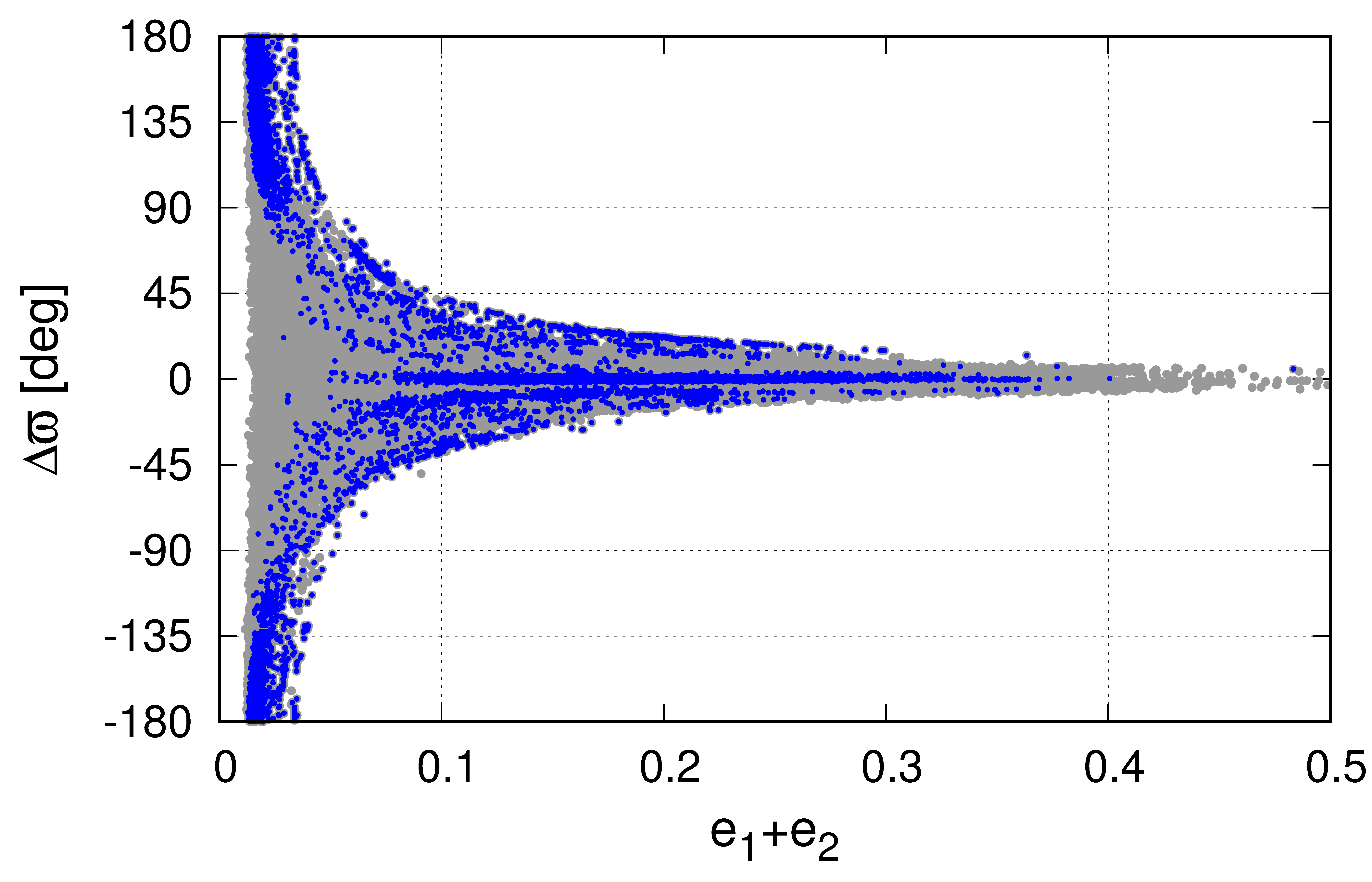}
\captionof{figure}{MCMC solutions with $L<0.0149$~d (grey filled dots) and with at least one
critical argument $\phi_j$, $j=1,2,3$ librating (blue filled dots). The Gaussian
prior for ($x_i,y_i$, $i=1,2$) with the zero mean value and the variance equal
to 0.1 has been set.}
\label{fig:fig3}
\end{minipage}
\end{center}

\subsection{The representative plane of initial conditions}

In order to address those issues, we apply the so called representative
(symmetric) plane of initial conditions $\Sigma$ \citep[e.g.,][]{Beauge2003a}
which makes it possible to illustrate global, qualitative 
features of the multi-dimensional and multi-parameter planetary systems.

Before introducing the $\Sigma$-plane, we recall some essential facts
about resonant configurations of two planets in coplanar orbits. Their dynamics
may be reduced through the
averaging over the fast angles to a two-degree-of-freedom Hamiltonian.
This Hamiltonian 
possesses two first integrals, i.e., the total angular momentum $C \equiv G_1 +
G_2$ and the so called spacing parameter 
$
K \equiv (p+q)\,L_1 + p\,L_2,
$
where
$L_i \equiv \beta_i \, \sqrt{\mu_i \, a_i}$, $G_i \equiv L_i \, \sqrt{1 -
e_i^2}$, $\beta_i \equiv (1/m_0 + 1/m_i)^{-1}$ and $\mu_i \equiv k^2 (m_0 +
m_i)$, $i=1,2$ \citep{Michtchenko2001,Beauge2003a}.

The averaged Hamiltonian can be expressed through
$
\overline{H} = \overline{H}(I_1, I_2, \sigma_1, \sigma_2; C, K),
$ 
where the canonical variables are $I_i = L_i - G_i$ and $\sigma_i = (1 +
s)\,\lambda_2 - s\,\lambda_1 - \varpi_i$ with $s \equiv p/q$. It can be shown
that $\partial \overline{H}/\partial \sigma_i = 0$ for critical values of
$(\sigma_1, \sigma_2) = \lbrace(0,0), (0,\pi), (\pm \pi/2, \pm \pi/2), (\pm
\pi/2, \mp \pi/2)\rbrace$. Therefore the equilibria of the averaged system, that
are periodic configurations of the full three-body problem, do exist for these
four pairs of angles. (Note that changing signs for both the angles
simultaneously does not lead to any change in the Hamiltonian.) Such
configurations are called apsidal corotation resonances (ACR). Here we consider
symmetric ACR only.

\begin{figure*}
\centerline{
\vbox{
\hbox{
\includegraphics[width=0.5\textwidth]{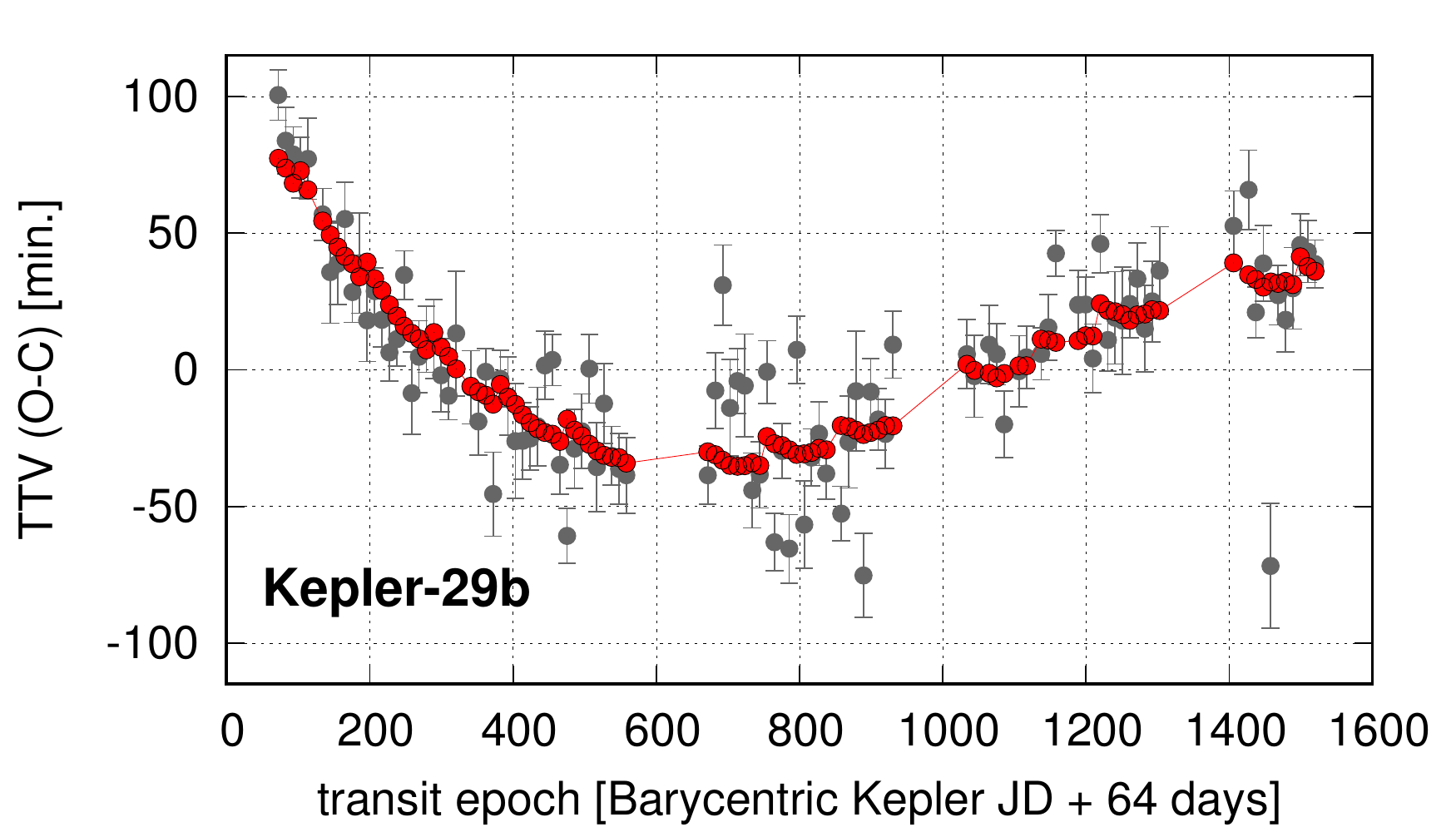}
\includegraphics[width=0.5\textwidth]{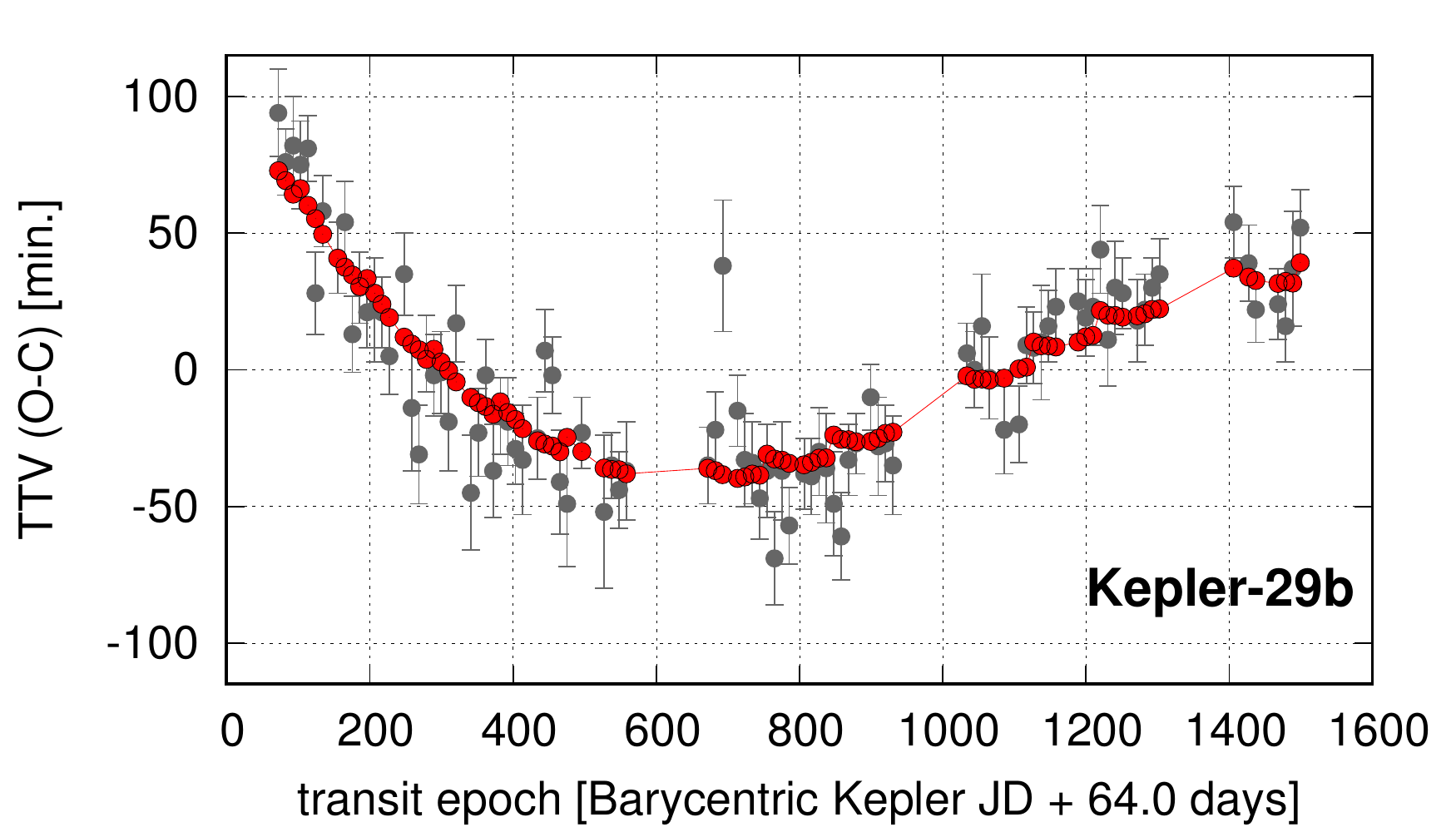}
}
\hbox{
\includegraphics[width=0.5\textwidth]{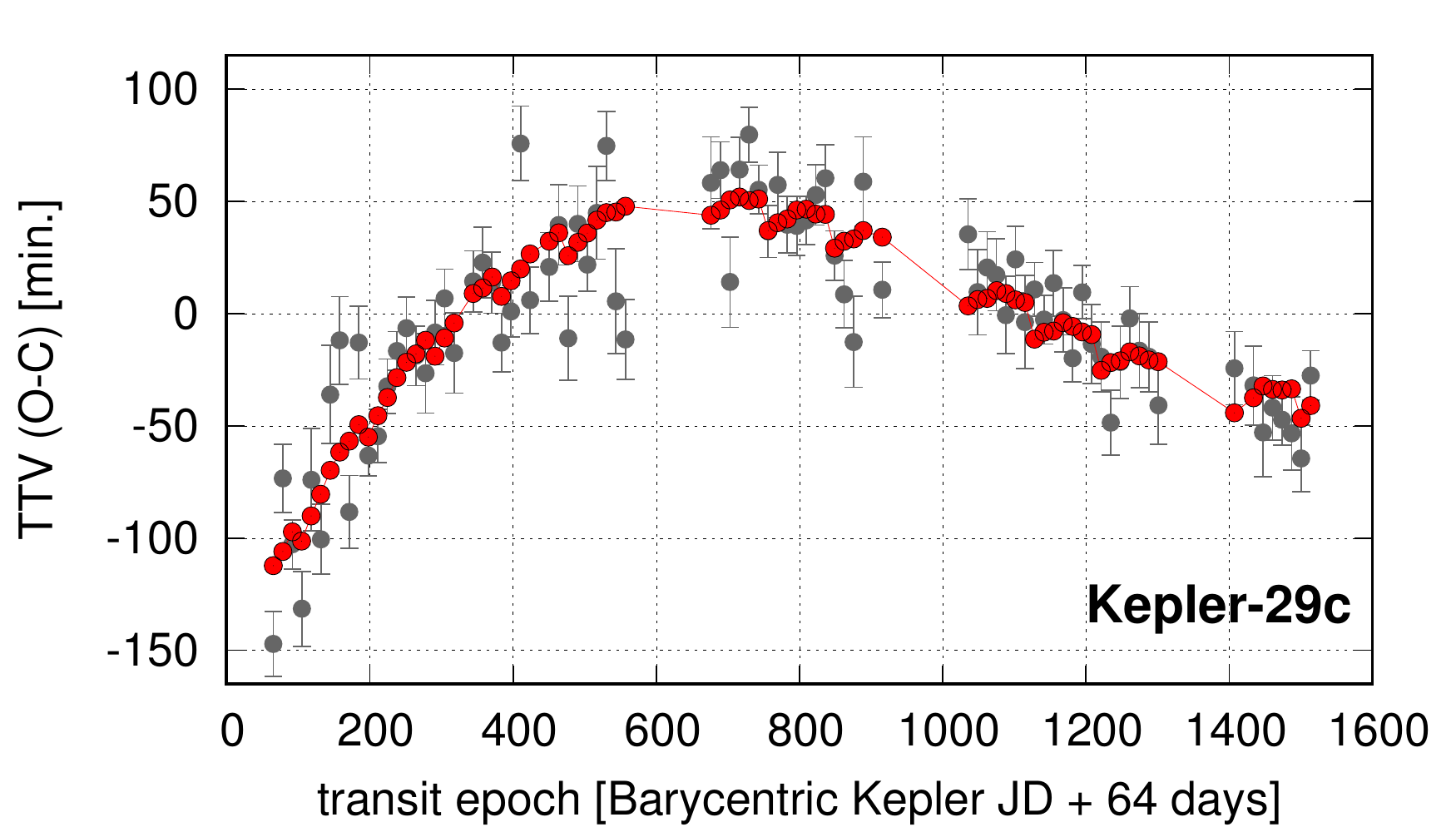}
\includegraphics[width=0.5\textwidth]{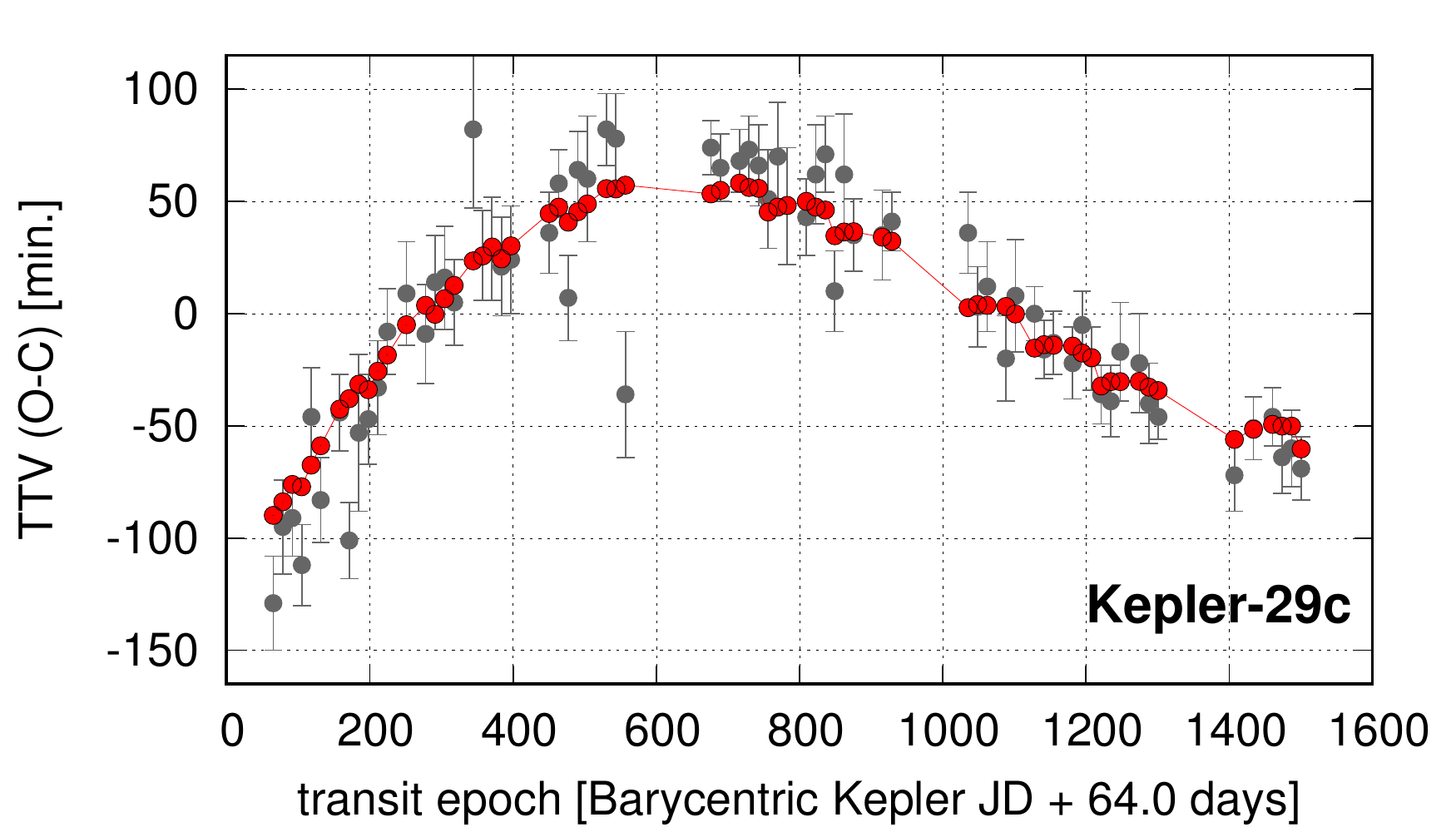}
}
}
}
\caption{
Synthetic TTVs of best-fitting low-eccentricity models for Kepler-29
(Tab.~\ref{tab:tab1}) over-plotted on the TTV measurements. (Red lines are shown
merely to guide the reader's eye). The left-hand column presents the fitting
results for the data set from \citep{Rowe2015}, while the right-hand column is
for an example best-fitting model to the data set from \citep{Holczer2016},
shown for a reference.
}
\label{fig:fig4}
\end{figure*}

\begin{figure*}
\centerline{
\hspace{-1mm}\vbox{
\hbox{
\includegraphics[width=0.5\textwidth]{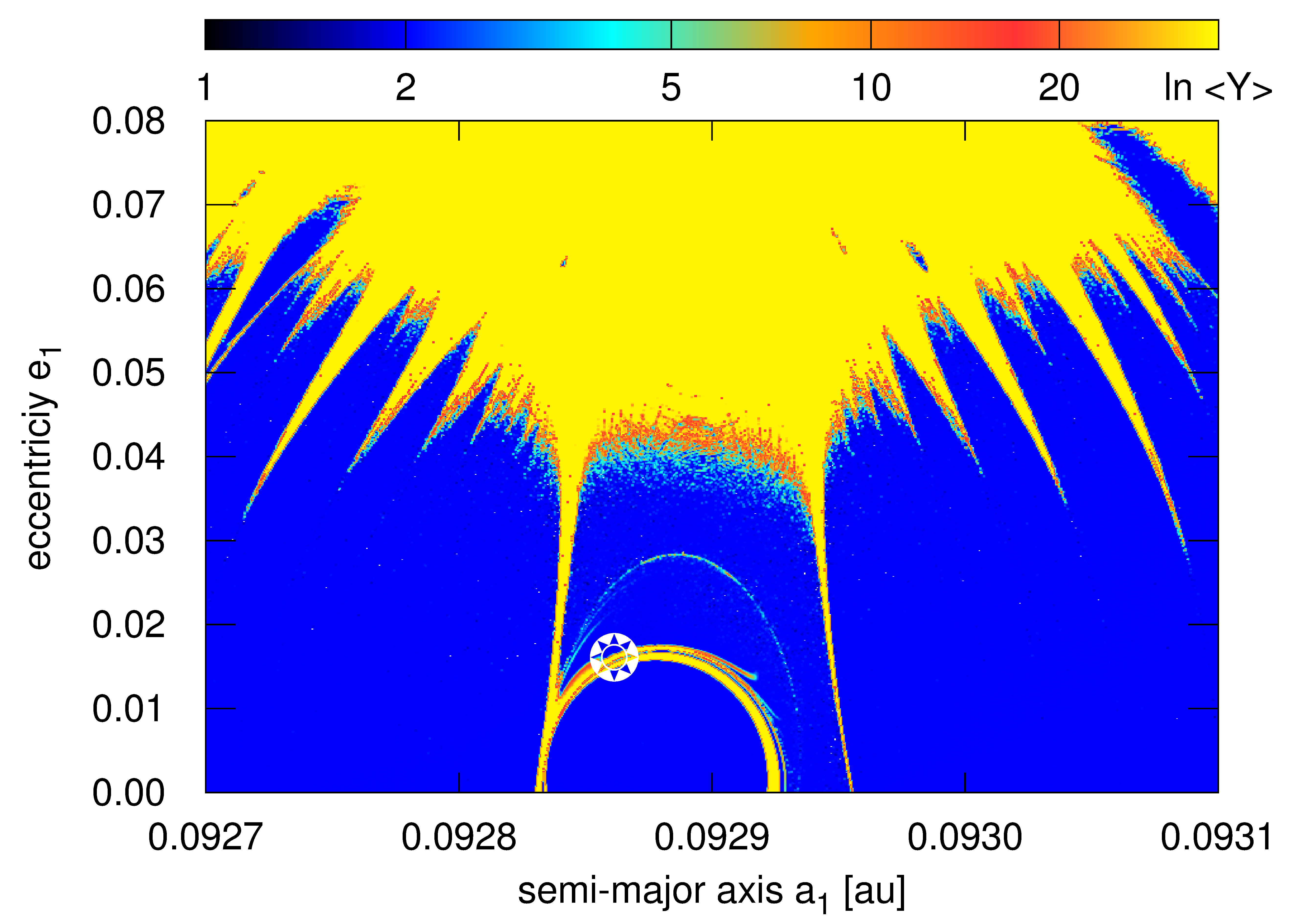}
\includegraphics[width=0.5\textwidth]{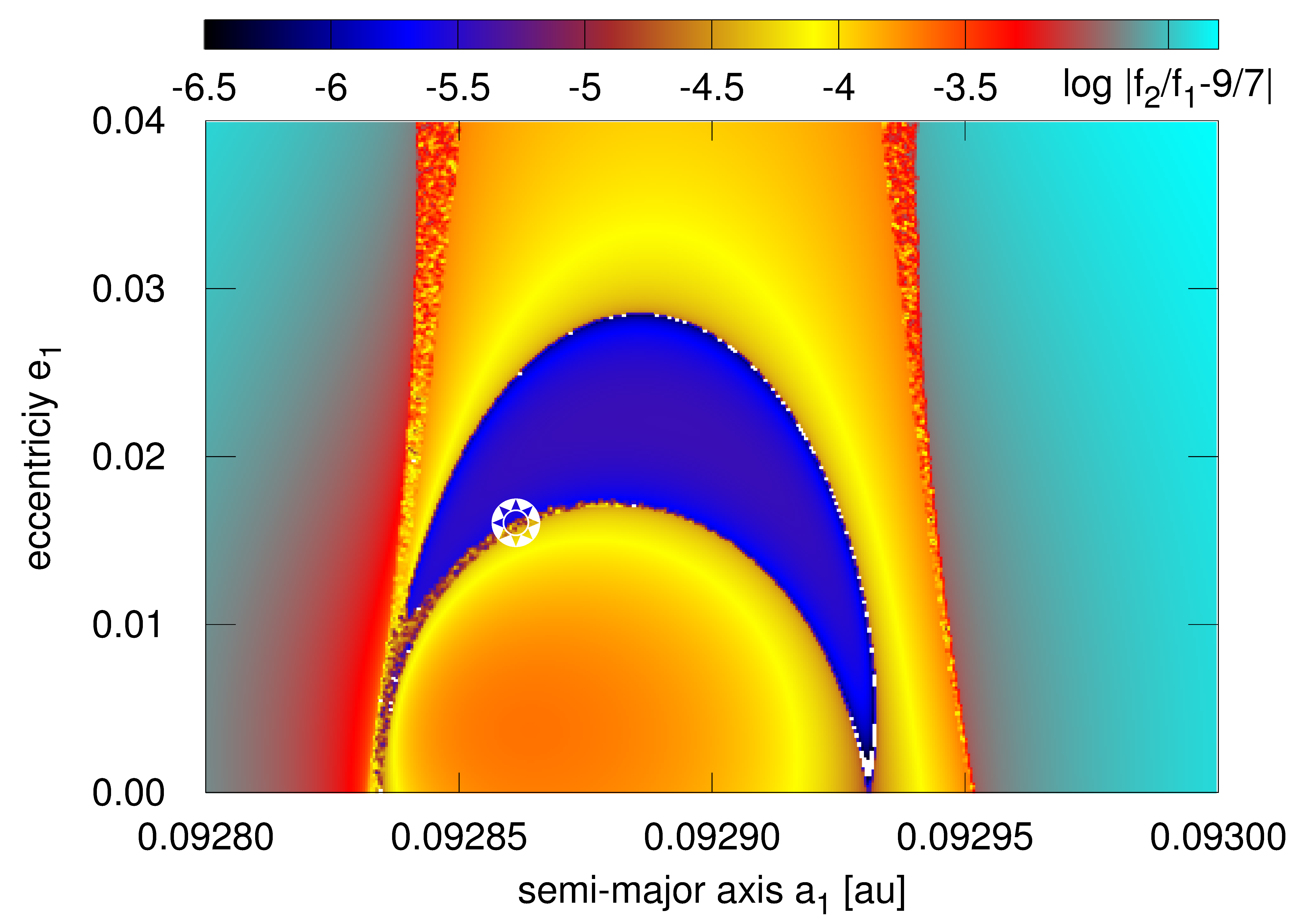}
}
\hbox{
\includegraphics[width=0.5\textwidth]{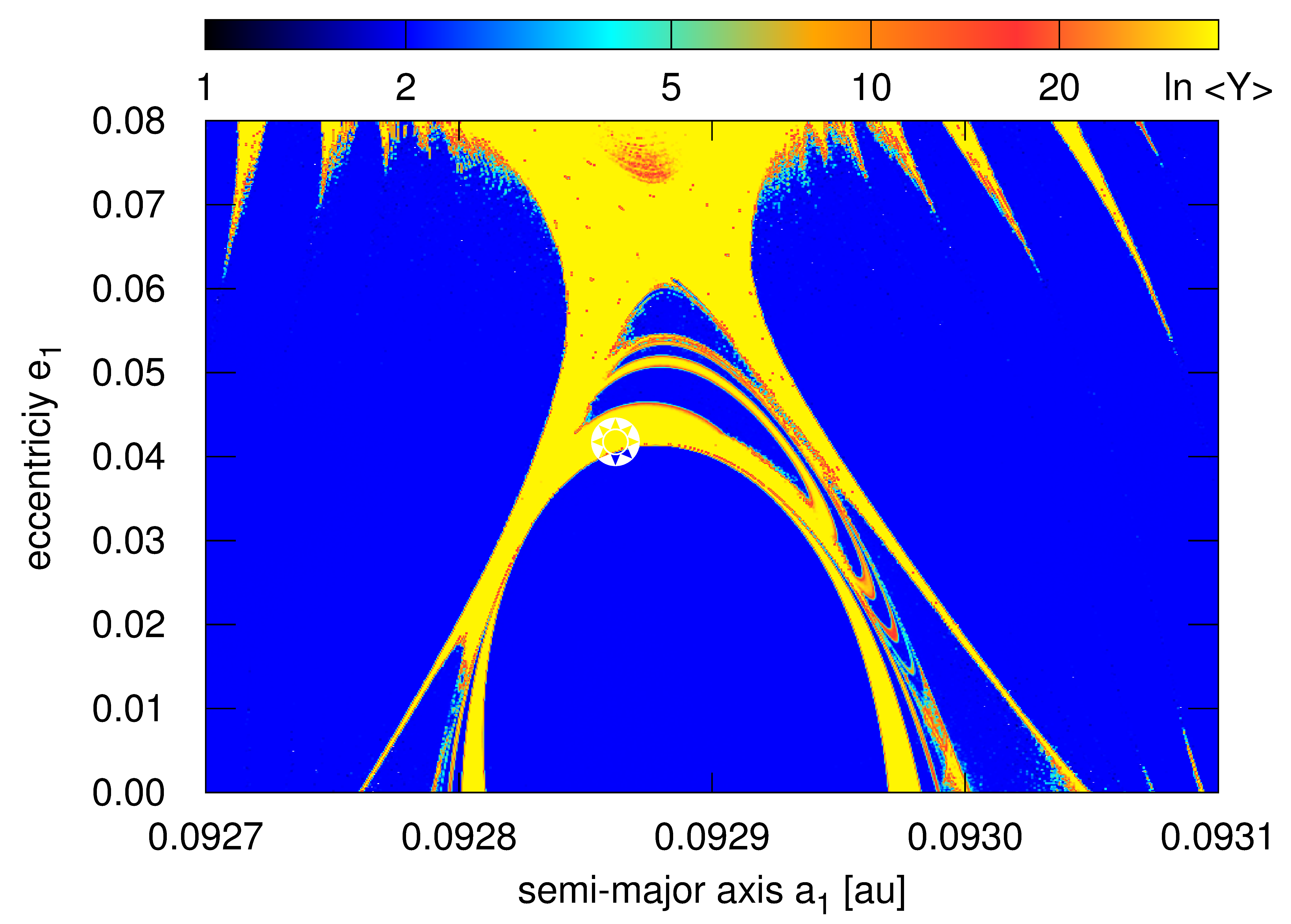}
\includegraphics[width=0.5\textwidth]{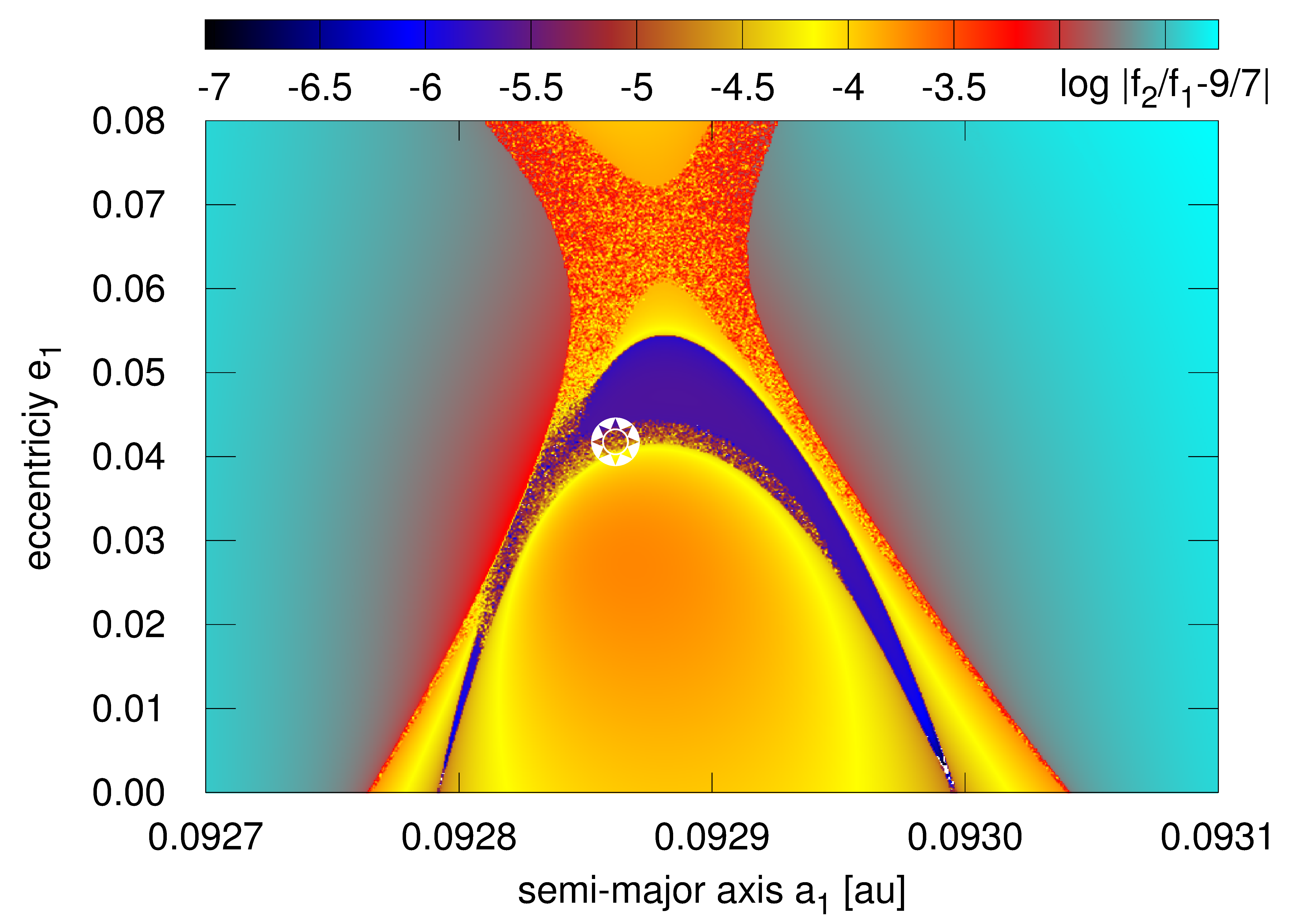}
}
}
}
\caption{
Dynamical maps for the best-fitting TTV models from the MCMC  in a
region of small eccentricities obtained with Gaussian priors imposed on
$(x_i,y_i)$ parameters with variances of 0.10 ({\em top row}, osculating
elements  of this model
at $t_0=\mbox{KBJD}+64$~days are displayed in Tab.~1.) and 0.25 ({\em
bottom row}), respectively.
{\em Left column}: dynamical maps in terms of the MEGNO indicator, $\Y \sim 2$
indicates a regular (long-term stable) solution marked with blue colour, $\Y$
much larger than $2$, up to $\gtrsim 32$ indicates a chaotic solution
(orange/red). Integrations done for 36~kyrs (roughly $1.2\times 10^6 P_2$).
{\em Right column}: deviation of the ratio of fundamental frequencies (mean
motions) w.r.t. the nominal value for the 9:7~MMR, computed for interval
spanning $10^{20}$ time steps of 0.5~days ($\simeq 4\times 10^4 P_2$). 
}
\label{fig:fig5}
\end{figure*}

\begin{figure*}
\centerline{
\vbox{
\hbox{
\includegraphics[width=0.33\textwidth]{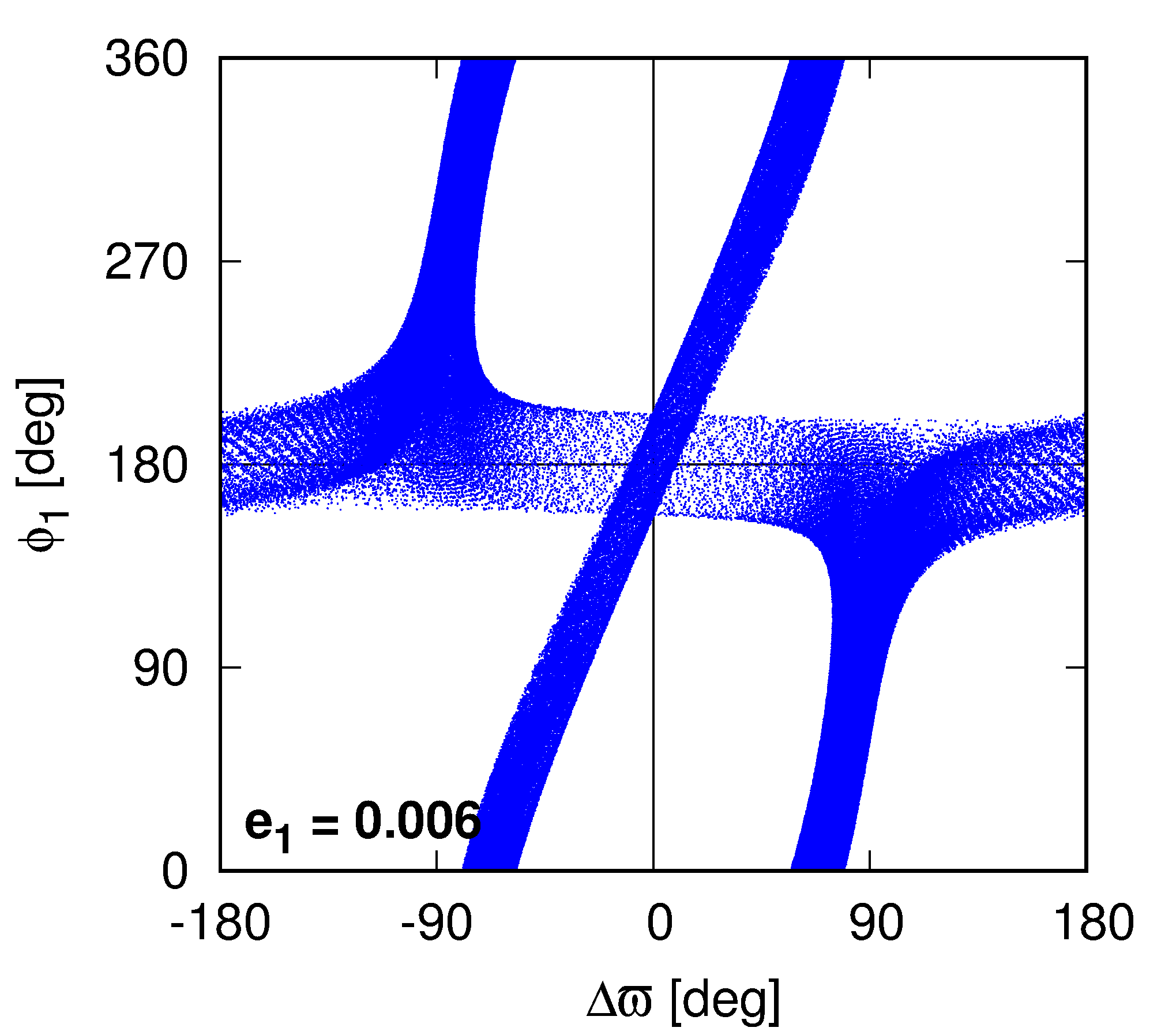}
\includegraphics[width=0.33\textwidth]{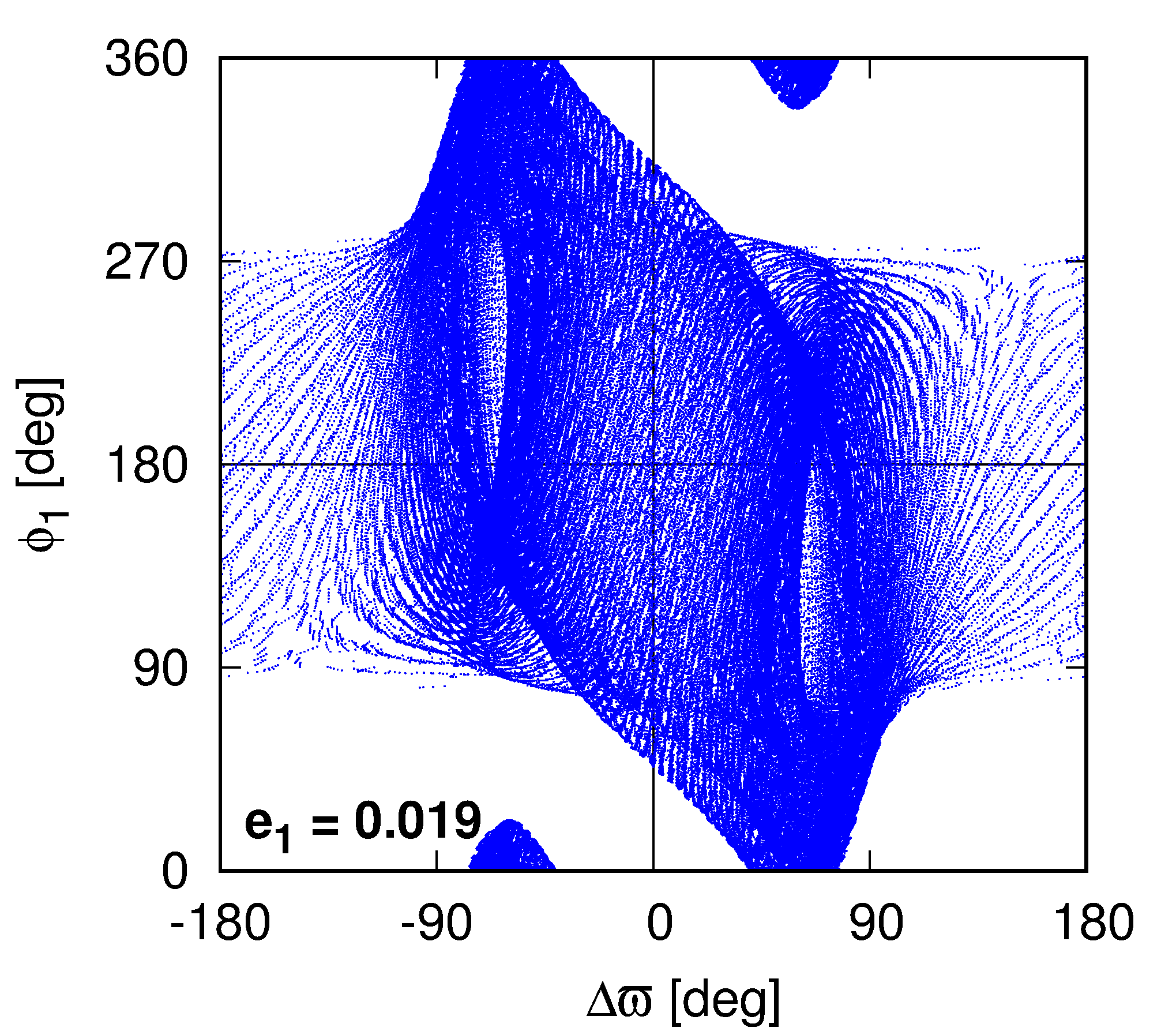}
\includegraphics[width=0.33\textwidth]{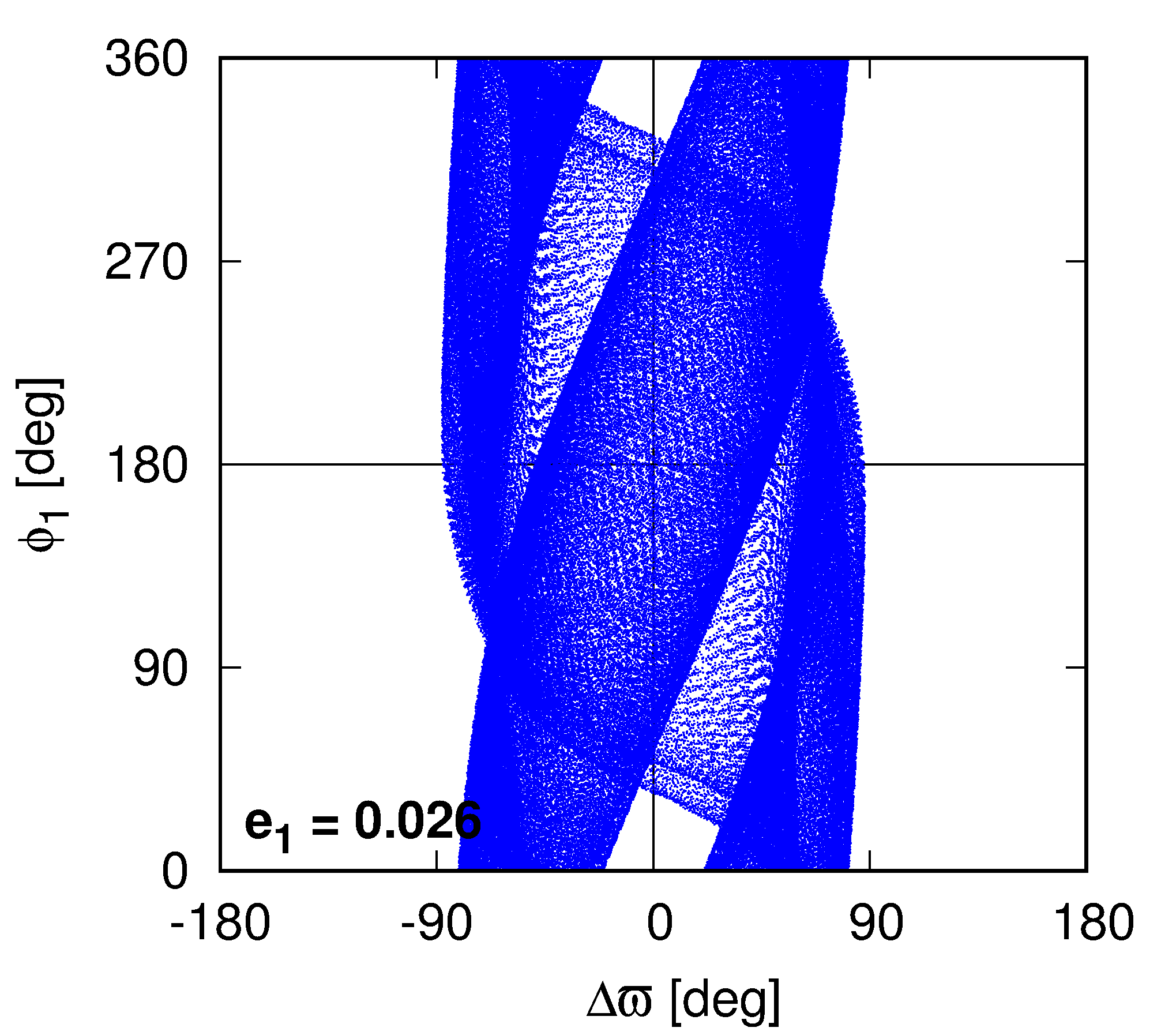}
}
\hbox{
\includegraphics[width=0.33\textwidth]{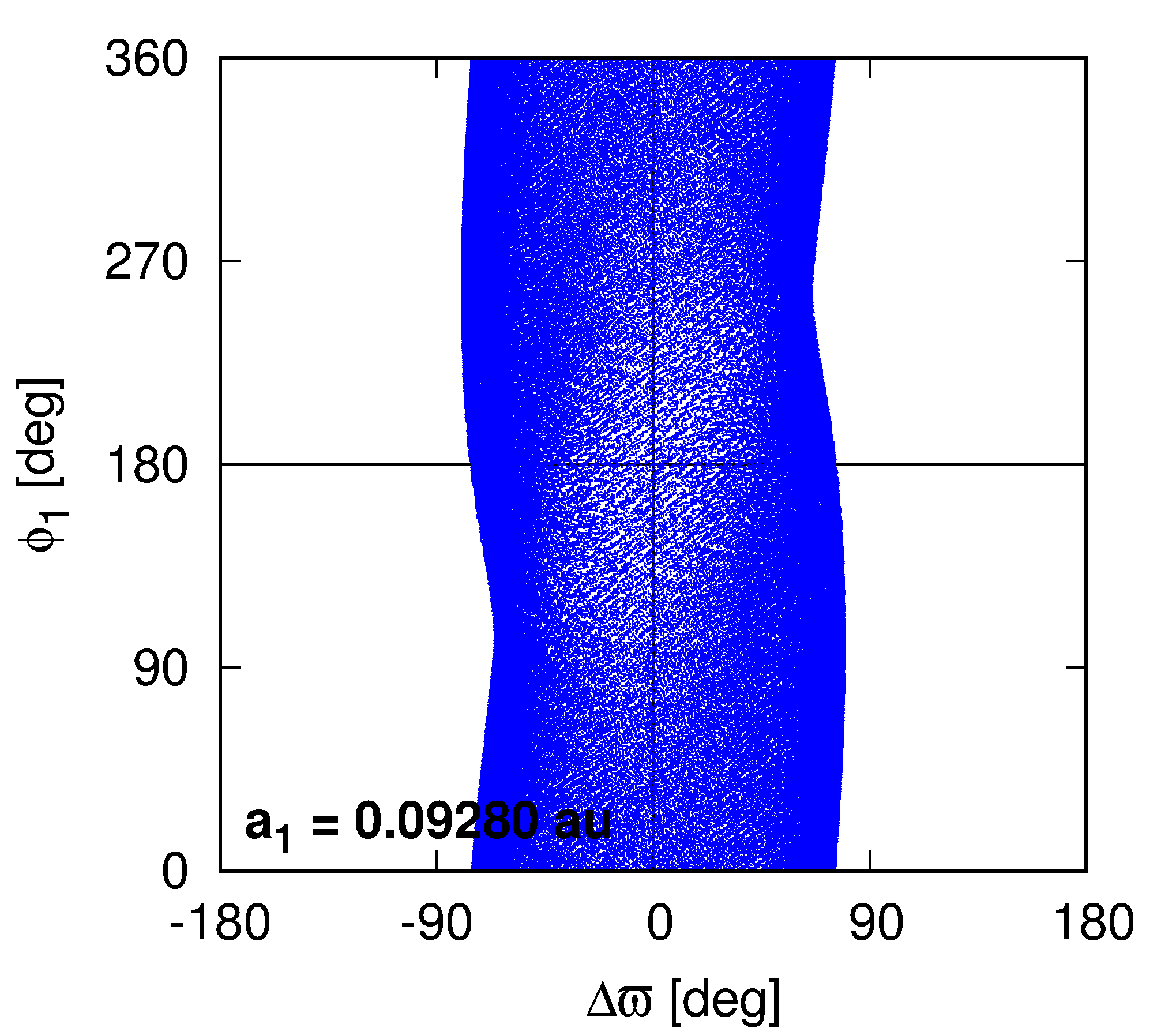}
\includegraphics[width=0.33\textwidth]{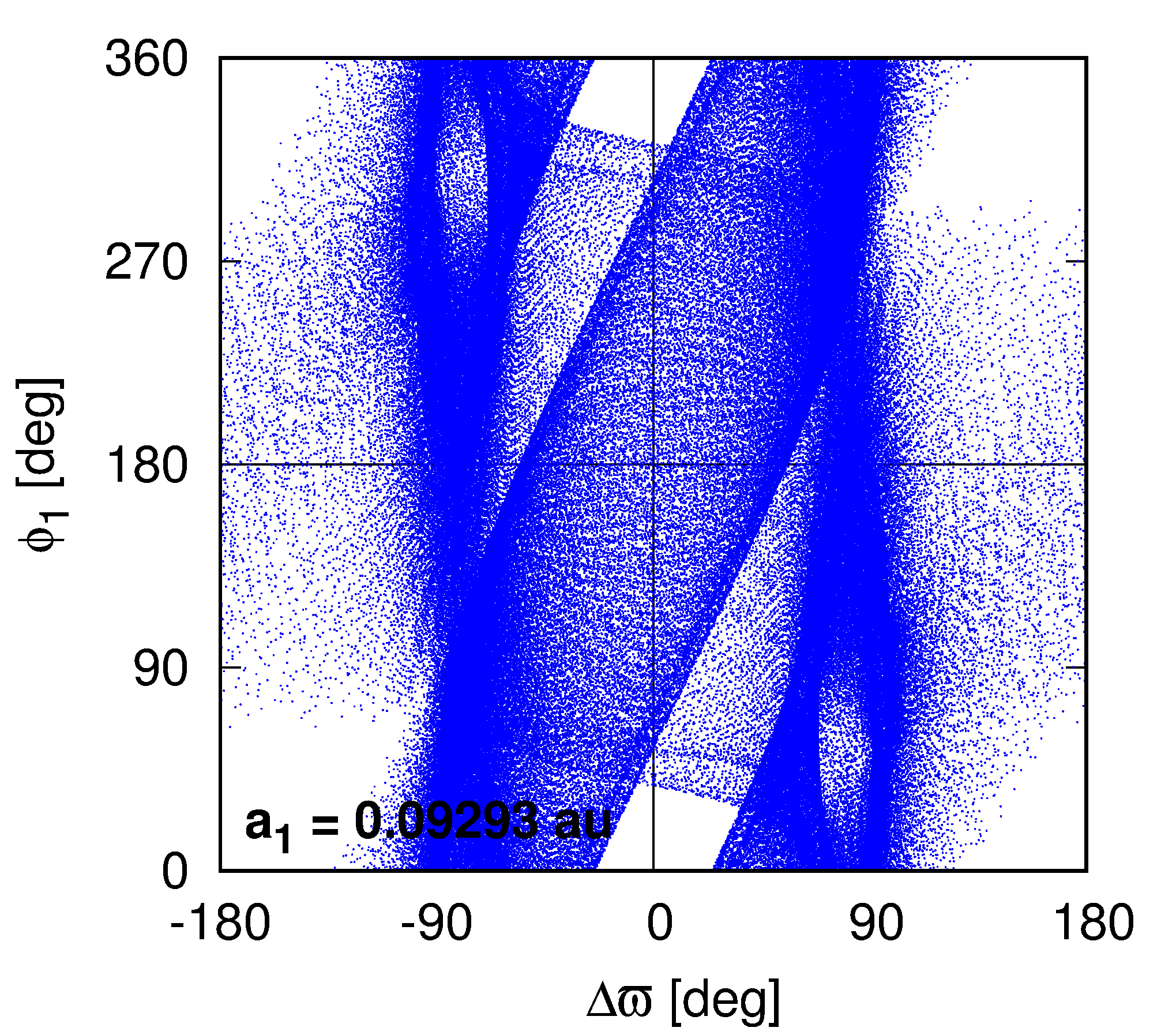}
\includegraphics[width=0.33\textwidth]{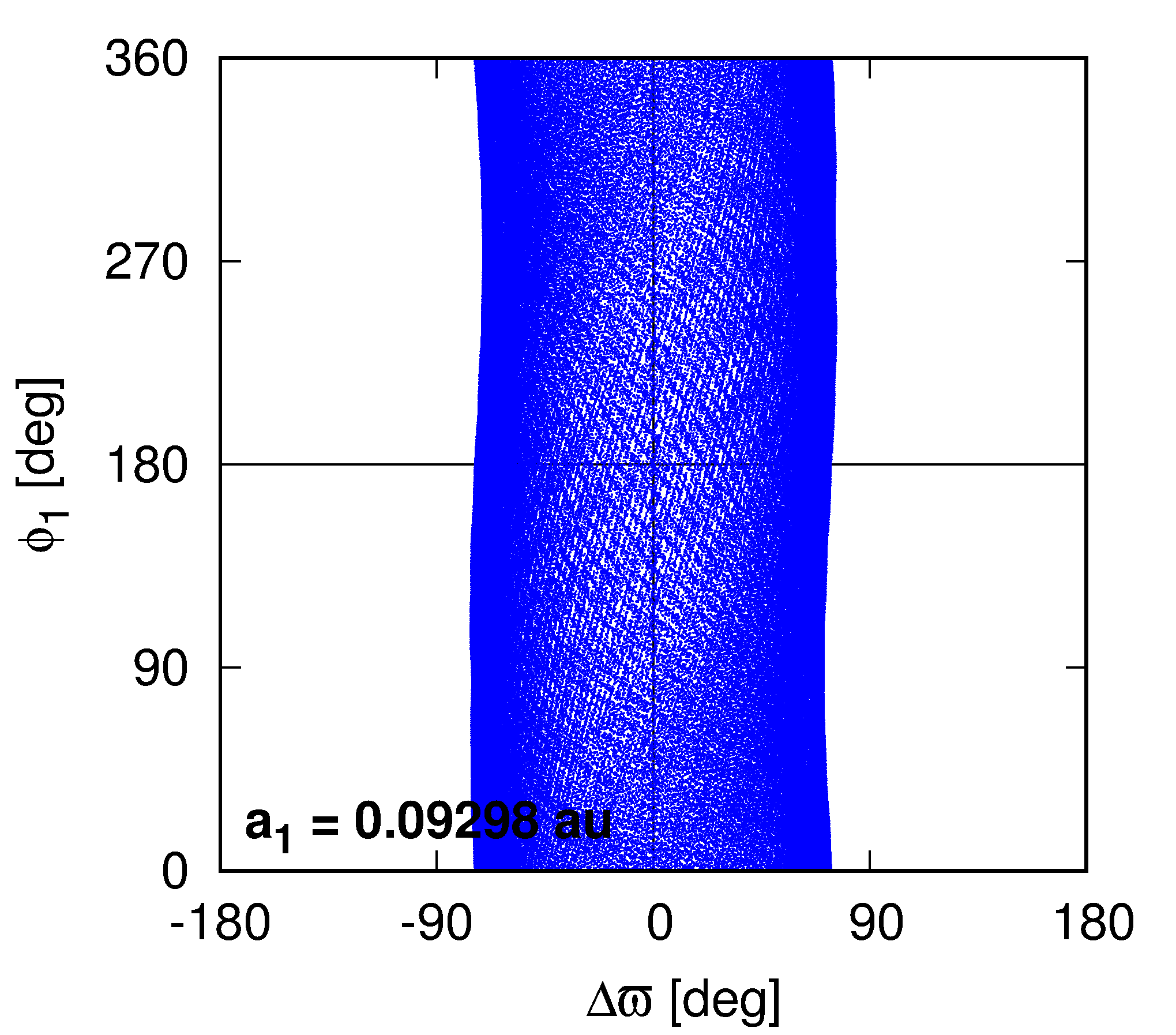}
}
}
}
\caption{
Evolution of critical angles $(\Delta\varpi,\phi_1)$ computed for six initial
conditions from a dynamical map in Fig.~\ref{fig:fig5}. Integrations interval is
36~kyrs. See the text for details.
}
\label{fig:fig6}
\end{figure*}

The canonical averaging can be done numerically or analytically. The explicit
form of the averaged Hamiltonian is 
\begin{displaymath}
\overline{H} = -\frac{\mu_1 \, \beta_1}{2\,a_1} - \frac{\mu_2 \, \beta_2}{2\,a_2} - \frac{k^2 \, m_1 \, m_2}{a_2} \, \overline{R},
\label{eq:Hamiltonian}
\end{displaymath}
where the disturbing function reads as follows:
\begin{equation}
\overline{R} = \frac{1}{2\,\pi} \int_0^{2\,\pi} \frac{a_2}{\| \vec{r}_1 - \vec{r}_2\|} \, dQ, \quad Q \equiv \frac{\lambda_1 - \lambda_2}{q}.
\label{eq:R}
\end{equation}
This integral can be evaluated numerically \citep{Michtchenko2006} or
analytically. For this operation the function under the integral must be expanded
in power series w.r.t. the small parameters, like eccentricities or semi-major
axes ratio, \citep[e.g.,][]{Beauge2003a}. Here, $\overline{R} =
\overline{R}_{\idm{sec}} + \overline{R}_{\idm{res}}$ is a sum of the secular
$\overline{R}_{\idm{sec}}$ and resonant $\overline{R}_{\idm{res}}$ terms,
following a~recipe in \citep{Murray1999a}. We selected terms up to the fourth
power in the eccentricities, which provides a very good approximation for
eccentricities $\lesssim 0.1$, and with no terms related to the inclinations and
nodal longitudes, since we assume a~coplanar configuration. The explicit form of
the averaged Hamiltonian is given in Appendix~A. 

The representative (or characteristic) plane of initial conditions $\Sigma$ is a
plane of eccentricities. A~point in this plane $(e_1, e_2)$ determines
semi-major axes $a_1, a_2$ through the first integrals $C$ and $K$. The initial
condition of a given system contains also angles $\sigma_1$ and $\sigma_2$
chosen from the critical set of 
$(\sigma_1, \sigma_2) = \lbrace(0,0), (0,\pi), (\pm \pi/2, \pm \pi/2), (\pm
\pi/2, \mp \pi/2)\rbrace$. 
Each orbital configuration with angles $(\sigma_1,
\sigma_2)$ that circulate or librate around pairs of critical values given above
must intersect this plane \citep{Michtchenko2001}. Therefore, in order to study
the dynamics of such systems globally, it is sufficient to consider the four
sets of initial resonant angles $(\sigma_1, \sigma_2)$ in the
$\Sigma$-plane. However, the $\Sigma$--plane defined in this way
is representative only for symmetric configurations, for which $\Delta\varpi =
\sigma_2 - \sigma_1$ equals $0$ or $\pi$. There could, in principle, also exist
asymmetric configurations with different libration centres \citep{Beauge2003b}.
Since the majority of the best-fitting configurations of the Kepler-29 system
are symmetric (apart from, possibly, a~few islands with large eccentricities,
Fig.\ref{fig:fig1}), we limit our analysis to the symmetric configurations only. 

Furthermore, instead of $\sigma_1, \sigma_2$ we choose their combinations,
the secular angle $\Delta\varpi$ and one of the critical arguments of
the 9:7~MMR, $\phi_1 = -2\,\sigma_1$,  which are more convenient for an
interpretation of their evolution. The representative angles at the
$\Sigma$-plane are then $(0, 0),
(0, \pi), (\pi, 0)$ and $(\pi, \pi)$, and the $\Sigma$-plane coordinates can be
defined as $(e_1 \, \cos\Delta\varpi, e_2 \cos\phi_1)$, where both cosines are equal
to $+1$ or to $-1$.

Although the $\Sigma$-plane has been defined for the averaged Hamiltonian with
two degrees of freedom, we can use the same concept also for the full
three-body, non-averaged system. It
is convenient to show orbital models that fit the observations w.r.t. the
periodic orbits represented as equilibria of the averaged system.

\subsection{TTV-constrained models in the $\Sigma$-plane}

We narrowed the set of dynamically stable initial conditions, as illustrated in
Fig.~\ref{fig:fig3}, by $L<0.0145$~days. To map this set on the $\Sigma$-plane, we integrated the
$N$-body equations of motion forward for $10^5\,\yr$. Each time when
$\epsilon \equiv \|\sin\Delta\varpi\| + \|\sin\phi_1\| \lesssim 0.02$ (expressed in radians; for numerical reasons of
limited integration time, we chose a~small limit of $\epsilon = 0.02$ 
for the 
intersection of the $\Sigma$-plane), osculating elements were transformed to
coordinates at the $\Sigma$-plane and presented in the left-hand panel of
Fig.~\ref{fig:fig8}. Usually, each initial configuration
evolving in time results in more than
one point at this plane. For the reduced system with two degrees of freedom one
obtains: i) one point at the $\Sigma$-plane if the system is in a stable
equilibrium; ii) two points if it is a stable periodic configuration of the
reduced system (fixed point at the Poincar\'e cross section); iii) four points
if the reduced system evolves along quasi-periodic orbit; iv) a continuum of
points for chaotic evolution. The full (non-averaged) system intersects the
$\Sigma$ plane in four groups of points if its reduced counterpart intersects
the plane in four points. 

As Fig.~\ref{fig:fig8} shows, stable models that fit the TTV data form a strip at the
$\Sigma$-plane along a line originating from $(0,0)$ and directed towards higher
eccentricities in the quarter of the plane with $\Delta\varpi=0$ and
$\phi_1=\pi$. There are points in the $(\Delta\varpi, \phi_1) = (\pi,
\pi)$--quarter as well, but no configurations intersect the upper half of the
$\Sigma$-plane ($\phi_1=0$).  

The second critical component of Fig.~\ref{fig:fig8} is a representation of
families of the periodic orbits. Green and red solid curves show families of
stable and unstable equilibria of the reduced system (periodic configurations of
the full system), respectively. The periodic system returns to its initial state
after a certain period $P$. For the periodic orbits in the 9:7~MMR it covers
nine revolutions of the inner planet, and seven revolutions of the outer planet.
Then the planets starting for instance from their pericenters will be back in
the pericenters after the period $P$. Also orbital elements $\vec{p} = (a_1,
a_2, e_1, e_2, \mathcal{M}_1, \mathcal{M}_2, \Delta\varpi)$ will retain their
initial values. Angles $\varpi_1$ and $\varpi_2$ will change, since the system
precesses as a whole. For a given point $(e_1, e_2)$ we integrate the equations
of motion and evaluate $\delta = \| \vec{p}(t=T) - \vec{p}(t=0) \|$. We search
for such $(e_1, e_2)$ (for a given $C, K$) that provides $\delta
\simeq 0$. In order
to check the stability of a given periodic configuration we integrate the system
for $\sim 10^5$~revolutions. A stable periodic configuration is preserved,
while for unstable one, sooner or later the system evolves into different
regions of the phase space.

Tracks of the periodic orbits in the $\Sigma$-plane correspond to the equilibria
of the averaged system. They may be found through solving the equations
\[
\partial\,\overline{H}/\partial\,I_i = 0, \quad i=1,2,
\]
since the partial derivatives over $\sigma_i$ are zero at the whole
$\Sigma$-plane. The stability of equilibria of two-degree-of-freedom Hamiltonian
systems can be verified through solving the eigenvalue problem for the
linearised equations of motion. One can also check if a given point is an
extremum of $\overline{H}$ for fixed $C$ and $K$. We verified that the branches
of periodic orbits of the full $N$-body equations of motion precisely coincide
with the equilibria of the averaged system in the region of the $\Sigma$-plane
we will be interested in. This is valid for the averaging done analytically and
numerically. However, the analytic averaging provides reliable results only for
non-crossing orbits (crossing orbits are shown with grey dots in
Fig.~\ref{fig:fig8}a). The numerically averaged Hamiltonian can be used to
describe the dynamics even for crossing orbits, provided the perturbation to the
Keplerian motions is sufficiently small. 

Apparently, the best-fitting solutions appear along a~family of unstable
equilibria. We found this result somehow unexpected from the point of view of
the formation of the system through the migration. Unstable periodic orbits in
the proximity of the observationally constrained configurations may be also a
factor provoking dynamical instability of the system. Although aligned
configurations are, in general, not impossible to form within the migration
formation scenario \citep[e.g.,][]{Ferraz-Mello2003,Beauge2003b}, the aligned
configurations studied in the cited papers are related to the branch of stable
equilibria. As shown in the previous Section, geometrical parameters of the
system like eccentricities and pericenter longitudes are not well constrained
due to low signal-to-noise ratio, narrow observational window and features of
the TTV method. This permits us to look for solutions fulfilling also the
migration constraints.

As \citep{Migaszewski2015} has demonstrated, two-planet systems that undergo
convergent migration end up in exact periodic configurations. Nevertheless, that
conclusion referred to the first-order resonances and might not be fully
applicable to the second-order MMRs. We will discuss this further in this paper.
After inspecting the O-C diagrams of Kepler-29 system (see Fig.~\ref{fig:fig4}),
one can conclude that it cannot be related to exactly periodic configurations.
In such a case there would be no secular TTV signal of a period longer than 
the resonant period of $\simeq 91$~days. However, such a signal of a few-year period is clearly visible in the
data. Therefore, the real Kepler-29 cannot be a strictly periodic configuration.
Nevertheless we show further that the migration is still very likely a way the system
has been formed.

\subsection{Particular solutions in the $\Sigma$-plane}

\begin{figure*}
\centerline{
\vbox{
\hbox{
\includegraphics[height=0.31\textwidth]{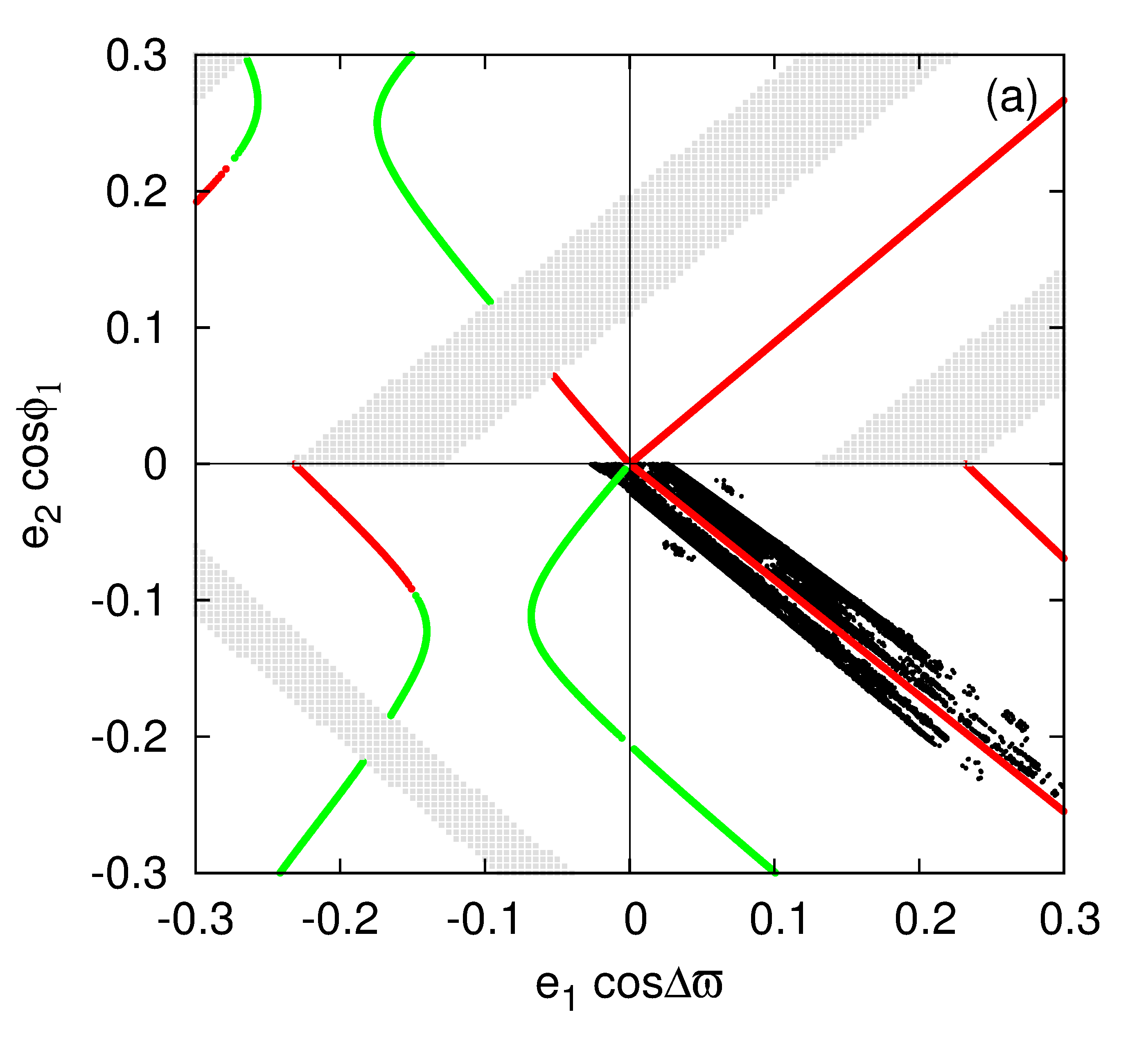}
\includegraphics[height=0.31\textwidth]{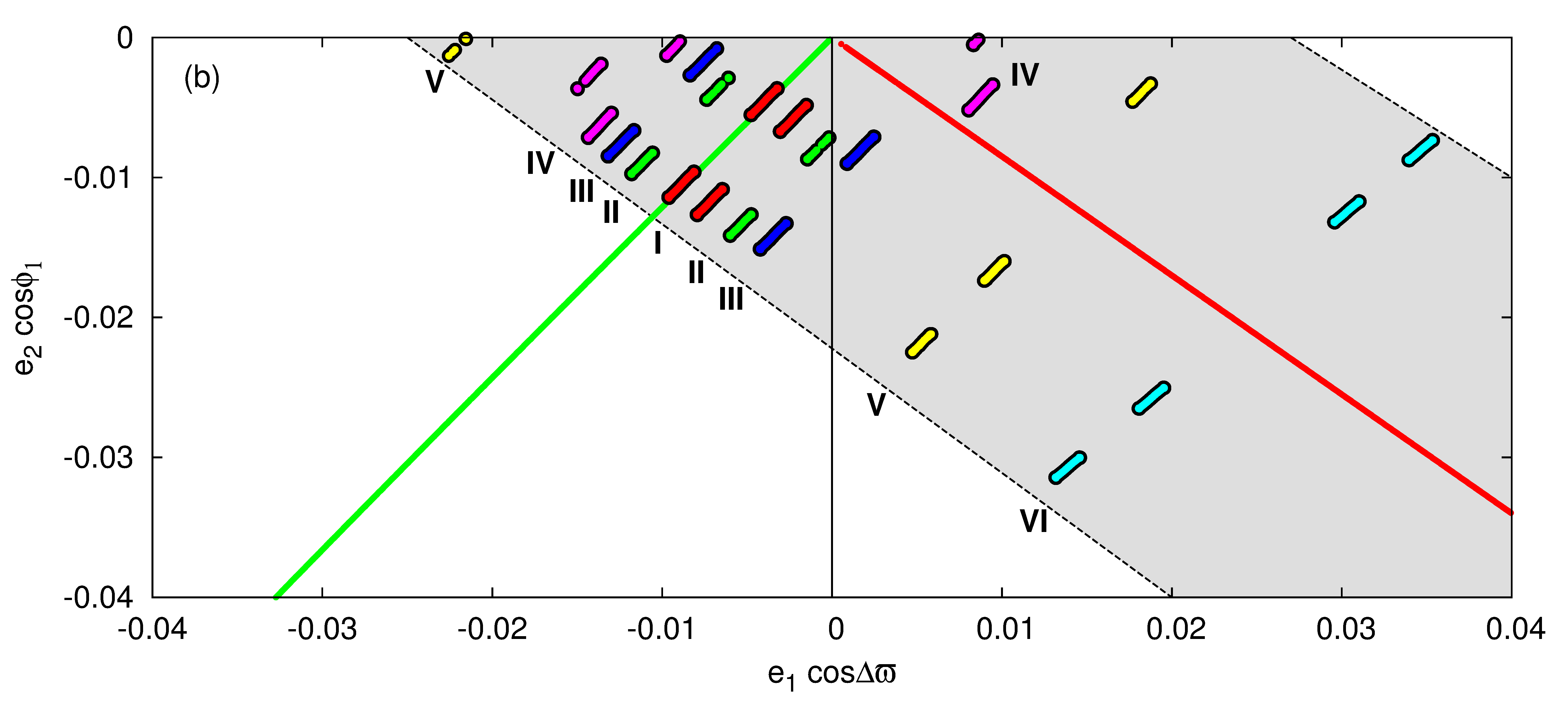}
}
}
}
\caption{
Panel~(a): The best-fitting solutions from the TTV analysis
(Figs.~\ref{fig:fig1} and~\ref{fig:fig3}) projected at the representative plane
(black points). We consider a given configuration as crossing the plane if
$\Delta\varpi$ and $\phi_1$ differ from the nominal values of $0$ or $\pi$ by
less than one degree. Green and red curves denote families of stable and
unstable periodic configurations of the N-body system, respectively. Grey
symbols denote configurations for which the closest encounter of the planets in
Keplerian orbits is smaller than $3$ Hill radii ($\approx 0.005\,\au$).
Panel~(b): Chosen configurations (whose parameters are listed in
Tab.~\ref{tab:tab2}) projected at the representative plane in the same manner as
for the whole statistics of systems presented in panel~(a). Each configuration
is plotted in different colour and labelled. Only bottom half of the
$\Sigma$--plane is shown. The grey filled area shows qualitatively the black
strip of points shown in panel~(a).
}
\label{fig:fig8}
\end{figure*}

The right-hand panel of Fig.~\ref{fig:fig8} illustrates models selected from the
set of solutions that well reproduce the observations. Those six models
(Tab.~\ref{tab:tab2}) were chosen as qualitative representatives for all the
statistics of the best-fitting models. Model~I (red points) intersects the
$\Sigma$-plane very close to the branch of stable equilibria and all four groups
of points where the system intersects the $\Sigma$--plane lie in the
$(\Delta\varpi, \phi_1) = (\pi, \pi)$ quarter. Model~III (blue points) lies
further from the stable branch than Model~I, and its phase trajectory intersects
both quarters with $\phi_1 = \pi$. Model~II (green points) is an intermediate
configuration between Models~I and~III. Its phase trajectory almost ``touches''
the quarter with $\Delta\varpi=0$. Model~VI (cyan points) lies in vicinity of
the unstable branch of equilibria in the $(0, \pi)-$quarter and intersects only
this quarter. Models~IV (magenta points) and~V (yellow points) are intermediate
states between Models~III and~VI. The sequence of models from I to VI shows
a~transition between configurations very close to the branch of stable equilibria
and configurations close to the branch of unstable equilibria.

Figure~\ref{fig:fig9} shows the orbital evolution of the selected models at
the $(\Delta\varpi, \phi_1)$-plane.  Both angles of Fit~I librate around
$\pi$, although the amplitude of $\phi_1$ libration is large. 
For the second model in the test sample (Fit~II), the $\phi_1$
libration amplitude reaches $2\pi$, while $\Delta\varpi$ librates with a
moderate amplitude.  Since the amplitude of $\phi_1(t)$ is actually greater
than $2\pi$, we could classify the behaviour of this angle as circulation.
Despite of the formal rotation of $\phi_1$, the phase-space trajectory of
Fit~II intersects only one quarter of $\Sigma-$plane with $(\Delta\varpi,
\phi_1) = (\pi, \pi)$.  The next Fit~III in the sample exhibits both
critical angles rotating and two quarters of $\Sigma-$plane are intersected
by the phase trajectory.  Two other quarters are being avoided.  Similarly,
the phase trajectory of Fit~IV intersects the same two quarters of
$\Sigma-$plane.  Yet the behaviour of the angles is different.  For Fit~IV,
$\phi_1$ seems to librate but with an amplitude larger than $2\pi$.  The
next system, Fit~V, shows both angles circulating, however in this case
$\Delta\varpi$ remains mainly around $0$ during the evolution, only
occasionally reaching $\pi$.  The last model, Fit~VI, has $\Delta\varpi$
librating around $0$ and $\phi_1$ circulating.

Similarly to Fig.~\ref{fig:fig8}, Fig.~\ref{fig:fig9} also reveals the
transition between two different types of configurations (from Fit~I to Fit~VI).
The sequence of the configurations can be also analysed in energy plots of the
averaged system presented in Fig.~\ref{fig:fig10}. Mean orbital parameters of
Fit~I to Fit~VI are displayed in Tab.~\ref{tab:tab3}. 

\begin{table}
\centering
\caption{
Parameters of six selected models that fit the TTV measurements and exhibit
different qualitative orbital behaviour (see text for details).  The
osculating Keplerian elements are given at the epoch of
$t_0=\mbox{BJKD}+64.0$~days.  The system is coplanar with $I=90^{\circ}$ and
$\Omega=0^{\circ}$.  Mass of the parent star is $1\,\msun$.
}
\begin{tabular}{l c c c c c}
\hline
\hline
model/pl & $m\,[\mE]$ & $a\,[\au]$ & $e$ & $\varpi$\,[deg] & $\Mmean$\,[deg]\\
\hline
I/b & $7.5173$ & $0.0928595$ &$0.00542$ &$18.30$ & $-40.97$\\
I/c & $6.5125$ & $0.1098237$ &$0.00809$ &$-148.75$ & $-334.13$\\
\hline
II/b & $6.2177$ & $0.0928608$ &$0.00629$ &$0.15$ & $-22.62$\\
II/c & $5.4607$ & $0.1098212$ &$0.00891$ &$-138.86$ & $-343.92$\\
\hline
III/b & $7.5940$ & $0.0928594$ &$0.00758$ &$58.44$ & $-81.28$\\
III/c & $6.6879$ & $0.1098237$ &$0.00776$ &$179.74$ & $57.30$\\
\hline
IV/b & $7.2520$ & $0.0928609$ & $0.01068$ &$9.75$ & $-31.75$\\
IV/c & $6.0265$ & $0.1098228$ & $0.00472$ &$-117.62$ & $-4.75$\\
\hline
V/b & $5.1439$ & $0.0928622$ &$0.01489$ &$84.22$ & $-107.33$\\
V/c & $4.4826$ & $0.1098190$ &$0.01388$ &$149.42$ & $87.16$\\
\hline
VI/b & $5.9161$ & $0.0928578$ &$0.01790$ &$155.65$ & $-181.05$\\
VI/c & $6.1775$ & $0.1098202$ &$0.02734$ &$177.86$ & $56.85$\\
\hline
\hline
\end{tabular}
\label{tab:tab2}
\end{table}

A given energy plot is constructed for values of the two integrals $C$ and
$K$ computed for each studied system.  The energy of each system is
determined by the averaged Hamiltonian (see Appendix~A).  Energy
levels are plotted in the $\Sigma-$plane.  The energy level for the nominal
system is plotted with blue solid curve, while black solid curves are for
other values of the energy, from the maximum of the energy that corresponds
to the stable equilibrium (black cross-circle symbol in the $(\pi,
\pi)$-quarter of $\Sigma-$plane), down to
smaller values.  The energy levels
are limited from the bottom by $\overline{H}(e_1=0, e_2=0)$. The levels could be plotted also for smaller
values, although these levels
would become subsequently denser and they would be placed further from the
centre of the plane.  Other three cross-circle symbols in the remaining
quarters represent positions of the unstable equilibria.

Let us recall that Fig.~\ref{fig:fig8} shows positions of equilibria of the
averaged system (periodic orbits of the full, non-average $N$-body
model of motion) computed not for
one particular value of $C$, like in Fig.~\ref{fig:fig10}, but for a series of
values. This leads to whole branches/families of equilibria shown with green
and red curves for stable and unstable equilibria, respectively. On contrary,
green and red curves in Fig.~\ref{fig:fig10} represents periodic orbits of the
averaged system, green are for stable, while red -- for unstable
configurations. Big red/black symbols point where the nominal systems intersect
the $\Sigma-$planes. The points of intersections can be compared with the ones
in the right-hand panel of Fig.~\ref{fig:fig8}. Small differences are
present between the results of the $N$-body and the averaged model,
which may be easily explained. The averaged model is of the first order w.r.t. the
perturbation, which do not have to be necessarily small for such a compact two-planet
system. Nevertheless, a sequence of models from Fit~I (close to the stable
equilibrium) to Fit~VI (close to the unstable equilibrium) is apparent here as
well.

A common feature of all the models is a close proximity of their nominal energy
curves to the bifurcation of the branches of periodic orbits of
the averaged system in the $(\pi, \pi)$-quarter (see the 
arrows in the top
left-hand panel of Fig.~\ref{fig:fig10}). Moreover, the energy values are just
below the critical energy of the saddle point in the $(0, \pi)$-quarter.
Naturally, those two characteristics of the energy for the nominal systems, are
not independent one from another, since the structure of periodic orbits is
determined by positions of the equilibria. This feature may be a "fingerprint"
of the migration scenario, which we discuss in the next Section.

\begin{table}
\centering
\caption{
Mean parameters of the selected configurations whose osculating Keplerian
elements are given in Tab.~\ref{tab:tab2}.
}
\begin{tabular}{l c c c c}
\hline
\hline
solution/planet & $m\,[\mE]$ & $a\,[\au]$ & $e$ & $\sigma$\,[deg]\\
\hline
I/b & $7.5173$ & $0.0928783$ & $0.00602$ & $53.16$\\
I/c & $6.5125$ & $0.1098163$ & $0.00871$ & $220.76$\\
\hline
II/b & $6.2177$ & $0.0928756$ & $0.00691$ & $68.06$\\
II/c & $5.4607$ & $0.1098110$ & $0.00927$ & $209.73$\\
\hline
III/b & $7.5940$ & $0.0928799$ & $0.00779$ & $12.87$\\
III/c & $6.6879$ & $0.1098067$ & $0.00862$ & $-111.90$\\
\hline
IV/b & $7.2520$ & $0.0928714$ & $0.01126$ & $58.13$\\
IV/c & $6.0265$ & $0.1098052$ & $0.00491$ & $193.85$\\
\hline
V/b & $5.1439$ & $0.0928691$ & $0.01470$ & $-16.53$\\
V/c & $4.4826$ & $0.1098080$ & $0.01450$ & $-84.06$\\
\hline
VI/b & $5.9161$ & $0.0928795$ & $0.01724$ & $89.52$\\
VI/c & $6.1775$ & $0.1098070$ & $0.02797$ & $68.22$\\
\hline
\hline
\end{tabular}
\label{tab:tab3}
\end{table}

\section{Planetary migration}

To reproduce the observational Kepler-29 system, and its features
discussed in Section~3,
we conducted migration simulations within a simple parametric model
of the force which mimics the planet-disc interactions
\citep[e.g.,][]{Beauge2006,Moore2013}
\begin{equation}
\vec{f}_i = -\frac{\vec{v}_i}{2\,\tau_{a,i}} - \frac{\vec{v}_i - \vec{v}_{c,i}}{\tau_{e,i}},
\end{equation}
where $\vec{v}_i$ is the astrocentric velocity of planet~$i$ ($i=1,2$),
$\vec{v}_{c,i}$ is the velocity at circular orbit at a distance of planet~$i$. The
time-scales of migration and circularisation of planet~$i$ are denoted by
$\tau_{a,i}$ and $\tau_{e,i}$, respectively. 

\begin{center}
\begin{minipage}[l]{\linewidth}
\hspace{-1mm}\includegraphics[width=1.03\textwidth]{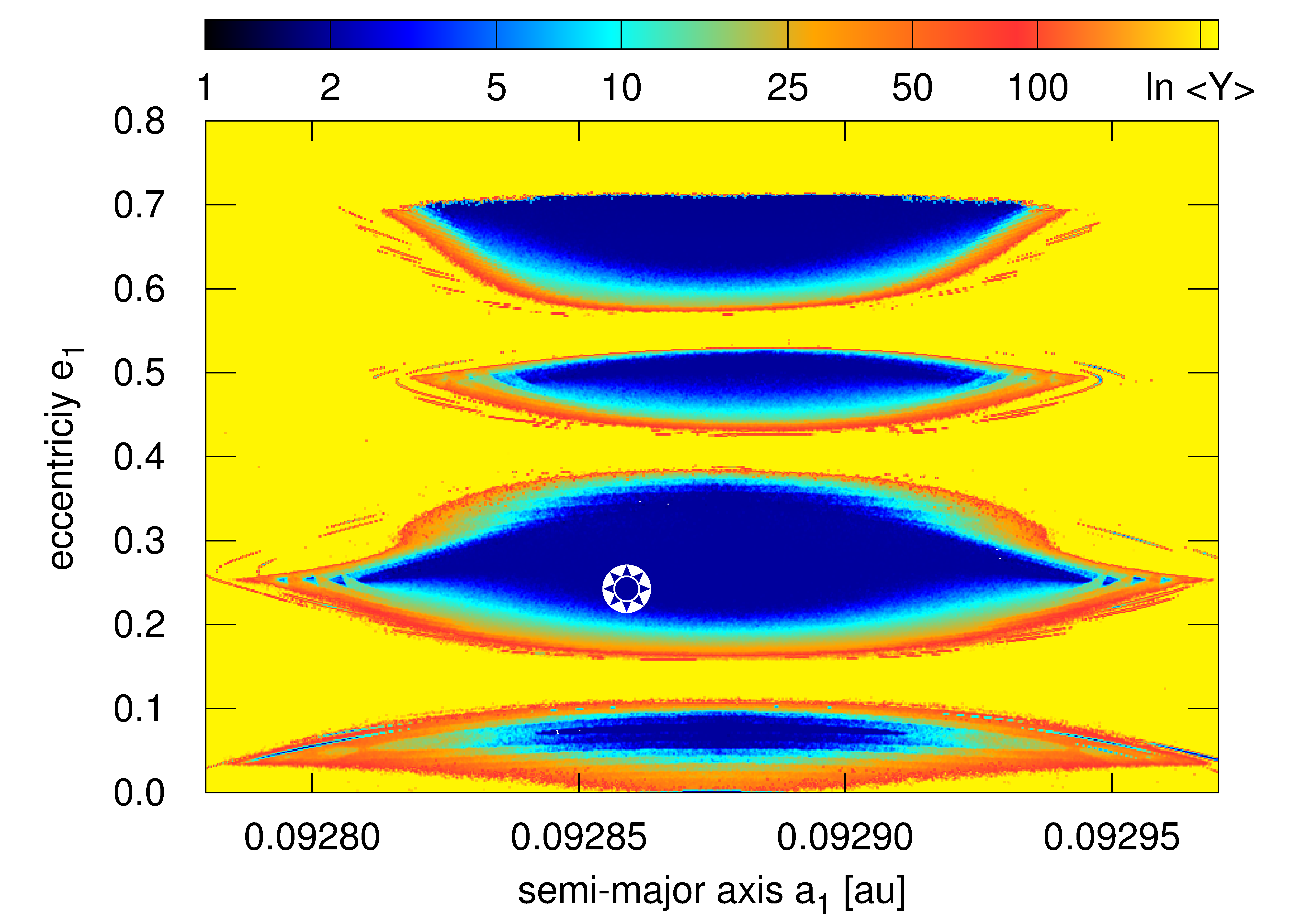} 
\captionof{figure}{The MEGNO dynamical map for a TTV model in a region of large eccentricities (see
Fig.~\ref{fig:fig1}) out of $\Delta\varpi=0$, computed for 64~kyrs
($\sim2.5\times 10^6 P_2$). The MEGNO indicator, $\Y \sim 2$ indicates a regular
(long-term stable) solution marked with blue colour, $\Y$ much larger than $2$,
up to $\gtrsim 256$ indicates a chaotic solution (light blue/red/yellow).
}
\label{fig:fig7}
\end{minipage}
\end{center}

We assumed that $\tau_{e,i} =
\tau_{a,i}/\kappa$, where $\kappa$ is constant and 
\begin{equation}
\tau_{a,i} = \tau_0 \, \left( \frac{r_i}{1\,\au} \right)^{-\alpha} \, \exp(t/T),
\end{equation}
where $\tau_0$, $\alpha$ and $T$ are constant. The free parameters of the model
were being changed in wide ranges, i.e., $\kappa \in [1, 300]$, $\alpha \in
[0.1, 1.5]$, $\tau_0 \in [10^3, 10^6]\,\yr$, $T \in [10^4, \infty]\,\yr$.
Initial orbits were chosen such that the period ratio was between $9/7$ and
$4/3$, the eccentricities $\sim 0$, and the angles were chosen to be $0$. The choice of the initial period ratio smaller than $4/3$ stems from the fact that for initial $P_2/P_1 > 4/3$, the system would very likely enter 4:3~MMR, taking the migration parameters from the ranges given above.

\begin{figure*}
\centerline{
\vbox{
\hbox{
\includegraphics[width=0.33\textwidth]{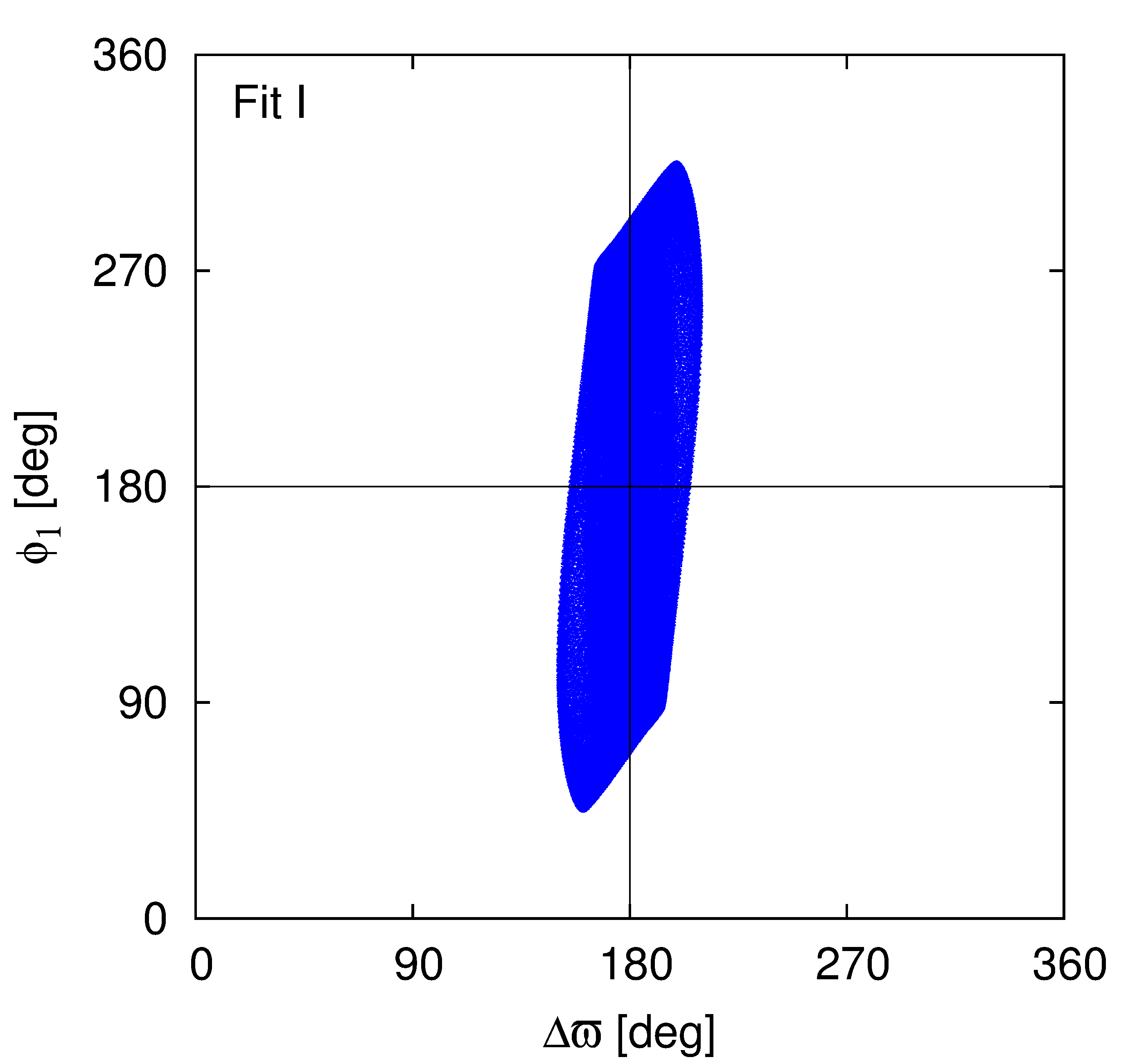}
\includegraphics[width=0.33\textwidth]{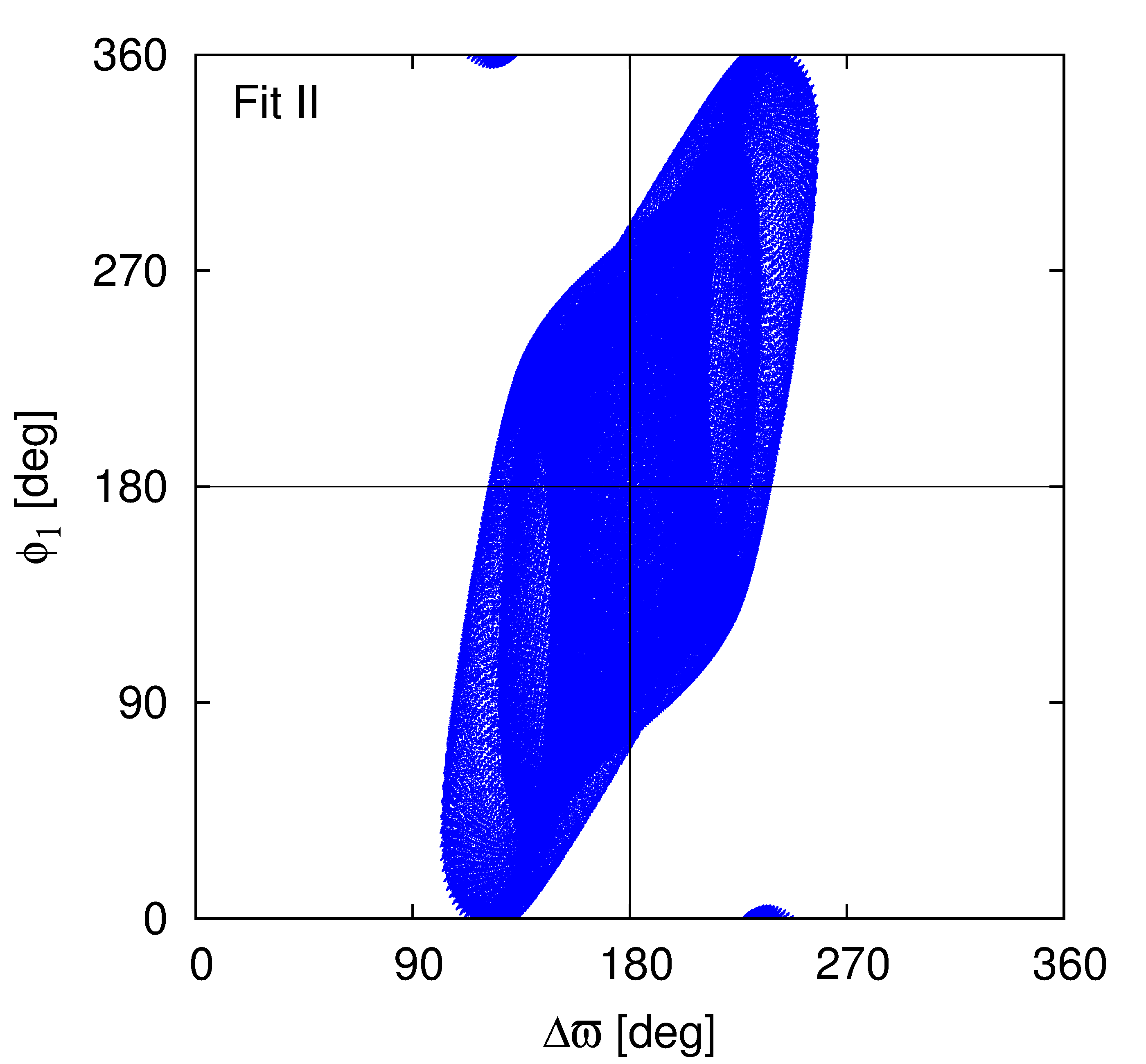}
\includegraphics[width=0.33\textwidth]{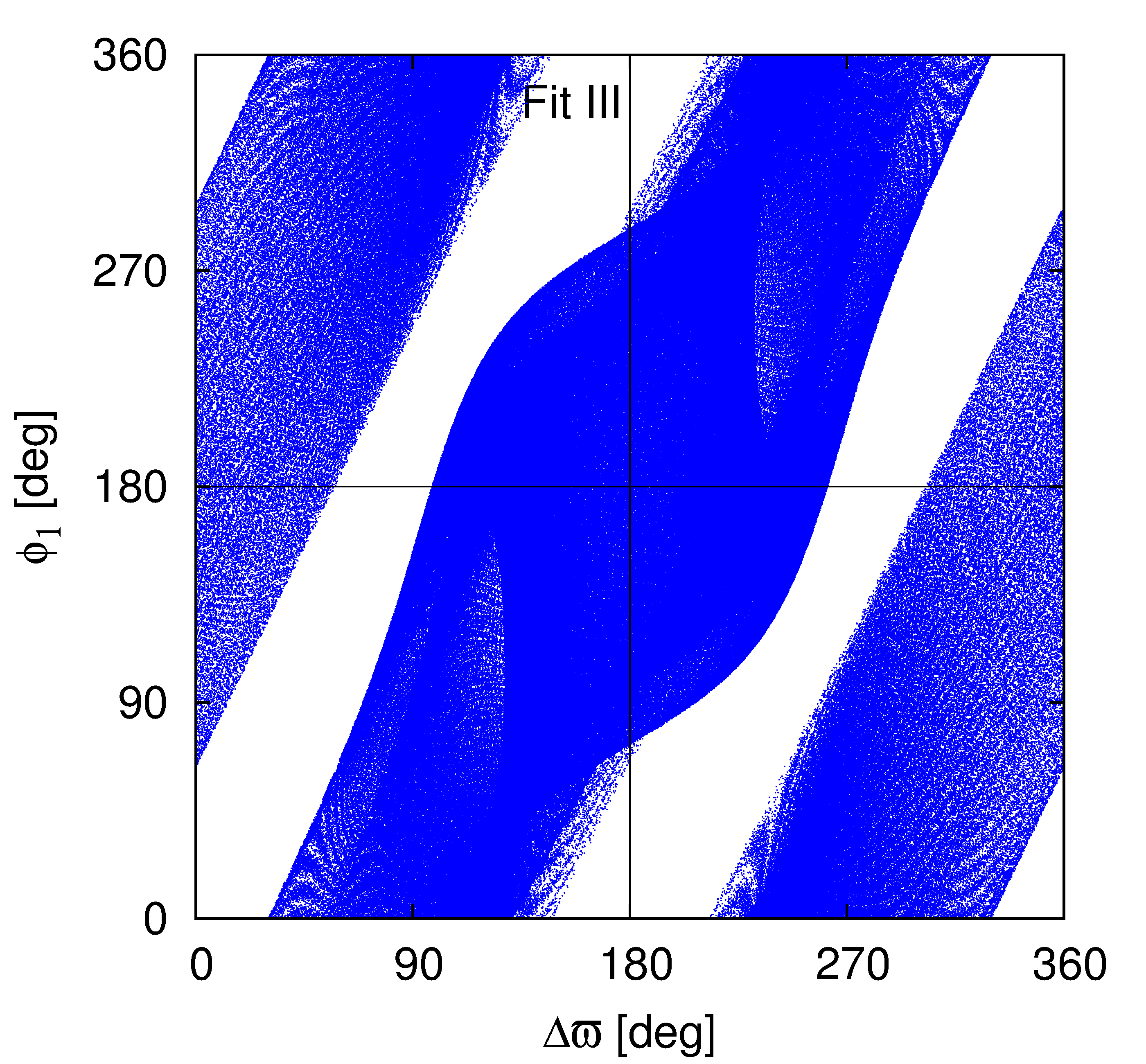}
}
\hbox{
\includegraphics[width=0.33\textwidth]{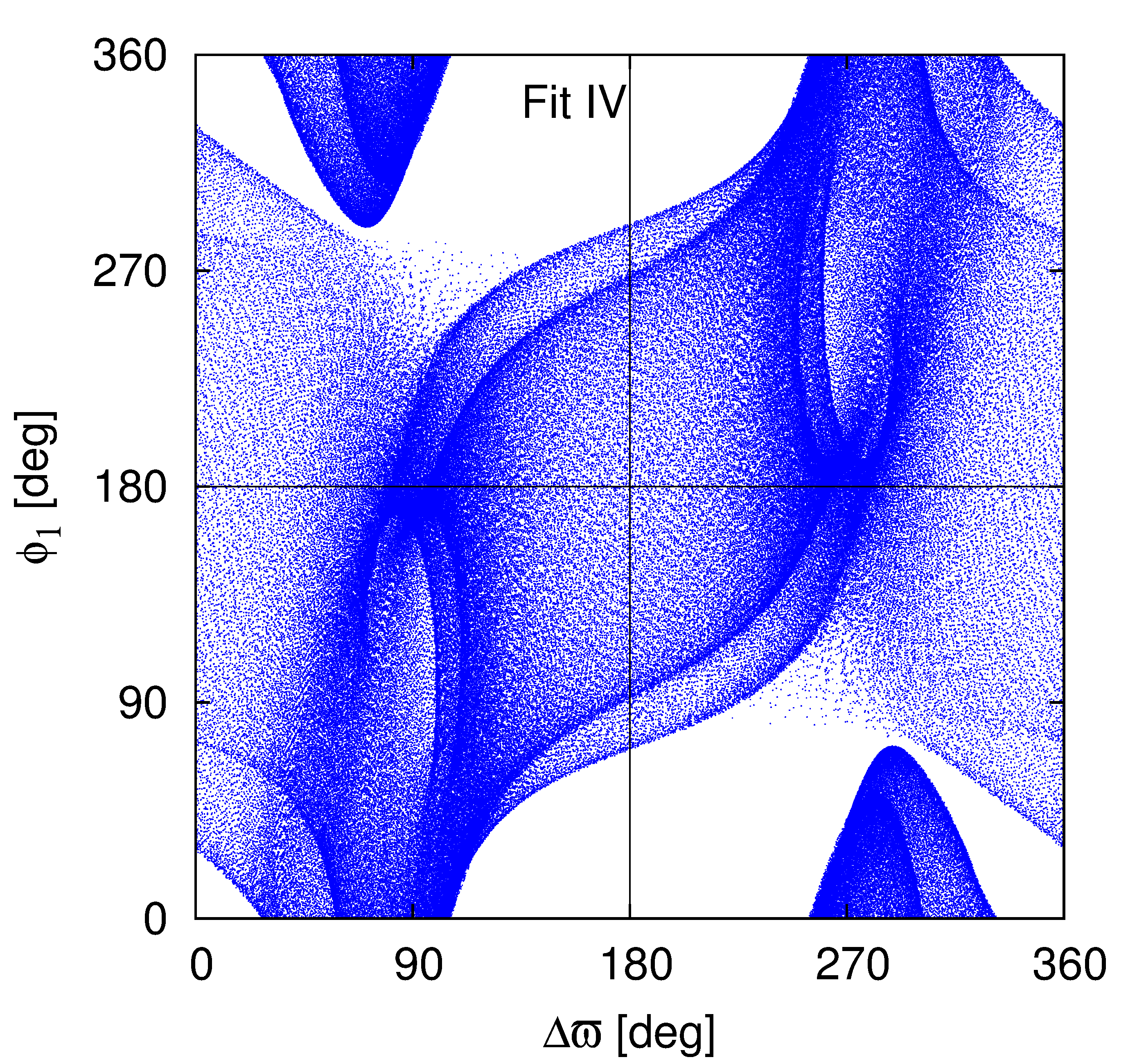}
\includegraphics[width=0.33\textwidth]{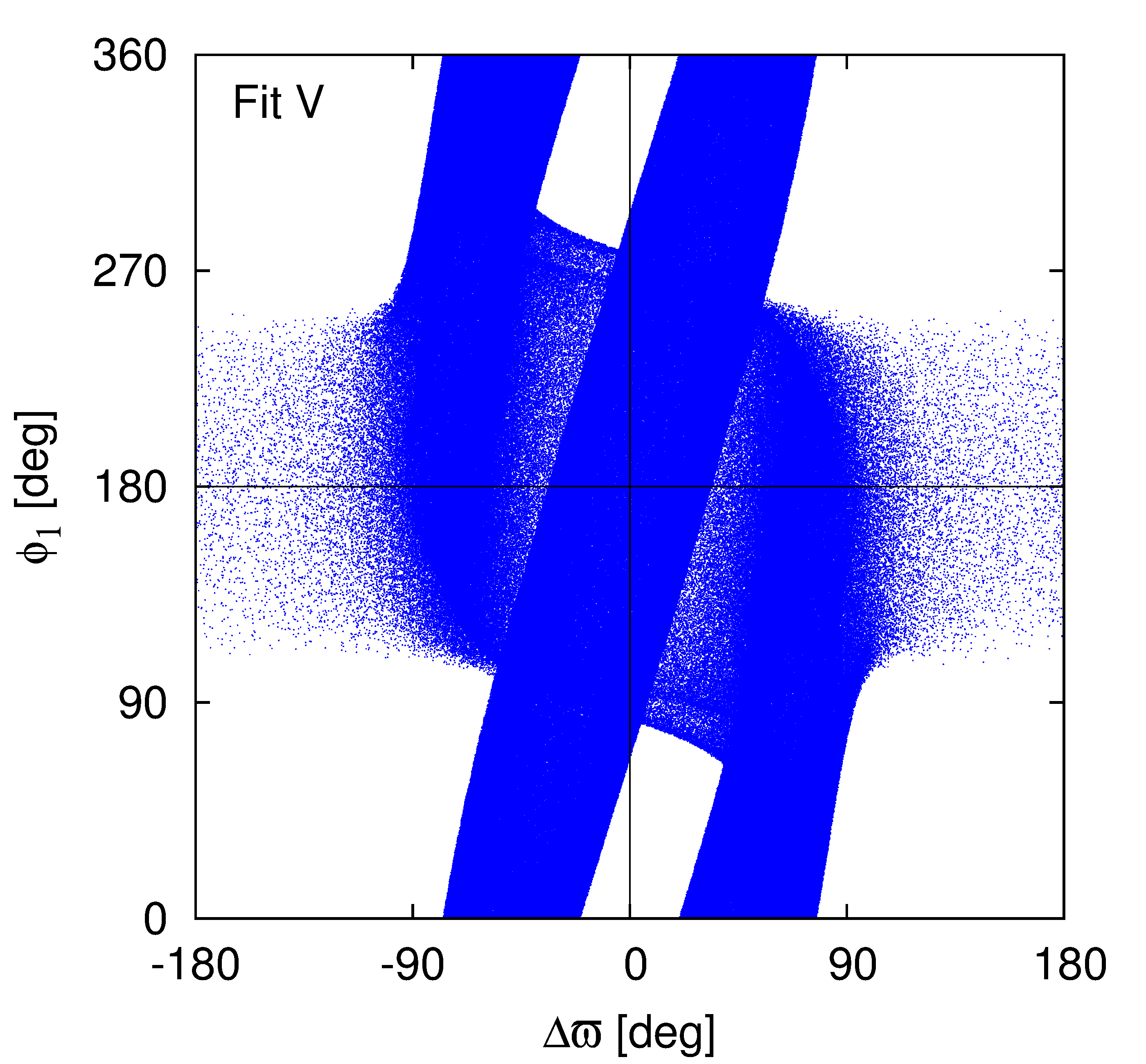}
\includegraphics[width=0.33\textwidth]{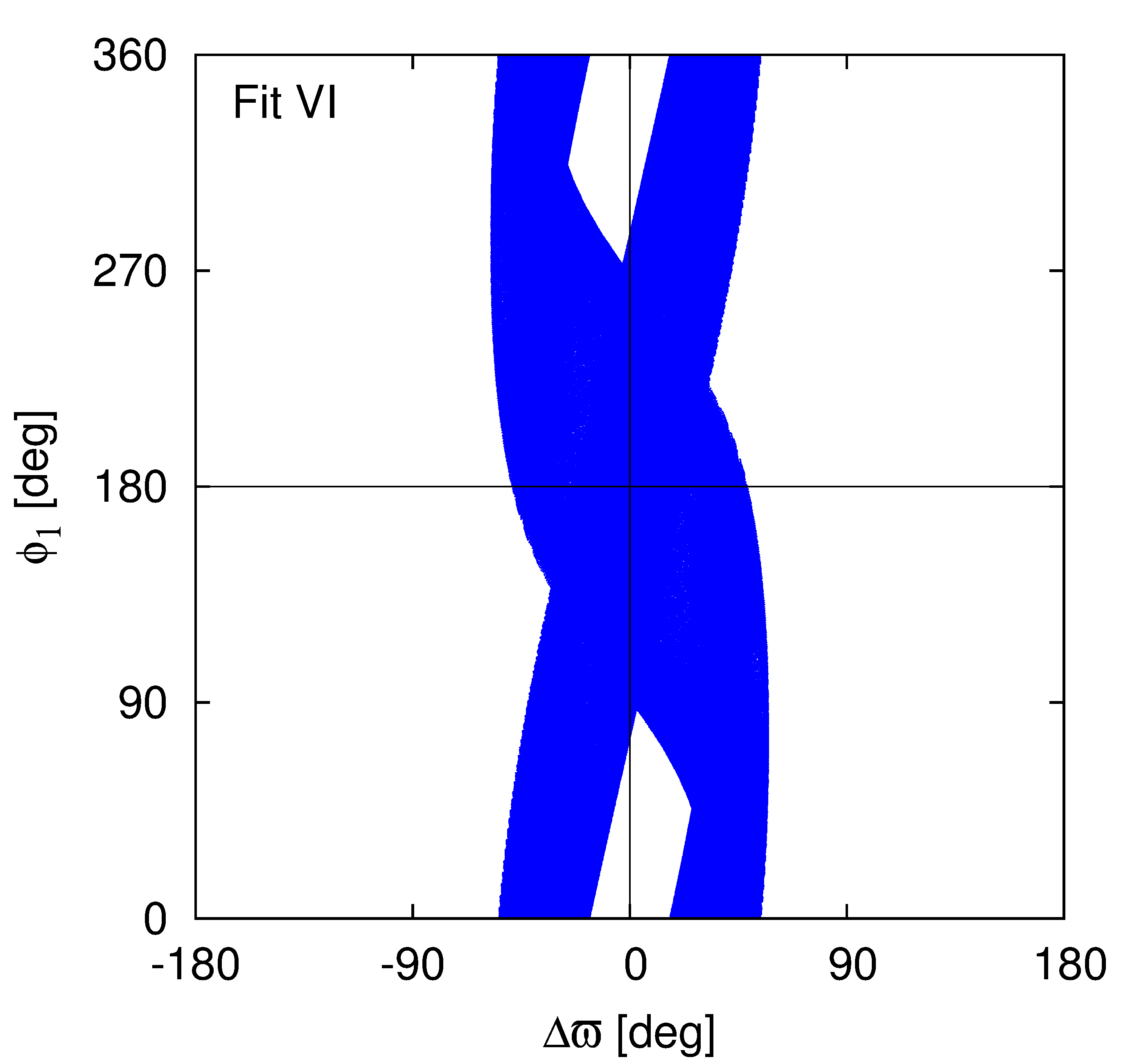}
}
}
}
\caption{
Evolution of example configurations that fit the data (see Table~\ref{tab:tab2}
for the masses and orbital parameters) presented at the $(\Delta\varpi,
\phi_1)$-diagram. The integration time is $10^5\,$yr.
}
\label{fig:fig9}
\end{figure*}

After a series of simulations we found the following properties of 
systems stemming
from the migration. For small $\kappa \lesssim 10$, moderate and high
eccentricities ($\gtrsim 0.03$) were possible to obtain, however $\Delta\varpi$
librates around $\pi$, which is opposite to the results of fitting the
data. Additionally, such systems have always $\phi_1$ librating around $\pi$.
Both the angles librate around $\pi$. For moderate $\kappa \sim 100$, small
eccentricities $\lesssim 0.01$ are obtained and $\Delta\varpi$ spans
the whole
range, revealing both librations around $\pi$ or circulations, which
agrees with the statistics of models that fit the data. 

Next, we tried to find configurations resulting from the convergent
migration, which could form the sequence of six models analysed in the
previous Section.  They should transform from one class of configurations
(close to the stable equilibrium) to another class (close to the unstable
equilibrium).

\begin{figure*}
\centerline{
\vbox{
\hbox{
\includegraphics[width=0.45\textwidth]{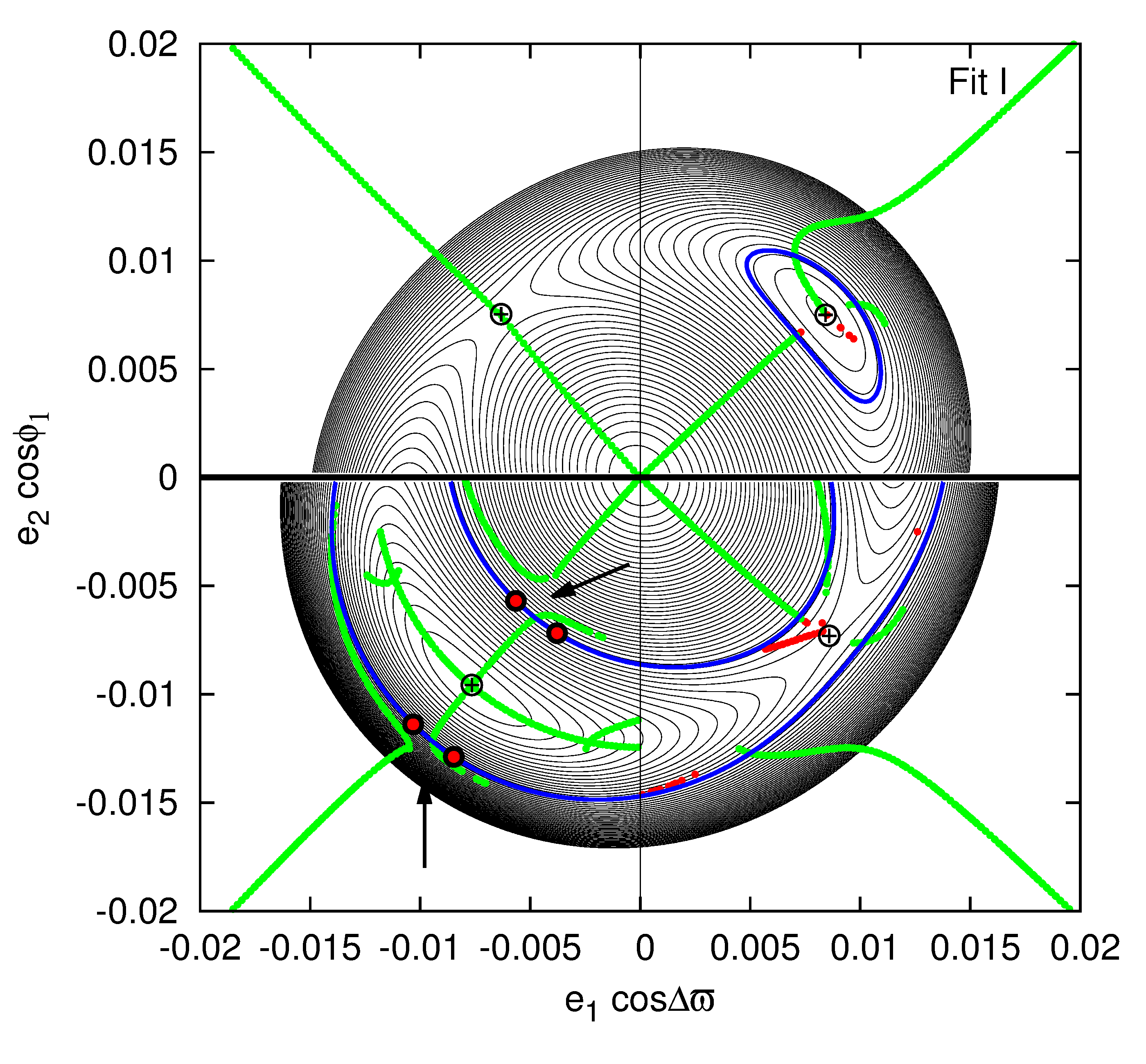}
\includegraphics[width=0.45\textwidth]{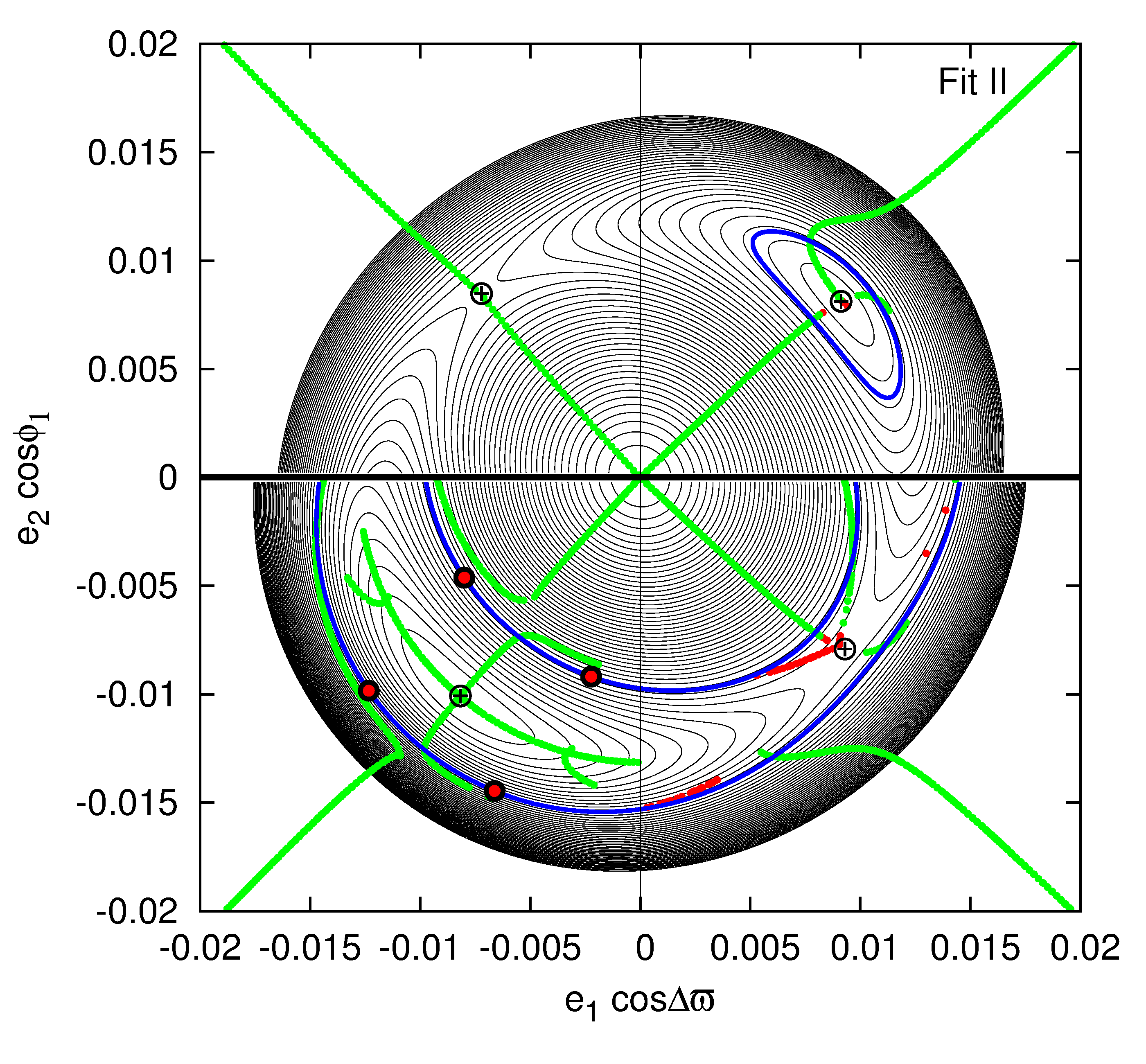}
}
\hbox{
\includegraphics[width=0.45\textwidth]{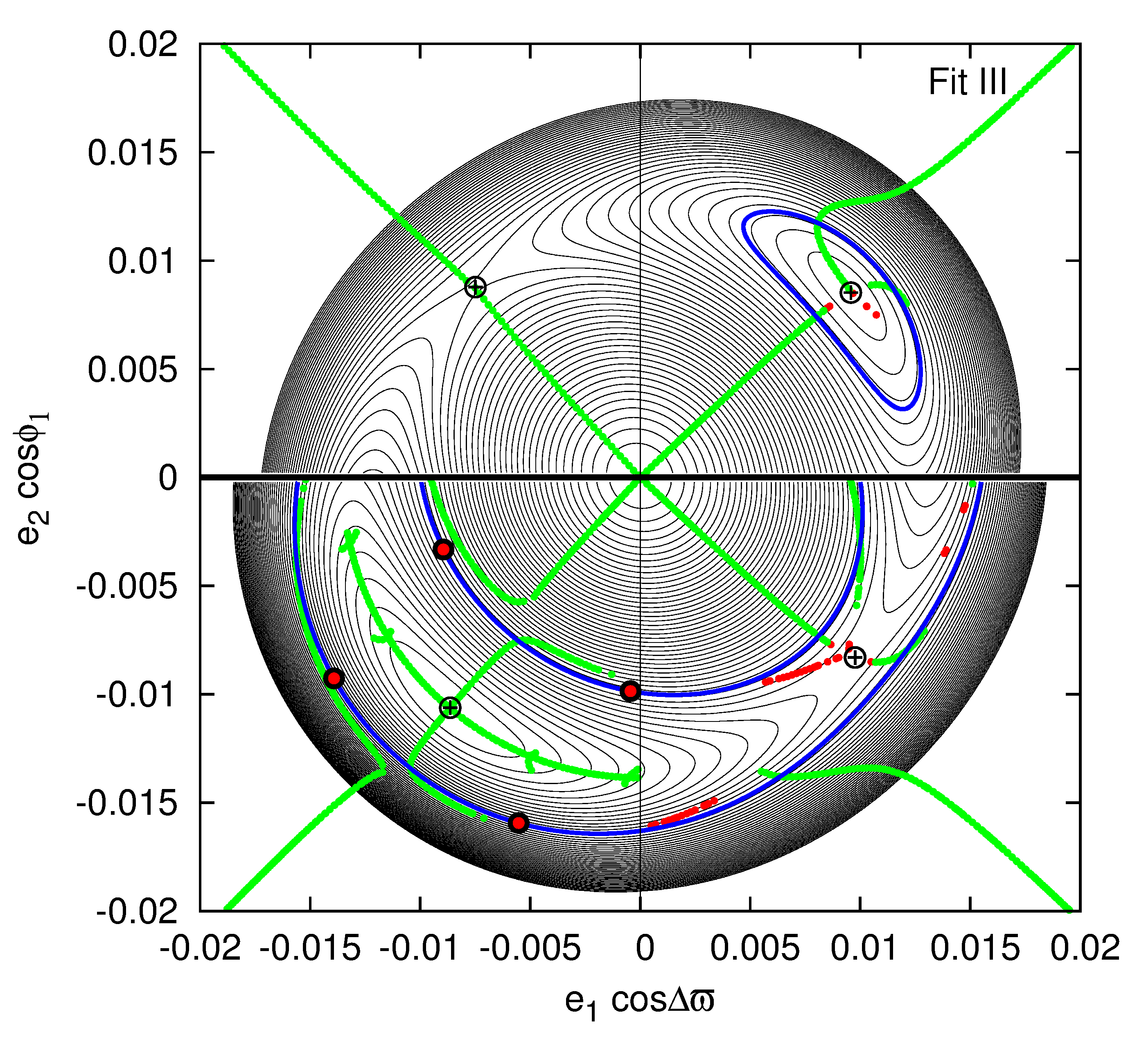}
\includegraphics[width=0.45\textwidth]{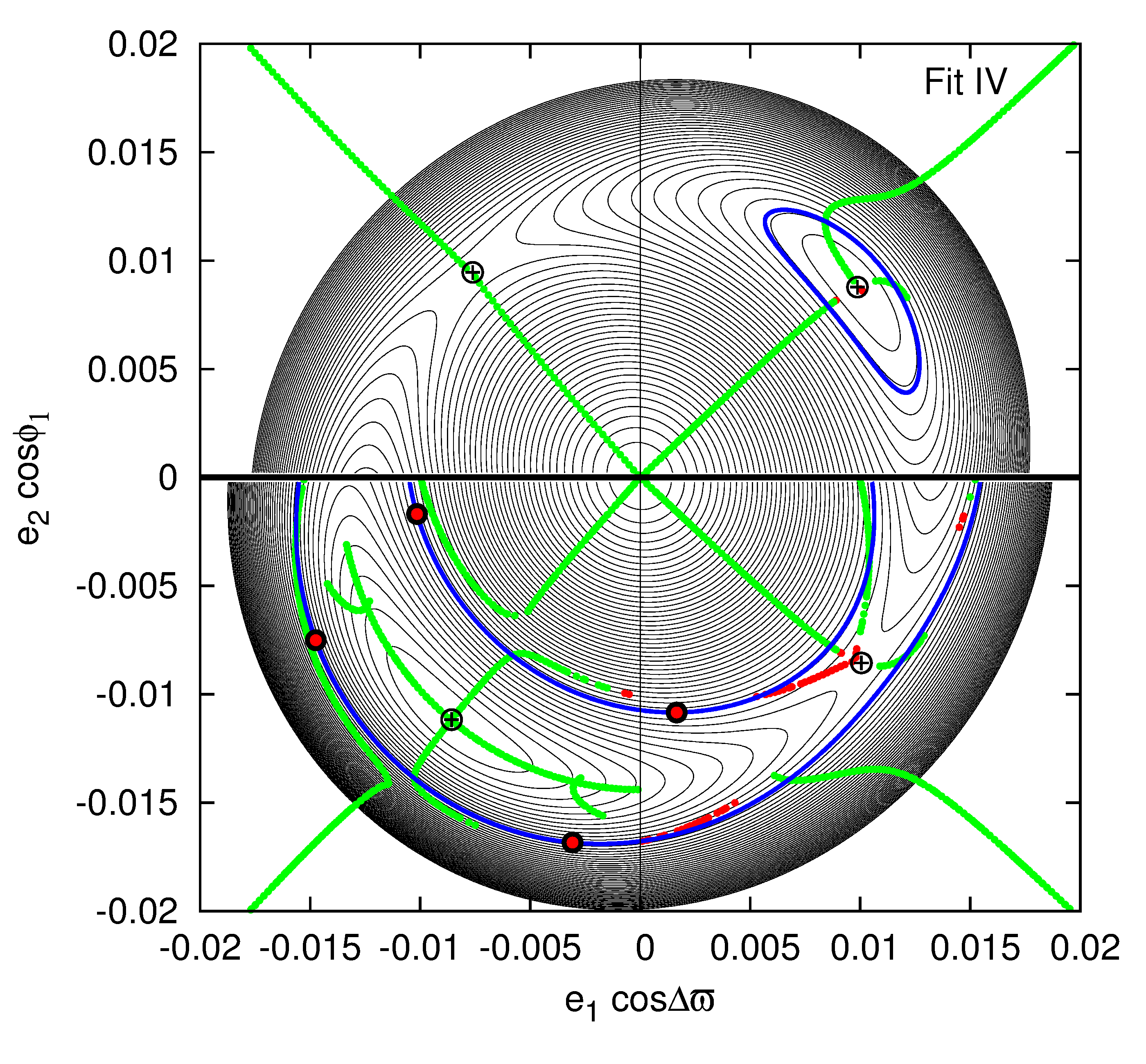}
}
\hbox{
\includegraphics[width=0.45\textwidth]{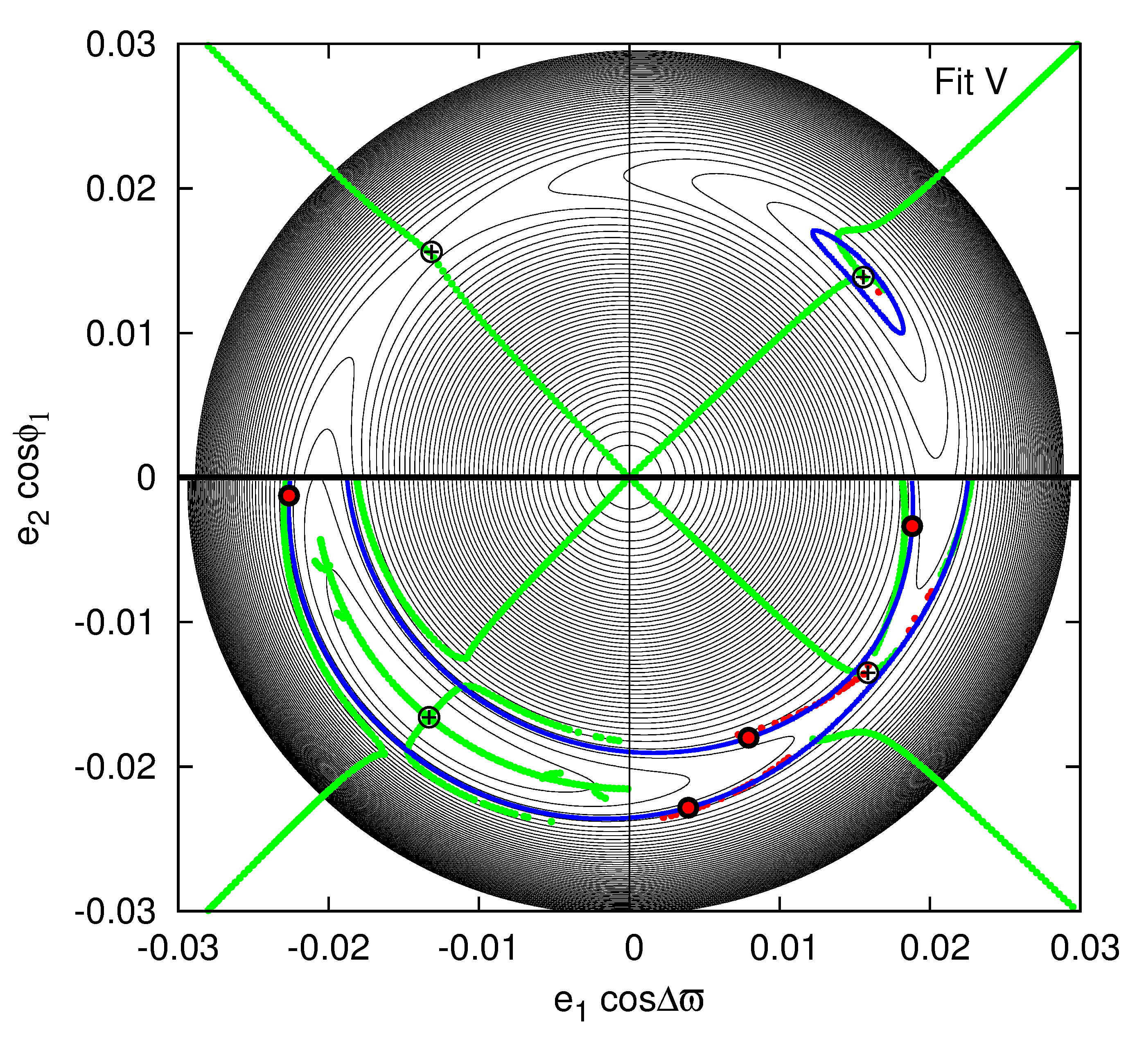}
\includegraphics[width=0.45\textwidth]{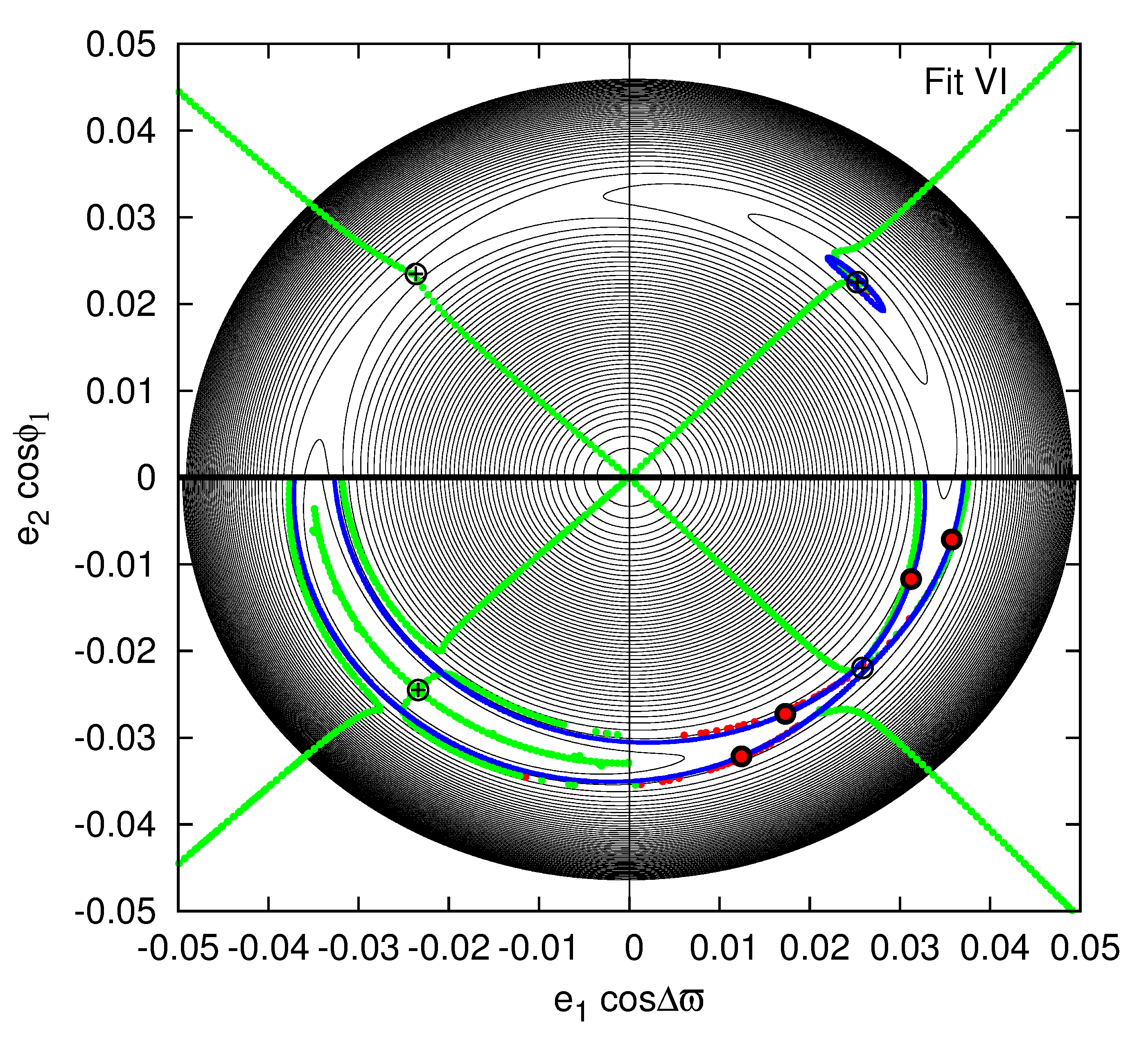}
}
}
}
\caption{
Energy levels (black curves) of the averaged system at the $\Sigma$--plane.
Masses and the $C$ and $K$ integrals values are for the selected configurations
(see Tab.~\ref{tab:tab3} for the parameters). Each panel is for one of the six
models chosen for the analysis. Blue curves are for energies of the nominal
systems. Green and red curves denote periodic configurations of the averaged
model, while black cross-circle symbols are for the periodic configurations of
the $N$-body (unaveraged) model, i.e., equilibria of the averaged system. Big
red symbols show intersections of the representative plane by the trajectories
of the nominal systems.
}
\label{fig:fig10}
\end{figure*}

\begin{figure}
\centerline{
\vbox{
\hspace{0,2mm}\hbox{
\includegraphics[width=0.48\textwidth]{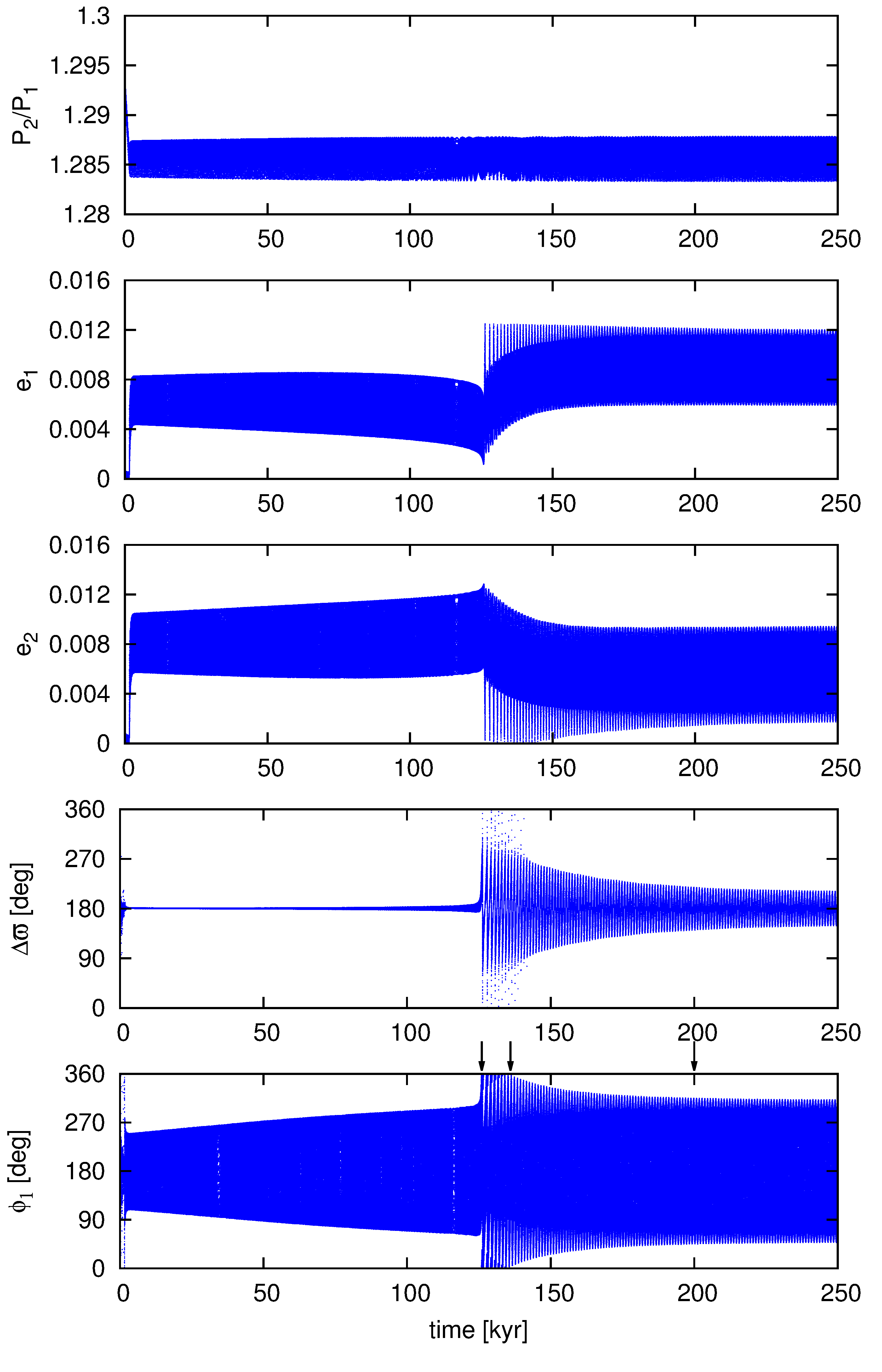}
}
}
}
\caption{
An example of the migration simulation which leads to systems similar to
Kepler-29. Initial semi-major axes are $a_1 = 0.2\,\au$ and $a_2 =
0.23723\,\au$, eccentricities $e_1 = e_2 = 0.0001$, both arguments of pericenter and
mean anomalies are set to $0$. Planets masses are $m_1 = 5.1\,\mE$ and $m_2 =
4.4\,\mE$. Parameters of the migration $\tau_0 = 12\,$kyr, $T = 40\,$kyr,
$\alpha = 1.3$, $\kappa = 130$.
}
\label{fig:fig11}
\end{figure}

After a series of experiments, we found that three out of six models could
be qualitatively reconstructed by a single migration simulation.  The
results of such a simulation are illustrated in Fig.~\ref{fig:fig11}. 
Its parameters as well as the initial orbital elements are given in the
caption to this Figure.  Subsequent panels, from the top to the bottom,
correspond to the period ratio, eccentricities, $\Delta\varpi$ and $\phi_1$
evolution in time.  Shortly after the system reaches the 9:7~MMR, the
eccentricities are excited, i.e., $e_1$ oscillates in a range of $[0.004,
0.008]$ and $e_2 \in [0.006, 0.01]$.  Both critical angles librate
around $\pi$, $\Delta\varpi$ with very small amplitude $<1^{\circ}$, while
$\phi_1$ with much larger amplitude of $\sim 70^{\circ}$.  Both the
amplitudes increase in the first part of the simulation $t < 125\,$kyr.  If
the migration has stopped for some reason at this stage of the
evolution, $\Delta\varpi$ would librate with too small amplitude, when
compared to the examples listed in Tab.~\ref{tab:tab2} and also to the
systems from the whole statistics of the TTV fits.

At $t \sim 125\,$kyr the system switches into a different regime of motion. Both
$\Delta\varpi$ and $\phi_1$ start to circulate, the ranges of the eccentricities
oscillations also change. After $\sim 10\,$kyr since the transition (at $t \sim
135\,$kyr) the angles start to librate again with decreasing amplitudes. One
should keep in mind, though, that it is not a generic situation. If $T$ had a
higher value (which would mimic slower disc dispersion), the system would leave
the resonance eventually. If, on the other hand, $T$ was smaller, the migration
could stop before the transition between the two discussed regimes of motion.

More details about the evolution of a system trapped in 9:7~MMR could be
found in our upcoming paper.  Here we only present an example showing that
systems listed in Tab.~\ref{tab:tab2} can be formed through the migration.
A critical issue is the transition described above. It occurs if
the energy of the migrating system reaches the bifurcation of the branches of periodic
orbits (see the arrows in the top left-hand panel of Fig.~\ref{fig:fig10}).

\begin{figure*}
\centerline{
\vbox{
\hbox{
\includegraphics[width=0.33\textwidth]{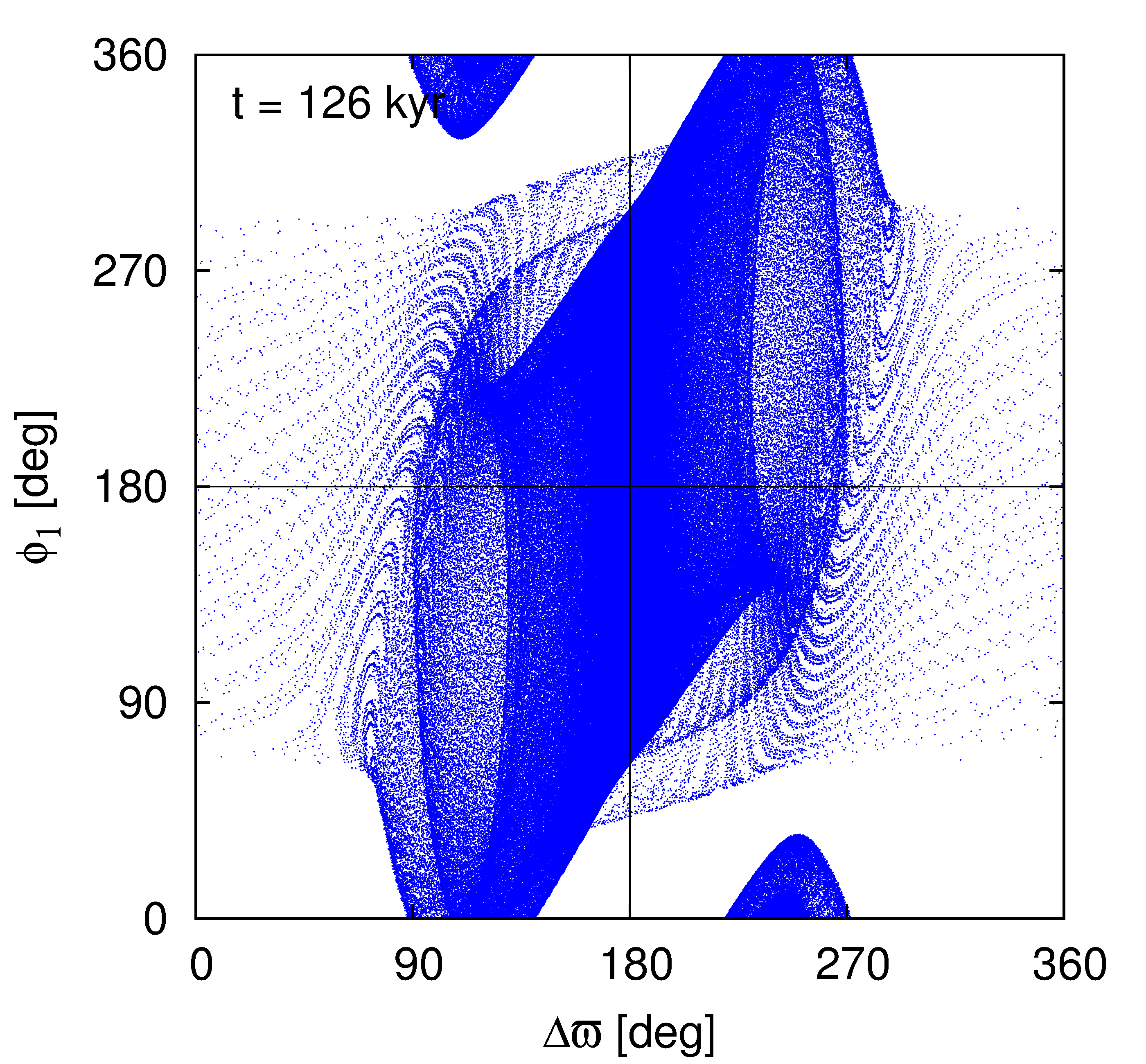}
\includegraphics[width=0.33\textwidth]{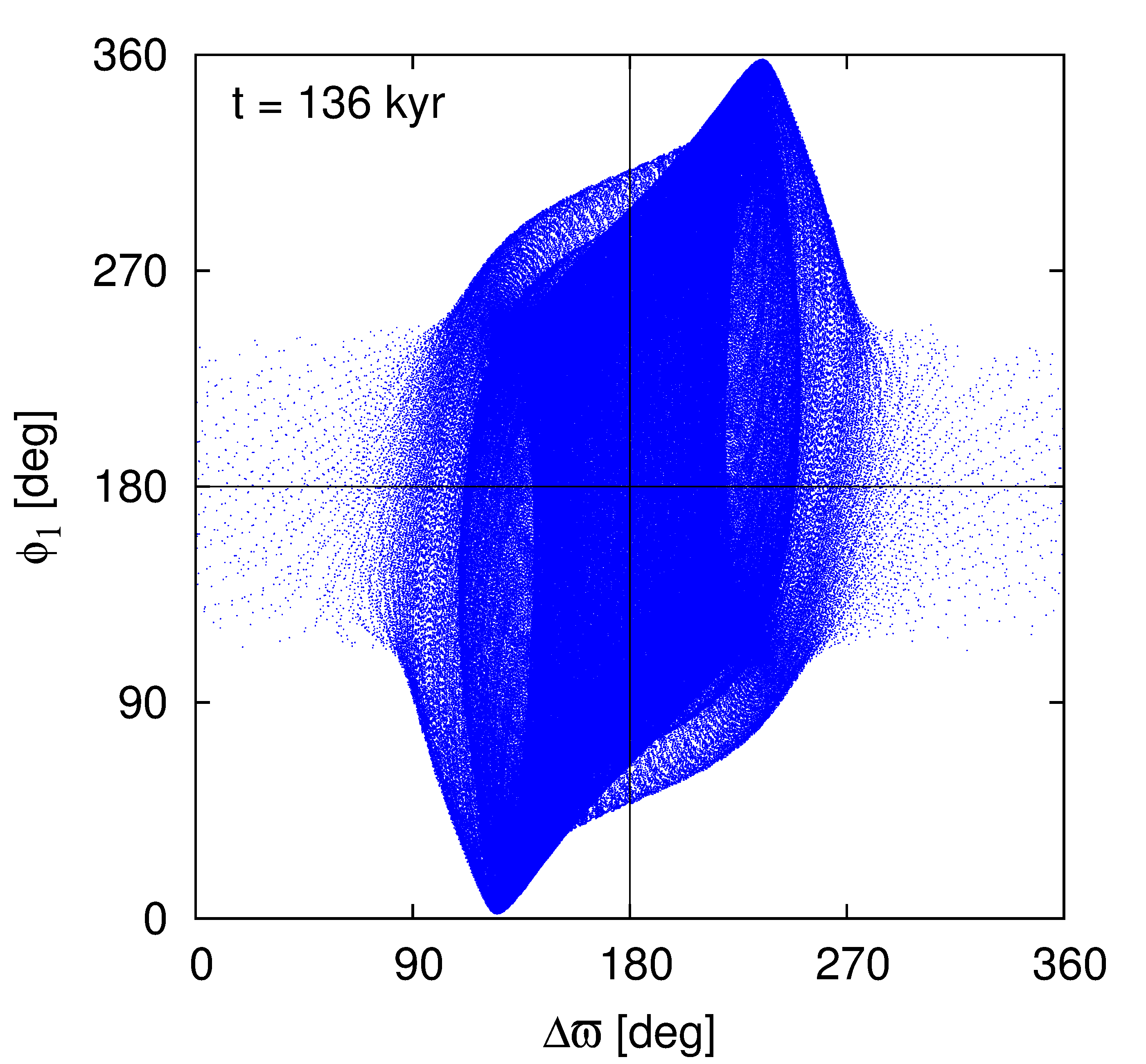}
\includegraphics[width=0.33\textwidth]{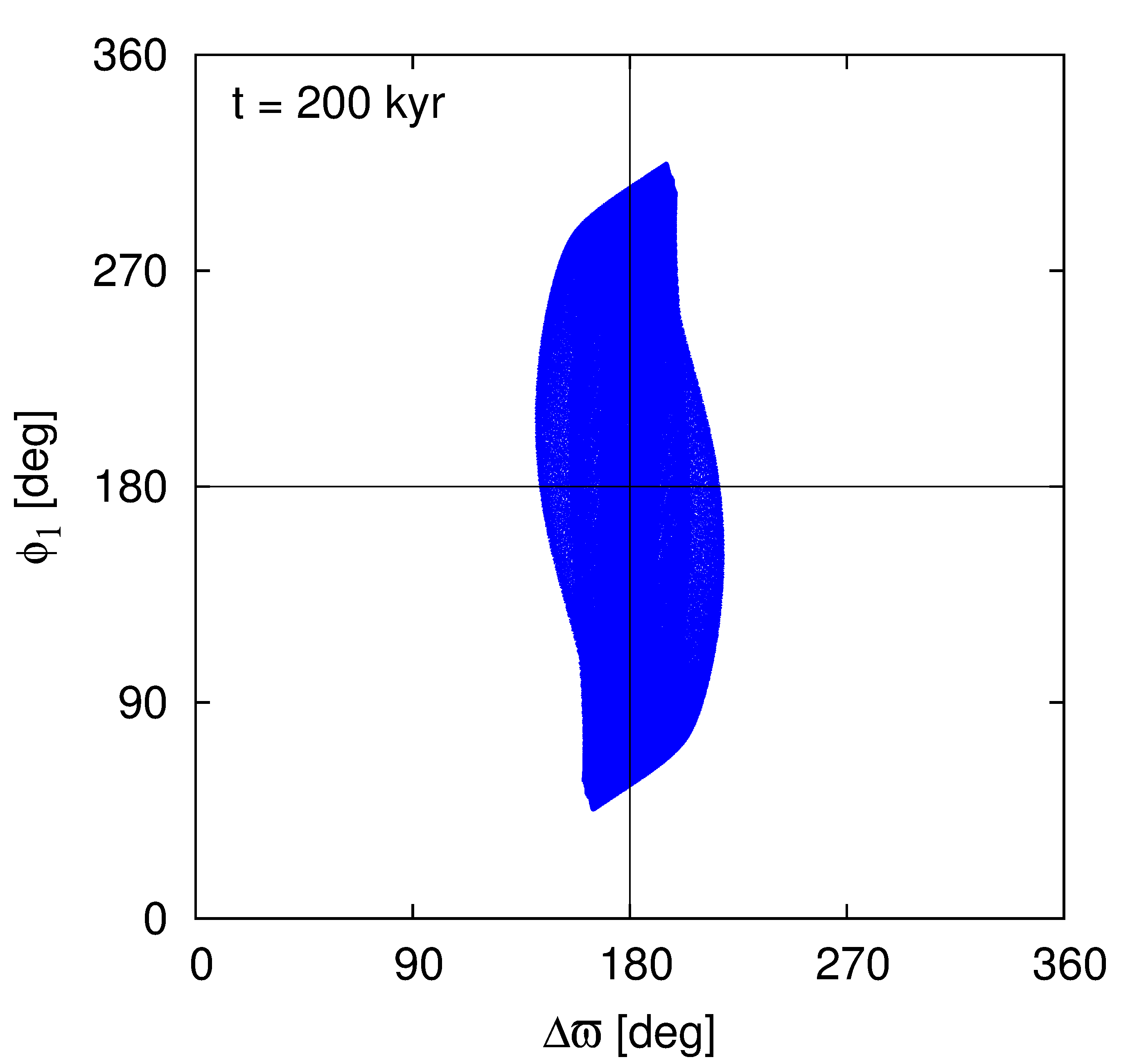}
}
}
}
\caption{
Evolution of example configurations stemming from the migration simulation
illustrated in Fig.~\ref{fig:fig11}, presented at the $(\Delta\varpi,
\phi_1)$-diagram. The integration time is $10^5\,$yr.
}
\label{fig:fig12}
\end{figure*}

We chose three moments of the simulation, $t = 126, 136$~and
$200$\,kyr, respectively, and we
integrated the $N$-body equations of motion for the three sets of the osculating
Keplerian elements at those epochs. Snapshots of their evolution
are presented at the $(\Delta\varpi, \phi_1)$-diagrams in Fig.~\ref{fig:fig12}.

The left-hand panel reminds the bottom left-hand panel of Fig.\ref{fig:fig9}
(Fit~IV) as well as the middle panel of Fig.~\ref{fig:fig5} for the
observational model close to the  best-fitting
solution in Tab.~\ref{tab:tab1}.  Both the angles formally circulate, however, similarly to
Fit~IV, $\phi_1$ librates with the amplitude greater than $2\pi$. The
evolution of the angles is not independent one from another, since the phase
trajectory avoids certain areas of the $(\Delta\varpi, \phi_1)$-diagram. 
The right-hand panel of Fig.~\ref{fig:fig12} reminds the top left-hand panel
of Fig.~\ref{fig:fig9} (Fit~I).  Both the angles librate and the amplitudes
for the simulated system correspond well to the amplitudes for Fit~I.  The
system in the middle panel can be interpreted as an intermediate state
between the systems illustrated in the left and right panels of
Fig.~\ref{fig:fig12}.  The amplitude of $\phi_1$ libration reaches $2\pi$,
which corresponds to the behaviour of Fit~II (the top middle panel of
Fig.~\ref{fig:fig9}).  The only difference between the simulated system and
Fit~II is the behaviour of $\Delta\varpi$. Yet we also note that
the evolution of critical angles for Fits~V and VI is very similar to
the observational models illustrated in the left and right panels of
Fig.~6.

The examples stemming from the migration simulation ensure us
that the Kepler-29 system could be formed by the planetary migration if its
orbits are close to circular.  On the other hand, the systems  with higher
eccentricities (and aligned orbits) which are also consistent with the
TTV observations are less likely to be formed in this way. 
Nevertheless, the migration induced formation of the 9:7~MMR as well as other
second- and higher-order resonances is a very complex mechanism. 
It needs to be studied in more details in order to bring a definitive 
solution.

%
\section{Conclusions}
%

We analysed the TTV data from \citep{Rowe2015} of the Kepler-29 system with two
low-mass planets of a period ratio very close to 9/7 \citep{JontofHutter2016}.
We confirmed that the masses of the planets are within a few Earth mass range,
i.e., $\sim 6\,\mE$ and $\sim 5\,\mE$ for the inner and the outer planet,
respectively. We demonstrated that, although the eccentricities as well as
longitudes of pericenters are not well determined, the system is very
likely in an exact 9:7~MMR. We found configurations with both aligned and
anti-aligned apsides, that are long-term stable and fit the data equally
well. The eccentricities may be as high as $0.3-0.4$ for models with aligned
orbits, while for anti-aligned configurations only low eccentric orbits are
allowed by the observational and stability constraints.

We demonstrated that the critical angles of the resonant configurations do not
necessarily librate. That implies also that the secular angle
$\Delta\varpi$ may both rotate or librate, around $0$ or $\pi$. The resonant
nature of such systems can be verified at the frequency maps (right-hand column
of Fig.~\ref{fig:fig5}) as well as at the $(\Delta\varpi, \phi_1)$-diagrams
(Fig.~\ref{fig:fig9}). The fundamental frequencies related to the mean motions
are very close to the nominal value of 9/7 for the systems whose resonant angles
rotate. Moreover, the evolution of the angles $\Delta\varpi$ and $\phi_1$ is
correlated.

We showed that the best-fitting solutions with low eccentricities (both with
aligned and anti-aligned apsides) are shifted with respect to the periodic
orbits (equilibria of the averaged system) of 9:7~MMRs, and demonstrated that it
is a natural outcome of the planetary migration. That holds even for
configurations that lie close to the branch unstable periodic orbits for
$\Delta\varpi=0$ (Fit~IV). On the other hand, we showed that configurations with
$e \gtrsim 0.03$ and $\Delta\varpi \sim 0$ are unlikely to be formed on the way
of migration. Systems with $e \gtrsim 0.03$ and $\Delta\varpi \sim \pi$ can form
this way, but configurations of this sort do not fit the TTV
observations. Therefore, we conclude that if the Kepler-29 system was formed through
the smooth migration, its orbits are low eccentric $e \lesssim 0.03$, but the
behaviour of $\Delta\varpi$ and the resonant angles can be hardly
determined on basis of the available TTV data.
%
\section{Acknowledgements}
%
We would like to thank the anonymous referee for helpful suggestions that helped us to improve the paper.
We thank Ewa Szuszkiewicz, John Papaloizou and Zija Cui for discussions
stimulating our interest in the Kepler-29 dynamics.
This work has been supported by Polish National Science Centre MAESTRO grant
DEC-2012/06/A/ST9/00276. K.G. thanks the staff of the Pozna\'n Supercomputer and
Network Centre (PCSS, Poland) for a generous support and 
computing resources (grant No. 195).
%
\appendix 
\section{Analytic 9:7 MMR Hamiltonian} \label{App:AppendixA}
%
The secular and the resonant parts of the averaged Hamiltonian (Eq.~\ref{eq:R}) read as follows \citep{Murray1999a}.

The secular Hamiltonian is expressed through,
\begin{eqnarray}
\overline{R}_{\idm{sec}} & = & f_1 + \left( e_1^2 + e_2^2 \right) f_2 + e_1^4 \, f_4 + e_1^2 \, e_2^2 \, f_5 + e_2^4 \, f_6 \nonumber\\
& & +\left( e_1 \, e_2 \, f_{10} + e_1^3 \, e_2 \, f_{11} + e_1 \, e_2^3 \, f_{12} \right) \cos\Delta\varpi\\
& & +e_1^2 \, e_2^2 \, f_{17} \, \cos 2\Delta\varpi.\nonumber
\end{eqnarray} 
where $f_k$ are functions of the semi-major axes ratio $X \equiv a_1/a_2$ through the Laplace coefficients $b_{1/2}^{(j)}(X)$ and their derivatives. They read
\begin{flalign*}
f_1 & = \frac{1}{2} D_{0,0},&
\end{flalign*}
\begin{flalign*}
f_2 &= \frac{1}{8} \left( 2 D_{1,0} + D_{2,0} \right),&
\end{flalign*}
\begin{flalign*}
f_4 &= \frac{1}{128} \left( 4 D_{3,0} + D_{4,0} \right),&
\end{flalign*}
\begin{flalign*}
f_5 &= \frac{1}{32} \left( 4 D_{1,0} + 14 D_{2,0} + 8 D_{3,0} + D_{4,0} \right),&
\end{flalign*}
\begin{flalign*}
f_6 &= \frac{1}{128} \left( 24 D_{1,0} + 36 D_{2,0} + 12 D_{3,0} + D_{4,0} \right),&
\end{flalign*}
\begin{flalign*}
f_{10} &= \frac{1}{4} \left( 2 D_{0,1} - 2 D_{1,1} - D_{2,1} \right),&
\end{flalign*}
\begin{flalign*}
f_{11} &= \frac{1}{32} \left( -4 D_{2,1} - 6 D_{3,1} - D_{4,1} \right),&
\end{flalign*}
\begin{flalign*}
f_{12} &= \frac{1}{32} \left( 4 D_{0,1} - 4 D_{1,1} - 22 D_{2,1} - 10 D_{3,1} - D_{4,1} \right),&
\end{flalign*}
\begin{flalign*}
f_{17} &= \frac{1}{64} \left( 12 D_{0,2} - 12 D_{1,2} + 6 D_{2,2} + 8 D_{3,2} + D_{4,2} \right),&
\end{flalign*}
where
\begin{equation}
D_{i,j} \equiv X^i \frac{d^i \, b_{1/2}^{(j)}}{d \, X^i}.
\end{equation}

Resonant terms that remain after averaging the expansion of the perturbing function,
\begin{eqnarray}
\overline{R}_{\idm{res}} & = & \left( e_1^2 \, f_{45} + e_1^4 \, f_{46} + e_1^2 \, e_2^2 \, f_{47} \right) \cos 2 \sigma_1 \nonumber\\
& & + \left( e_1 \, e_2 \, f_{49} + e_1^3 \, e_2 \, f_{50} + e_1 \, e_2^3 \, f_{51} \right) \cos(\sigma_1 + \sigma_2) \nonumber\\
& & + \left( e_2^2 \, f_{53} + e_1^2 \, e_2^2 \, f_{54} + e_2^4 \, f_{55} \right) \cos 2 \sigma_2 \\
& & + e_1^3 \, e_2 \, f_{68} \cos(3 \sigma_1 - \sigma_2) \nonumber \\
& & + e_1 \, e_2^3 \, f_{69} \cos(3 \sigma_2 - \sigma_1) \nonumber,
\end{eqnarray}
where coefficients are specified as follows:
\begin{flalign*}
f_{45} &= \frac{1}{8} \left( 279 D_{0,9} + 34 D_{1,9} + D_{2,9} \right),&
\end{flalign*}
\begin{flalign*}
f_{46} &= \frac{1}{96} \left( -66222 D_{0,9} - 8174 D_{1,9} + 69 D_{2,9} + 36 D_{3,9} + D_{4,9} \right), &
\end{flalign*}
\begin{flalign*}
f_{47} &= \frac{1}{32} \left( -90396 D_{0,9} - 10390 D_{1,9} + 97 D_{2,9} + 40 D_{3,9} + D_{4,9} \right), &
\end{flalign*}
\begin{flalign*}
f_{49} &= \frac{1}{4} \left( -272 D_{0,8} - 34 D_{1,8} - D_{2,8} \right), &
\end{flalign*}
\begin{flalign*}
f_{50} &= \frac{1}{32} \left( 59024 D_{0,8} + 7126 D_{1,8} - 139 D_{2,8} -38 D_{3,8} - D_{4,8} \right), &
\end{flalign*}
\begin{flalign*}
f_{51} &= \frac{1}{32} \left( 80528 D_{0,8} + 9130 D_{1,8} - 173 D_{2,8} - 42 D_{3,8} - D_{4,8} \right), &
\end{flalign*}
\begin{flalign*}
f_{53} &= \frac{1}{8} \left( 263 D_{0,7} + 34 D_{1,7} + D_{2,7} \right), & 
\end{flalign*}
\begin{flalign*}
f_{54} &= \frac{1}{32} \left( -51548 D_{0,7} - 6070 D_{1,7} + 209 D_{2,7} + 40 D_{3,7} + D_{4,7} \right),&
\end{flalign*}
\begin{flalign*}
f_{55} &= \frac{1}{96} \left( -70422 D_{0,7} - 7878 D_{1,7} + 249 D_{2,7} + 44 D_{3,7} + D_{4,7} \right),&
\end{flalign*}
\begin{flalign*}
f_{68} &= \frac{1}{96} \left( 99940 D_{0,10} + 11642 D_{1,10} - 21 D_{2,10} - 38 D_{3,10} - D_{4,10} \right),&
\end{flalign*}
\begin{flalign*}
f_{69} &= \frac{1}{96} \left( 43884 D_{0,6} + 5022 D_{1,6} - 279 D_{2,6} - 42 D_{3,6} - D_{4,6} \right).&
\end{flalign*}

%
\bibliographystyle{mn2e}
\bibliography{ms}
\label{lastpage}
%
\end{document}